\documentclass{aa}

\usepackage{graphicx}
\usepackage{txfonts}

\usepackage{subcaption}
\usepackage{lscape}
\usepackage{placeins}
\usepackage{hyperref}

\newcommand{\unc}[3]{\mbox{$#1^{\scriptscriptstyle+#2}_{\scriptscriptstyle-#3}$}}
\usepackage{booktabs}

\usepackage{url}

\begin{document}

   \title{Stellar evolution models and asteroseismic predictions for red giants in the Galactic nuclear star cluster}

       \author{
    C.~Machado\inst{1}
    \and D.~Capelo\inst{1}
    \and I.~Lopes\inst{1}
    }

    \institute{
    Centro de Astrofísica e Gravitação - CENTRA,
    Departamento de Física, Instituto Superior Técnico,
    Universidade de Lisboa}

   \date{Received September 30, 20XX}

    \abstract
    {The nuclear star cluster (NSC) of the Milky Way is the closest galactic nucleus in which individual stars can be resolved, yet the formation history of its population remains poorly understood.
    Upcoming near-infrared missions will make oscillations of NSC stars observable for the first time.}
    {We aim to model individual NSC red giants with high-resolution spectroscopy and predict their asteroseismic properties as a reference for future detections.}
    {We derived the initial mass, metallicity, and age of 23 stars (27 entries from four studies) with a Bayesian Markov chain Monte Carlo calibration against the PARSEC isochrones. With these parameters, we computed a track for each star with MESA, which provides the internal structure needed, adopting a solar-calibrated mixing-length parameter, $\alpha_{\rm MLT}$, and extracted the frequency of maximum oscillation power ($\nu_{\rm max}$), the large frequency separation ($\Delta\nu$, validated with GYRE), and the dipole gravity-mode period spacing ($\Delta\Pi$). For stars whose position in the Hertzsprung--Russell diagram is not reproduced, we fitted a star-specific mixing-length parameter, $\alpha_{\rm MLT}^*$, and considered more advanced evolutionary phases.}
    {The nominal models reproduce 19 stars, with initial masses of $1.07$--$1.47$\,$M_\odot$, ages of $3.3$--$8.3$\,Gyr, metallicities of $-0.31$ to $+0.43$\,dex, $\nu_{\rm max}\approx0.23$--$0.97\,\mu$Hz, $\Delta\nu\approx0.10$--$0.32\,\mu$Hz, and $\Delta\Pi\approx29$--$38$\,s. Of the other four, three are reproduced with a super-solar $\alpha_{\rm MLT}^*$; for the fourth, a red-giant-branch solution with a super-solar $\alpha_{\rm MLT}^*$ is about four times more probable than an asymptotic-giant-branch solution, which predicts a different $\nu_{\rm max}$.}
    {The predicted $\nu_{\rm max}$ correspond to periods of about 12--50\,d, within reach of JASMINE and the Roman Space Telescope, which can therefore test the calibration; $\Delta\Pi$ remains a theoretical diagnostic. Since $\alpha_{\rm MLT}^*$ can absorb both modelling deficiencies and temperature-scale offsets, it should be regarded as a diagnostic rather than a measurement; asteroseismology will be needed to break this degeneracy.}


   \maketitle
\nolinenumbers
\section{Introduction}
\label{section_1}
The nuclear star cluster (NSC) of the Milky Way is a dense, massive stellar system surrounding the supermassive black hole Sagittarius~A$^{*}$ (Sgr~A$^{*}$), located at a distance of
$8.275 \pm 0.034$~kpc from the Sun \citep{GRAVITY_2021}, with a half-light radius of
$\sim\!4$--5~pc \citep{Gallego-Cano_2026} and a total stellar mass of $\sim\!(2.5$--$4) \times 10^{7}$~M$_\odot$
\citep{Schodel_2014, Fritz_2016}.
Its proximity makes it the only galactic nucleus in the Universe in which individual stars can be resolved down to milliparsec scales \citep{Nogueras-Lara_2018, Gallego-Cano_2026}, giving the Galactic NSC a unique role as a laboratory for understanding (extra)galactic stellar nuclei and their place in the broader context of galaxy evolution \citep{Nogueras-Lara_2018, Thorsbro_2020}.
Despite this privileged observational access, the formation and evolutionary history of the NSC still remain poorly understood \citep{FK_2017, Chen_2023}. About 80\% of its stellar mass formed 10~Gyr ago or longer, and an episode about 3~Gyr ago contributed about 15\% \citep{Schodel_2020}; the NSC may also be connected to the surrounding nuclear stellar disc (NSD; \citealt{Nogueras-Lara_2023}). Its study is limited by the high and variable extinction, with a mean $A_{K_s} = 2.74 \pm 0.30$~mag in the central parsec \citep{Schodel_2010}, and by extreme crowding \citep{Nogueras-Lara_2018}.
Near-infrared spectroscopic surveys over the past decade have revealed a stellar population of evolved late-type giants spanning a wide range of metallicities, from significantly sub-solar to highly super-solar values
\citep[e.g.][]{Do_2015, FK_2017}.
This broad metallicity dispersion is difficult to reconcile with a single-population formation scenario and instead raises fundamental questions about the star formation history of the NSC: whether the population formed in situ at the Galactic centre, whether it assembled through the infall and disruption
of globular clusters, or whether it reflects a combination of both channels \citep{FK_2017, Chen_2023}.
Establishing the metallicity distribution of the NSC with
sufficient precision to distinguish between these scenarios is therefore one of the central open problems in Galactic centre astrophysics \citep{Rich_2017, Chen_2023}.
Progress has been complicated by the fact that the largest available stellar samples come from medium-spectral-resolution surveys ($R\sim 4000$-$5000$) \citep[e.g.][]{Do_2015, FK_2017}, whose inferred surface gravities are systematically unreliable for cool, evolved giants, limiting the physical interpretation of the derived
parameters \citep{Rich_2017}.
High-spectral-resolution observations ($R \gtrsim 23\,000$) overcome this limitation but are restricted to much smaller samples
\citep[e.g.][]{Rich_2017, Do_2018, Thorsbro_2020, Ryde_2025}.
The observational data sets used in this work are described in detail in Sect.~\ref{section_2}.

A powerful and largely unexplored approach to characterising
the NSC stellar population is asteroseismology.
Solar-like oscillations in red giants are stochastically excited
by near-surface convection and produce an oscillation spectrum whose global properties, such as the frequency of maximum
oscillation power, $\nu_{\rm max}$, and the large frequency
separation, $\Delta\nu$, are directly related to the stellar
surface gravity and mean density through scaling
relations \citep{Brown_1991, Kjeldsen_1995}:
\begin{equation}
    \frac{\nu_{\rm max}}{\nu_{\rm max,\odot}} \simeq
    \frac{g/g_\odot}{(T_{\rm eff}/T_{\rm eff,\odot})^{1/2}},
    \qquad
    \frac{\Delta\nu}{\Delta\nu_\odot} \simeq
    \left(\frac{\bar{\rho}}{\bar{\rho}_\odot}\right)^{1/2},
    \label{eq:scaling}
\end{equation}
where $\bar{\rho}$ denotes the mean stellar density and solar
reference values are $\nu_{\rm max,\odot} = 3090~\mu$Hz and
$\Delta\nu_\odot = 135.1~\mu$Hz \citep{Huber_2011}.
These relations provide mass and radius estimates that are independent of stellar evolution models, making them
a valuable complement to isochrone-based methods
\citep{Asteroseismology_book}.
The scaling relation for $\Delta\nu$ follows from its asymptotic expression, the inverse of twice the sound-crossing time of the star \citep{Tassoul_1980},
\begin{equation}
    \Delta\nu \simeq \left(2\int_0^R \frac{\mathrm{d}r}{c_s}\right)^{-1},
    \label{eq:dnu_asymp}
\end{equation}
where $c_s$ is the adiabatic sound speed.
The mixed-mode period spacing, $\Delta\Pi$, arising from
gravity-dominated oscillation modes that propagate into the
stellar core,
is extremely
sensitive to the internal structure, allowing for a direct
distinction to be made between hydrogen-shell-burning red giant branch
(RGB) stars and helium-core-burning red clump stars, which
occupy nearly the same place in the Hertzsprung-Russell (HR) diagram
but differ substantially in age and evolutionary history
\citep{Bedding_2011, Mosser_2012}.
It is related to the Brunt-V\"{a}is\"{a}l\"{a} frequency
profile, $N(r)$, via the asymptotic relation
\citep{Tassoul_1980}:
\begin{equation}
    \Delta\Pi_{\ell=1} = \frac{2\pi^2}{\sqrt{2}}
    \left(\int_{r_1}^{r_2}
    \frac{N}{r}\,\mathrm{d}r\right)^{-1},
    \label{eq:DeltaPi}
\end{equation}
where the integral extends over the gravity-wave propagation
cavity $[r_1,\, r_2]$.
Space-based photometric surveys such as Kepler/K2
\citep{Borucki_2010, Howell_2014} and TESS
\citep{Ricker_2015} have enabled the systematic measurement
of stellar masses, radii, and ages across the Galaxy
\citep{Chaplin_2013, Miglio_2013}, but none have observed
the Galactic centre, whose extreme line-of-sight extinction
and crowding lie beyond their reach.
This is set to change with the near-infrared missions
JASMINE \citep{kawata2024_jasmine} and the Nancy Grace Roman
Space Telescope \citep{Huber_2023, Weiss_2025}, both capable of reaching stars behind this extinction
(Sect.~\ref{sec:disc_prospects}).
Theoretical predictions of $\nu_{\rm max}$, $\Delta\nu$, and
$\Delta\Pi$ for a well-characterised sample of NSC red
giants are therefore a necessary preparatory step to define
the expected signal properties and the most promising
targets.

In this work, we compute individual stellar evolution models for a sample of red giant stars in the inner few parsecs of the Milky Way, using the stellar evolution code MESA \citep{Paxton2011, Paxton2013, Paxton2015, Paxton2018, Paxton2019, Jermyn2023}, calibrated against the PARSEC isochrone grid \citep{Bressan_2012}.
For each star, we derive the initial mass, $M_{\rm ini}$, and initial
metallicity, $Z_{\rm ini}$, through Bayesian inference, combining the spectroscopic observables from the high-resolution catalogues with photometric and distance constraints (Sect.~\ref{section_3}).
From the calibrated MESA models we extract predictions of
$\nu_{\rm max}$, $\Delta\nu$, and $\Delta\Pi$ via 
Eqs.~(\ref{eq:scaling}), (\ref{eq:dnu_asymp}), and (\ref{eq:DeltaPi}), respectively, implemented in MESA.
We validate our $\Delta\nu$ predictions by computing oscillation-mode frequencies directly with the stellar
oscillation code GYRE \citep{Townsend_2013} for the full
sample.
We further investigate stars for which the nominal solar mixing length does not yield satisfactory model fits, exploring the effects of varying the convective mixing-length parameter $\alpha_{\rm MLT}$ \citep{Joyce_2018} and, for stars whose mismatch this does not resolve, testing whether relaxing the RGB-only assumption of the isochrone calibration to allow a post-RGB evolutionary phase accounts for the discrepancy.
The paper is organised as follows: Sect.~\ref{section_2}
describes the observational data sets; Sect.~\ref{section_3}
presents the calibration and modelling methodology;
Sect.~\ref{section_4} presents the results for the
well- and poorly calibrated stars; Sect.~\ref{sec:discussion} discusses their interpretation; and Sect.~\ref{section_5}
summarises our findings and future prospects.
\section{Observational data}
\label{section_2}
The stellar sample studied in this work draws on six independent near-infrared spectroscopic catalogues of late-type giant stars located within the inner few parsecs of the Galactic centre.
We refer to the two medium-spectral-resolution surveys
\citep{Do_2015, FK_2017} collectively as the medium-resolution data sets, and to the four high-spectral-resolution surveys \citep{Rich_2017, Do_2018, Thorsbro_2020, Ryde_2025} as the
high-resolution data sets.
\subsection{Medium-resolution data sets}
\label{sec:medium_res}
\citet{Do_2015} present adaptive-optics (AO) assisted observations of 83 red-giant stars within a projected radius of $\sim 1$~pc from Sgr~A$^{*}$, and covering a total surface area of $\sim 0.15$~pc$^{2}$ at a spectral resolution of $R \sim 5000$, obtained in the $K$ band with the NIFS integral-field spectrograph \citep{McGregor_2003} at the Gemini North telescope.
The resulting sample spans a broad metallicity range of
$-1.27 < [{\rm M/H}] < 0.96$, with a mean measurement uncertainty of $\sigma_{[{\rm M/H}]} = 0.32$~dex.
Obvious foreground stars were excluded on the basis of their colours, but the membership of individual stars was not established (Sect.~\ref{sec:membership}).

\citet{FK_2017} extended the spatial coverage of NSC metallicity studies with a seeing-limited spectroscopic survey of more than 700 late-type stars within a projected radius of $\sim 1.5$~pc from Sgr~A$^{*}$, at a spectral resolution of $R \sim 4000$, obtained in the $K$ band with the KMOS integral-field spectrograph \citep{Sharples_2013} at the ESO Very Large Telescope,
covering a total field area of $\sim 4$~pc$^{2}$.
The resulting metallicity distribution spans
$-1.25 < [{\rm M/H}] < 1.00$, consistent with the range found by \citet{Do_2015}, with a mean uncertainty of $\sigma_{[{\rm M/H}]} = 0.27$~dex. Foreground stars were rejected on the basis of their extinction-corrected $H-K_s$ colours (Sect.~\ref{sec:membership}).
The spatial footprints of the two medium-resolution surveys occupy distinct but overlapping regions of the NSC \citep{Chen_2023}.

While the two medium-resolution surveys provide valuable statistics on the NSC stellar population, a well-known limitation of spectral analysis at $R \sim 4000$-$5000$ is that surface gravities cannot
be reliably determined from near-infrared K-band spectra alone \citep{FK_2017, Rich_2017}.
This is reflected in the catalogued $\log g$ values: in
\citet{Do_2015} the surface gravities reach as high as
$\log g \approx 3.98$, with a sample mean of $\langle \log g \rangle \approx 3.42$; in \citet{FK_2017} extreme values as high as $\log g \approx 5.6$ are present, although the majority of the sample is more consistent with the expected values for evolved giants
\citep{Rich_2017}, with a sample mean of
$\langle \log g \rangle \approx 0.5$.
Such high gravity values are physically inconsistent with the standard picture of the luminous, cool red giants with extended envelopes that populate these catalogues: inserting them into the standard mass-luminosity-gravity relation, where $\log g_\odot = 4.437$~dex is the solar surface gravity,
\begin{equation}
    \frac{M}{M_\odot} = \frac{L}{L_\odot}
    \left(\frac{T_{\rm eff,\odot}}{T_{\rm eff}}\right)^{4}
    10^{\log g - 4.437}
    \label{eq:mass_logg}
\end{equation}
yields inferred masses of order $10^{2}$-$10^{3}~M_\odot$, which are not physically plausible for single evolved stars.
As \citet{Rich_2017} note, deriving reliable stellar parameters from low-S/N, heavily blended cool-star spectra at this resolution is an extremely challenging task: the strong molecular lines that dominate the spectra are simultaneously saturated and blended to a degree that precludes trustworthy traditional abundance analysis.
For this reason, and as further discussed in
Sect.~\ref{section_3}, the physical modelling and calibration presented in this work relies primarily on the high-resolution catalogues.
\subsection{High-resolution data sets}
\label{sec:high_res}
High-spectral-resolution ($R \gtrsim 23\,000$) observations
remove the blending problem that affects the medium-resolution catalogues, enabling the accurate simultaneous
determination of $T_{\rm eff}$, $\log g$, $[{\rm Fe/H}]$, and individual elemental abundances.
In this work we only considered the stars labelled as NSC members in the source catalogues; the membership criteria are described in Sect.~\ref{sec:membership}.
The four catalogues collectively supply 27 catalogue entries, corresponding to 23 distinct physical stars once duplicate measurements of the same star across catalogues are accounted for (see Sect.~\ref{sec:mcmc}).

\citet{Rich_2017} present the first high-resolution
($R \approx 24\,000$) near-infrared spectroscopic study targeting the NSC specifically, for 18 M-giant stars, obtained in the $K$ band with NIRSPEC \citep{McLean_1998} at the Keck~II telescope
with a projected radius of $\sim 3.6$ pc from Sgr~A$^{*}$.
In this work we excluded GC7688, identified as a foreground star by \citet{Rich_2017}, and GC7104, GC10812, and GC11473, which \citet{Thorsbro_2020} classify as members of the NSD rather than the NSC.
The most metal-rich members of their sample reach $[\mathrm{Fe/H}] \lesssim 0.6$, with the majority of stars at or below solar metallicity, substantially lower than the high-metallicity tail inferred from the medium-resolution surveys.
\citet{Rich_2017} report two sets of stellar parameters for each star in their Table~2, with the same CO-based effective temperature but different surface gravities, derived either photometrically or from a theoretical isochrone, and correspondingly different [Fe/H]. We treated the two sets as separate calibration inputs, labelled with the suffixes $_\text{phot}$ and $_\text{iso}$, respectively.
This doubles the number of entries from the
\citet{Rich_2017} catalogue to 28, representing
14 physical stars.
Since all 14 NSC members of the \citet{Rich_2017}
sample are also present in \citet{Thorsbro_2020} and \citet{Ryde_2025}
catalogues,
the latter measurements constitute the primary calibration input for these stars in this work, and all subsequent modelling and asteroseismic predictions (Sects.~\ref{sec:mesa_models} and~\ref{sec:difficult_cases}) are computed exclusively from them.
The \citet{Rich_2017} variants are included as a
secondary diagnostic: as noted in
Sect.~\ref{sec:bad_fits}, several \citet{Rich_2017}
targets have comparatively high effective temperatures
and surface gravities, similar to the
Feld31 and Feld84 stars from \citet{Ryde_2025}, and we wish to assess whether the
same calibration failure modes appear in
\citet{Rich_2017} stars.

\citet{Do_2018} targeted two specific stars from the \citet{Do_2015} catalogue (NE1-1\,001 and NE1-1\,002) that had been flagged as outliers with unusually high template-fitted metallicities, at a spectral resolution of $R \approx 25\,000$.
\citet{Do_2018} obtained $K$-band spectra at $R \sim 25\,000$ with NIRSPEC behind the laser-guide-star adaptive optics system of the Keck~II telescope. \citet{Do_2018} confirm that NE1-1\,002 is indeed a high-metallicity giant,
while NE1-1\,001 is found to be only slightly above solar metallicity, suggesting that the most extreme values in \citet{Do_2015} could be partly due to the difficulties of medium-resolution fitting at high metallicity; such metallicities can, however, be calibrated against high-resolution measurements to about 0.2--0.3~dex \citep{Fritz_2021, Schultheis_2026}.
We did not consider NE1-1\,002 in the analysis because a fixed value for [M/H] is not catalogued, only a lower boundary. For NE1-1\,001, \citet{Do_2018} fix $\log g = 1.0$ to a
typical value for giants at the observed luminosity and
temperature, rather than measuring it spectroscopically; in this work
we adopted a correspondingly conservative uncertainty of
$\sigma_{\log g} = 0.5$ dex to reflect this. \citet{Thorsbro_2020} identify NE1-1\,001
as the same physical star as GC25.
Since GC25 is in the well-calibrated primary sample with a satisfactory MESA model fit, we did not treat NE1-1\,001 as an independent entry in the analysis.

\citet{Thorsbro_2020} present a detailed chemical abundance analysis of 17 M-giant stars in the NSC, based on $K$-band NIRSPEC spectra ($R \approx 23\,000$) from the Keck~II telescope, including the 14 stars from \citet{Rich_2017},
at $R \approx 23\,000$.
The derived metallicities range from $[{\rm Fe/H}] = -0.32$ to $+0.49$.
\citet{Thorsbro_2020} conclude that the Galactic centre does not host stars with extreme chemical compositions.

Finally, \citet{Ryde_2025} present the highest-resolution
spectroscopic study of red giant stars in the NSC
currently available, delivering spectra at
$R \approx 45\,000$ for nine M-giant stars, obtained in the $H$ and $K$ bands with IGRINS \citep{Park_2014} at the Gemini South telescope.
\citet{Ryde_2025} find predominantly super-solar metallicities and enhanced $\alpha$-element abundances, leading them to conclude that the NSC stellar population likely shares a similar chemical enrichment history with the inner Galactic bulge, challenging the hypothesis of a recent, dominant in situ star-formation burst as the primary channel for NSC assembly.
In this work, we included all nine stars from \citet{Ryde_2025} without exclusion, as their NSC membership is supported by several criteria (Sect.~\ref{sec:membership}).

\begin{figure}[h!]
    \centering
    \includegraphics[width=\columnwidth]
        {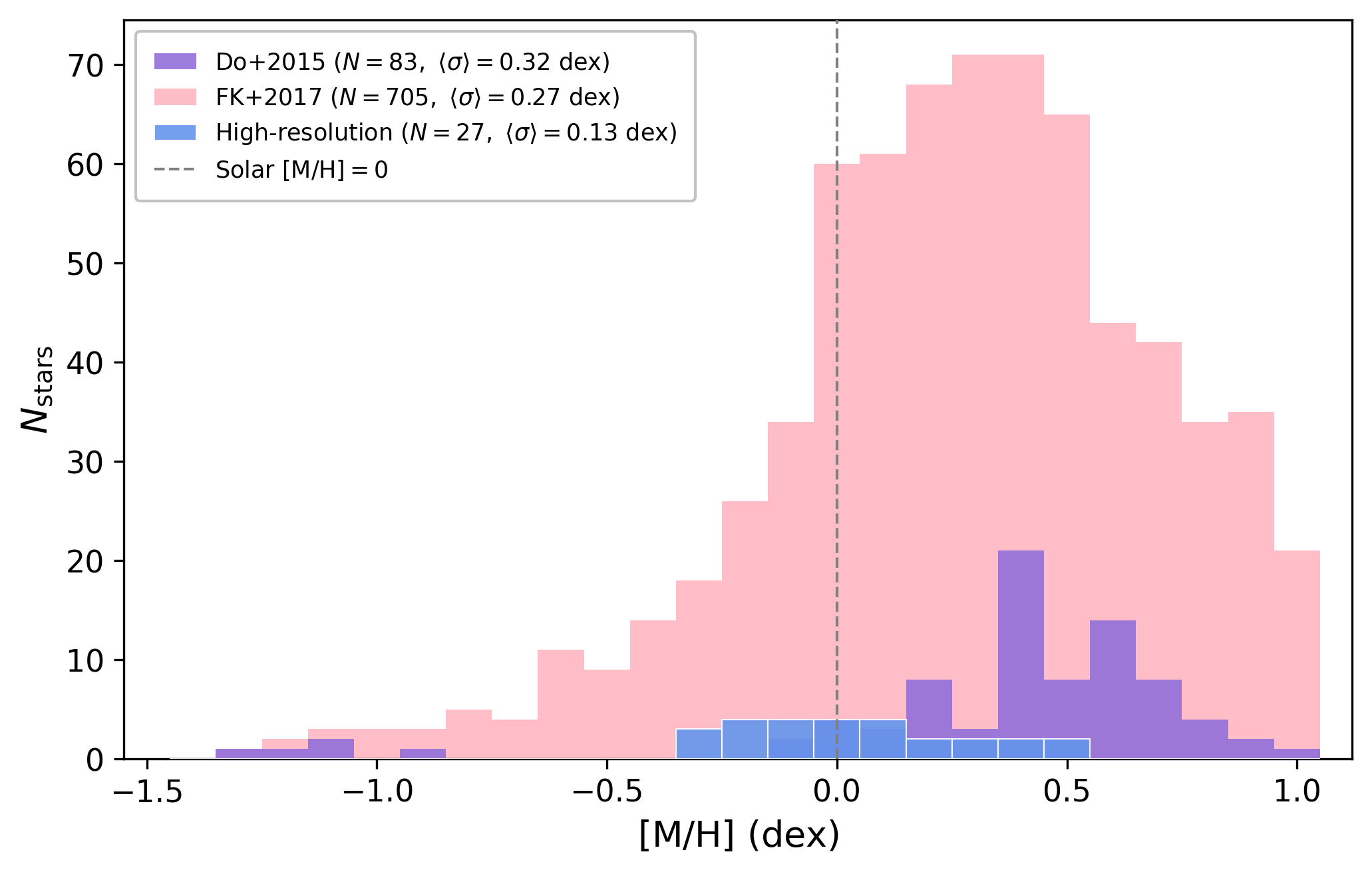}
    \caption{Spectroscopic metallicity distributions of
             the NSC stellar samples used in this work.
             The pink and purple histograms show the medium-resolution
             sample from \citet{FK_2017} and
             \citet{Do_2015}, respectively.
             The blue histogram shows the high-resolution
             sample from \citet{Rich_2017},
             \citet{Do_2018}, \citet{Thorsbro_2020}, and
             \citet{Ryde_2025}.
             The medium-resolution values are total metallicities, [M/H], and the high-resolution values iron abundances, [Fe/H]; no conversion is applied (Sect.~\ref{sec:mcmc}).
             The vertical dashed line marks the solar value.
             Bin width is 0.1~dex.
             }
    \label{fig:mh_hist_obs}
\end{figure}

Figure~\ref{fig:mh_hist_obs} shows the distribution of spectroscopic metallicities across all catalogues used in
this work.
The medium-resolution surveys together comprise 788
stars from \citet{Do_2015} and \citet{FK_2017}
and span a broad metallicity range of $-1.25 \leq [\mathrm{M/H}] \leq
+1.00$ with a mean measurement uncertainty of
$\langle \sigma_{[\mathrm{M/H}]} \rangle = 0.27$~dex.
The high-resolution sample of 23 stars from
\citet{Rich_2017}, \citet{Do_2018}, \citet{Thorsbro_2020},
and \citet{Ryde_2025} spans the narrower range
$-0.32 \leq [\mathrm{Fe/H}] \leq +0.49$.
The majority of stars in both samples are at or above solar
metallicity, with a pronounced super-solar peak near
$[\mathrm{M/H}] \approx +0.1$--$+0.3$ in the
medium-resolution distribution.
This peak agrees with the analyses of \citet{Schultheis_2021} and \citet{Nogueras-Lara_2022}, who found the NSC to host a metal-rich population at $[\mathrm{M/H}] \sim 0.3$~dex and a broader one slightly below solar.
This comparison is only qualitative: the catalogues differ in resolution, spectral grids, and fitting methods, and the metallicities of \citet{FK_2017} were calibrated only for $[\mathrm{M/H}] < 0.3$~dex, so that higher values are likely overestimated \citep{Schultheis_2021, Nogueras-Lara_2022}; \citet{Do_2015} likewise caution that their metallicities are most uncertain above solar.

The broad metal-poor tail extending to $[\mathrm{M/H}] \lesssim -1.0$ in the medium-resolution surveys is not reproduced in the high-resolution sample and may in part reflect systematic uncertainties in the spectral fitting at low signal-to-noise ratio and moderate spectral resolution \citep{Rich_2017}. Other effects may also contribute: the selection of luminous M giants with strong CO band heads in the high-resolution programmes, the exclusion of the most metal-poor star of \citet{Thorsbro_2020}, GC10812 ($[\mathrm{Fe/H}] \approx -1.0$), as an NSD member, the small size and uncharacterised completeness of the high-resolution samples (Sect.~\ref{sec:membership}), and the possible inclusion of less metal-rich NSD and bulge stars in the medium-resolution samples \citep{Nogueras-Lara_2022}.
We note that the metallicity values shown in
Fig.~\ref{fig:mh_hist_obs} are the spectroscopic
measurements reported in each catalogue, that is, total metallicities, [M/H], for the medium-resolution surveys and iron abundances, [Fe/H], for the high-resolution studies. Throughout this paper, [Fe/H] denotes spectroscopic iron abundances and [M/H] total metallicities, including those obtained from our isochrone calibration.
The metallicities obtained from our isochrone calibration (Sect.~\ref{sec:mcmc}) are discussed in Sect.~\ref{sec:good_fits} for the 19 distinct stars whose observed position is reproduced by the nominal models.
\subsection{NSC membership, contamination, and sample selection}
\label{sec:membership}
The line of sight towards the NSC also crosses the NSD and the Galactic bulge and bar, and none of the catalogues used here provides membership probabilities. In the medium-resolution samples, foreground stars were removed by colour \citep{St_stad_2015, FK_2017}; \citet{Do_2015} estimated that about 8 of their 83 stars may be bulge contaminants, and \citet{Nogueras-Lara_2022} that $\lesssim$10\% of the KMOS stars belong to the NSD. \citet{Rich_2017} and \citet{Thorsbro_2020} assigned each high-resolution star to the NSC, the NSD, or the Galactic disc from its $H-K_s$ colour, radial velocity, and proper motion, flagging GC6227 and GC10195 as possible background stars, and \citet{Ryde_2025} combined the extinction, an $H-K_s$ criterion, a diffuse interstellar band, and orbit integrations. We retained only NSC members.

As a homogeneous check, we used the GALACTICNUCLEUS extinctions (Sect.~\ref{sec:mcmc}), since NSD and NSC stars have $A_{K_s} \approx 1.7$ and $\approx 2.3$~mag, respectively \citep{Nogueras-Lara_2021c, Nogueras-Lara_2022}. The three NSD stars of \citet{Thorsbro_2020} have 1.70--1.86~mag and 18 of our 23 stars 2.3--3.0~mag; the other five (GC25, GC16763, FK48, FK87, and FK5020265), the only stars of our sample in the first extinction layer of GALACTICNUCLEUS (Sect.~\ref{sec:mcmc}), have 1.9--2.2~mag, but their kinematics support NSC membership \citep{Thorsbro_2020, Ryde_2025}. Excluding these five stars, together with GC6227 and GC10195, leaves the ranges of all calibrated and predicted quantities of the well-calibrated sample unchanged.

Also, the high-resolution sample is small and was not designed to be representative of the NSC: \citet{Rich_2017} selected bright M giants with measurable CO band heads, avoiding the central 30\arcsec, known asymptotic-giant-branch (AGB) stars, long-period variables, supergiants, and stars cooler than about 3300~K. We therefore do not use it to infer population properties of the NSC, but provide star-by-star predictions for its brightest asteroseismic targets.

\begin{figure}
    \centering
    \includegraphics[width=\columnwidth]{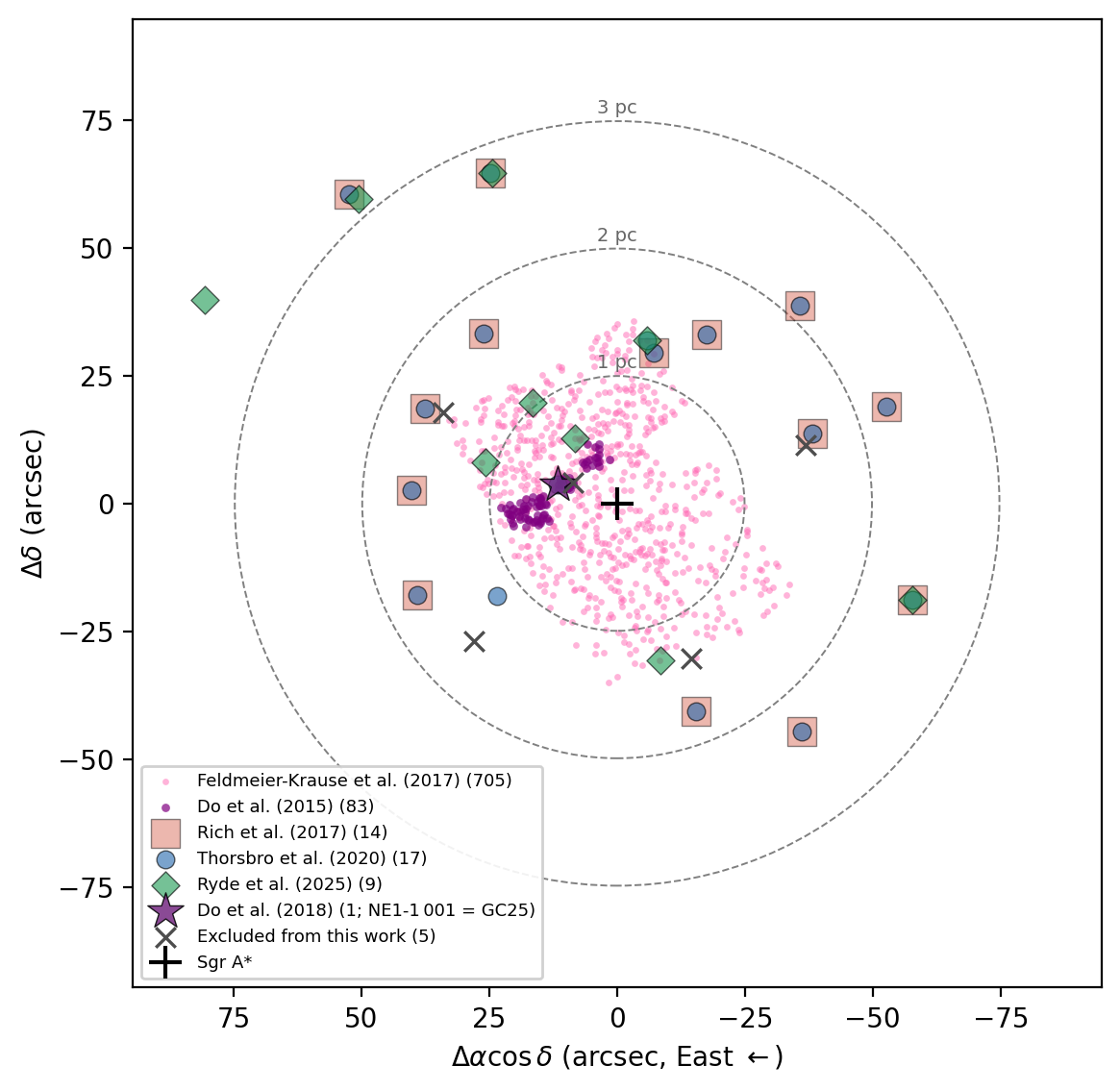}
    \caption{Projected positions, relative to Sgr~A$^*$ (black cross), of the stars in each catalogue, with the colours and symbols of Fig.~\ref{fig:parsec_grid_coverage}. Grey crosses mark the five excluded entries (Sect.~\ref{sec:high_res}). Dashed circles indicate projected distances of 1, 2, and 3~pc.}
    \label{fig:spatial_coverage}
\end{figure}

Figure~\ref{fig:spatial_coverage} shows the projected positions of the stars: the medium-resolution samples cover the central $\sim$1.5~pc (median 0.6 and 0.8~pc), and the high-resolution stars 0.5--3.6~pc (median 1.7~pc). Our calibrated [M/H] shows no significant trend with projected distance (slope $+0.04 \pm 0.06$~dex\,pc$^{-1}$; Spearman $\rho = 0.00$), about $2\sigma$ from the radial gradient of $-0.10 \pm 0.02$~dex\,pc$^{-1}$ reported by \citet{Schultheis_2026} \citep[see also][]{Nogueras-Lara_2023}. Such a gradient would make the medium-resolution samples only $\sim$0.1~dex more metal-rich than ours, which does not affect our results.

\section{Methodology}
\label{section_3}
First, we calibrated the initial stellar parameters of
each star by fitting a Bayesian Markov chain Monte Carlo
(MCMC) model to the spectroscopic and photometric
observables, using the PARSEC isochrone grid as a
theoretical reference
(Sect.~\ref{sec:parsec_calibration}).
Second, we performed a solar calibration of the MESA input
physics to determine a self-consistent mixing-length
parameter, $\alpha_{\rm MLT,\odot}$
(Sect.~\ref{sec:solar_calibration}).
Third, we constructed individual stellar evolution models
with the MESA code using the calibrated parameters and extracted
asteroseismic predictions from them
(Sect.~\ref{sec:mesa_models}), validating the
reliability of the resulting $\Delta\nu$ values against
oscillation-mode frequencies computed directly with the
stellar oscillation code GYRE.
Fourth, for stars whose nominal calibration does not
produce a satisfactory model fit, we explored two possible
explanations: a star-specific mixing-length parameter
(Sect.~\ref{sec:difficult_cases}), and, for stars this
does not resolve, a more advanced evolutionary phase than
assumed by the standard RGB-only calibration
(Sect.~\ref{sec:bad_fits}).

\subsection{Isochrone calibration with PARSEC}
\label{sec:parsec_calibration}
Theoretical reference isochrones were drawn from the
PARSEC stellar evolution library \citep{Bressan_2012},
version~1.2S \citep{Tang_2014, Chen_2014, Chen_2015},
downloaded via the CMD~3.9 web
interface\footnote{\url{http://stev.oapd.inaf.it/cmd}}
for the metallicity range $-1.2 \leq [\mathrm{M/H}]
\leq +0.70$ (steps of 0.1~dex) and ages
$9.0 \leq \log(\mathrm{Age/yr}) \leq 10.12$ (steps of
0.04~dex), for a total of 580 isochrones. This lower age limit is consistent with the star formation history of the NSC \citep{Schodel_2020} and does not truncate any posterior (the 16th percentile of every age posterior lies above $\log(\mathrm{Age/yr}) = 9.2$).
Mass loss on the RGB follows the Reimers prescription
\citep{Reimers_1975} with $\eta_{\rm Reimers}=0.2$,
matched in our MESA models
(Sect.~\ref{sec:mesa_models}) for consistency between
the calibration and evolution stages; photometry was computed in the 2MASS $JHK_s$ system \citep{Cohen_2003}, and the resulting $K_s$ magnitudes were compared with the observed absolute magnitudes in the calibration likelihood (Sect.~\ref{sec:mcmc}), where they provide the luminosity constraint;
with dust and long-period variability treated following
\citet{Marigo_2008} and \citet{Trabucchi_2021},
respectively, and the \citet{Kroupa_2001} initial mass
function (IMF) entering the calibration likelihood as a
measure (Sect.~\ref{sec:mcmc}).
From the downloaded grid we retained only evolutionary
points with PARSEC label~$=3$ (RGB), except for a
small subset of stars discussed in
Sect.~\ref{sec:bad_fits} for which this restriction
is relaxed.
Full details of the isochrone interpolation scheme are
given in Appendix~\ref{appendix:interpolation}.

Figure~\ref{fig:parsec_grid_coverage} shows the
distribution of all catalogues used in this work in
the $(T_{\rm eff},\, \log g)$ plane, together with the
PARSEC RGB grid (label~$= 3$).
The high-resolution sample (Sect.~\ref{sec:high_res})
lies almost entirely within the RGB locus of the grid,
while the medium-resolution sample contains a large
fraction of stars with surface gravities well above the
RGB sequence, reflecting the known overestimation of
$\log g$ in medium-resolution K-band spectroscopy
\citep{Rich_2017, Do_2018} and the reason
why the calibration presented here cannot be applied to
these catalogues. The medium-resolution samples are therefore used only for context and do not enter the calibration; applying the method to them would require dropping $\log g$ from the likelihood, at the cost of weaker constraints, which we leave for future work.
The four high-resolution stars for which the nominal MESA track does not reproduce the observed position under the standard
solar mixing length and RGB-only assumption are: Feld31, Feld84,
GC8631, and GC8623.
Although all four lie inside the grid, they are the
anomalous cases mentioned in Sect.~\ref{sec:difficult_cases},
discussed further in Sect.~\ref{sec:bad_fits}.
NE1-1\,001 is
physically outside the grid but corresponds to the same
physical star as GC25 \citep{Thorsbro_2020}, and is
therefore not treated as an independent case.
\begin{figure*}[h!]
    \centering
    \includegraphics[width=\textwidth]
        {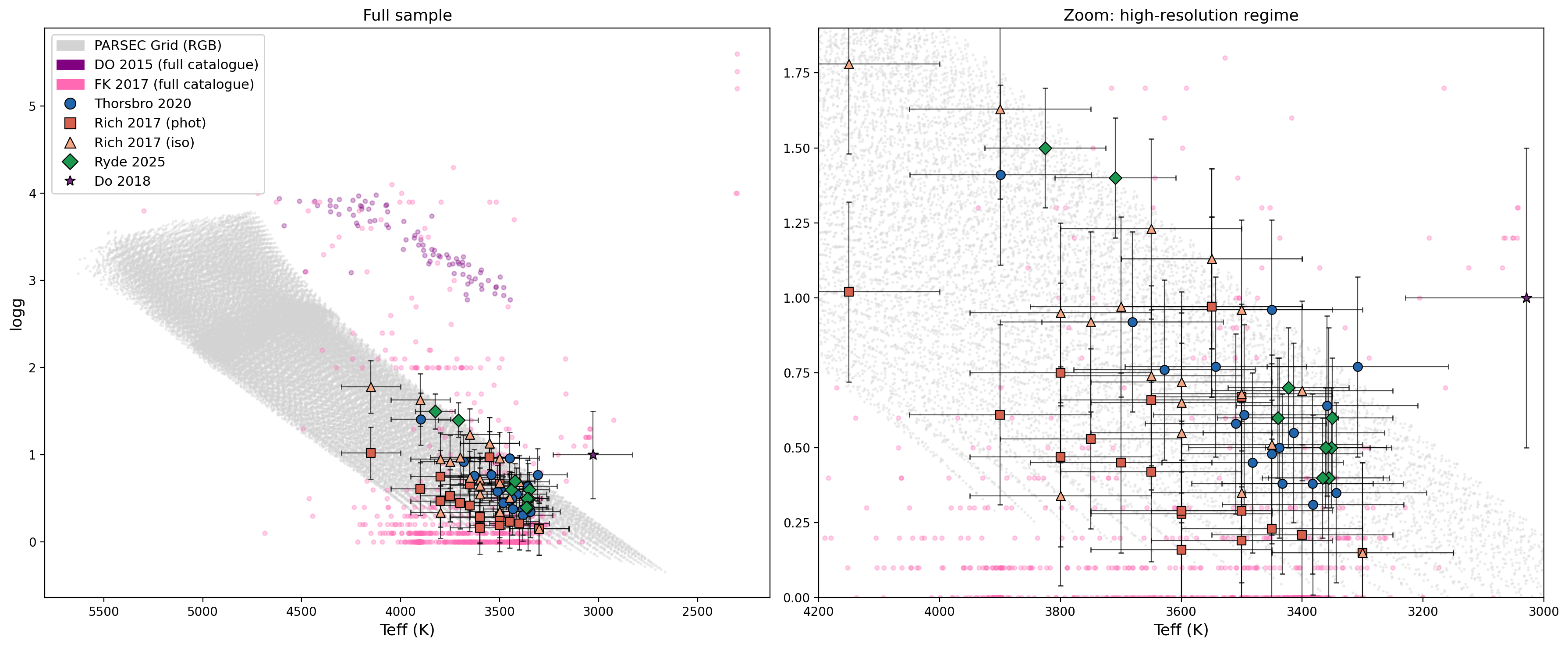}
    \caption{Distribution of all catalogues in the
             $(T_{\rm eff},\, \log g)$ plane, overlaid
             on the PARSEC grid (label~$= 3$) used for the isochrone
             calibration.
             Left: Full sample.
             Right: Zoom on the high-resolution regime.
             Purple circles and pink circles represent the \citet{Do_2015} and
             \citet{FK_2017} samples.
             High-resolution stars are shown by
             coloured symbols: blue circles for
             \citet{Thorsbro_2020}, red squares for
             \citet{Rich_2017} (photometric parameters),
             salmon triangles for \citet{Rich_2017}
             (isochrone parameters), green diamonds for
             \citet{Ryde_2025}, and a purple star for
             \citet{Do_2018}.
             Error bars show $1\sigma$ uncertainties
             in $T_{\rm eff}$ and $\log g$.}
    \label{fig:parsec_grid_coverage}
\end{figure*}
As a consistency check between the isochrone-based
calibration and the MESA evolutionary models used
throughout this work, we compared, for each of the 27
catalogue entries, the MESA-predicted $T_{\rm eff}$ on
the RGB with the $T_{\rm eff}$ of the closest PARSEC grid
point of identical $(M_{\rm ini},\, Z_{\rm ini})$, at
matched luminosity.
For the 15 well-calibrated catalogue entries in Table~\ref{tab:mcmc_results} (out of the 22 entries corresponding to the 19 well-calibrated distinct stars; Sect.~\ref{sec:good_fits})
for which a common luminosity exists on the MESA track and on the closest PARSEC RGB isochrone
(label~$=3$), we find a median offset of
$+16$~K, with $|\Delta T_{\rm eff}| \leq 50$~K for 13 of them, well below the spectroscopic $T_{\rm eff}$ uncertainties of 100--150~K, confirming that the MESA
models reproduce the PARSEC isochrones used in their own
calibration.
\subsection{Bayesian MCMC calibration}
\label{sec:mcmc}
For each star, the two free parameters are the
logarithmic age, $\log \mathrm{Age}$, and the overall
metallicity, $[\mathrm{M/H}]$.
For each proposed $(\log \mathrm{Age},\, [\mathrm{M/H}])$,
the likelihood is obtained by marginalising over the
initial mass along the RGB isochrone, with the IMF, $\xi(m)$, as the prior measure:
\begin{equation}
   \ln\mathcal{L} = \ln \sum_{i=1}^{N_{\rm iso}}
      w_i\,\exp\!\left(-\tfrac{1}{2}\,\chi^2_{{\rm total},i}\right),
   \qquad
   w_i = \xi\!\left(M_{{\rm ini},i}\right)\,
         \bigl|\Delta M_{{\rm ini},i}\bigr|,
   \label{eq:lnL}
\end{equation}
where the sum runs over the $N_{\rm iso}=200$ grid points
along the isochrone, $\xi(m)\propto m^{-2.3}$ is the
canonical \citet{Kroupa_2001} IMF over
the mass range spanned by the sample ($m>0.5\,M_\odot$),
$|\Delta M_{{\rm ini},i}|$ is the initial-mass interval
subtended by point $i$ (evaluated numerically along the
non-uniform $\log g$ grid), and
\begin{equation}
\begin{split}
   \chi^2_{{\rm total},i} =\ &
   \left(\frac{\log T_{{\rm eff},i}-\log T_{\rm eff,obs}}{\sigma_{\log T_{\rm eff}}}\right)^{\!2}
   + \left(\frac{\log g_i-\log g_{\rm obs}}{\sigma_{\log g}}\right)^{\!2} \\
   & + \frac{\left([\mathrm{M/H}]-[\mathrm{Fe/H}]_{\rm obs}\right)^2}
          {\sigma_{[\mathrm{Fe/H}]}^2}
   + \frac{\left(K_{s,{\rm mag},i}-M_{K,\rm obs}\right)^2}
          {\sigma_{M_K}^2}.
\end{split}
\label{eq:chi2_total}
\end{equation}
The catalogue
$T_{\rm eff}$ and $\log g$ (the latter estimated from isochrones by \citealt{Thorsbro_2020} and \citealt{Ryde_2025}) fix the position
along the isochrone, while $[\mathrm{M/H}]$ and $M_K$ select which isochrone the star belongs to.
Since the PARSEC isochrones adopt a scaled-solar mixture, the third term assumes $[\mathrm{M/H}] = [\mathrm{Fe/H}]$, whereas super-solar NSC stars are mildly $\alpha$-enhanced \citep{Thorsbro_2020, Ryde_2025}. Repeating the calibration with [Fe/H] raised by 0.14~dex, as for $[\alpha/\mathrm{Fe}] = +0.2$ \citep{Salaris_1993}, changes the masses and ages by 0.31 and 0.19 times their uncertainties on average (at most 0.51 and 0.40), and $\nu_{\rm max}$ and $\Delta\nu$ by about 5 and 2 percent; we regard this as an upper limit to the corresponding systematic uncertainty. A top-heavy IMF, $\xi(m) \propto m^{-1.8}$ \citep{Hosek_2019}, changes them by only 0.07 times their uncertainties.
The posterior follows by multiplying $\mathcal{L}$ by
flat priors in age and $[\mathrm{M/H}]$.
We adopted a prior that is flat in linear stellar age
, which introduces a
Jacobian factor $p(\log \mathrm{Age}) \propto
10^{\log \mathrm{Age}}$; combined with the bounds
$9.0 < \log(\mathrm{Age/yr}) < 10.12$ and
$-1.2 < [\mathrm{M/H}] < 0.70$ (reflecting the parameter
space covered by the downloaded PARSEC grid), this gives
$-\infty$ outside these limits and
$\log \mathrm{Age}\cdot\ln 10$ within them.
We computed $\log T_{\rm eff}$ from the isochrone's
tabulated $\log T_{\rm eff}$ directly (rather than
converting to linear $T_{\rm eff}$), for consistency
with the logarithmic parametrisation of the grid, and
propagate the spectroscopic uncertainty as
$\sigma_{\log T_{\rm eff}} =
\sigma_{T_{\rm eff}}/(T_{\rm eff}\ln 10)$.
The absolute magnitude is
$M_{K,\rm obs} = K_{s,\rm obs} - \mu - A_K$, where $K_{s,\rm obs}$ is not derived from the spectra but is the apparent $K_s$ magnitude listed in the source catalogues (Table~\ref{tab:mcmc_inputs}), with
distance modulus $\mu = 5\log(d/10\,\mathrm{pc})$.
We adopted a Galactic centre distance of
$d = 8275 \pm 34$~pc from the GRAVITY Collaboration
astrometric determination of $R_0$
\citep{GRAVITY_2021}.
We adopted individual $K_s$-band extinctions from the de-reddened GALACTICNUCLEUS catalogue \citep{Nogueras-Lara_2021}, which assigns to each source the value of its $H-K_s$-based extinction map, for one of two extinction layers selected according to the colour of the source; each cross-match was checked against the catalogued $K_s$ magnitude. The two layers are defined from the detection of the reference stars in the $J$ band and from their position in the $K_s$ versus $H-K_s$ colour--magnitude diagram, so that the first layer contains the less reddened sources; they do not necessarily correspond to independent physical layers \citep{Nogueras-Lara_2021}. Eighteen of our 23 stars belong to the second layer, and the other five (GC25, GC16763, FK48, FK87, and FK5020265; Sect.~\ref{sec:membership}) to the first. \citet{Nogueras-Lara_2021} recommend not combining de-reddened stars from different layers, to avoid systematics related to the cuts that define them. We do not combine de-reddened photometry in any population analysis: the extinction of each star enters only its own calibration and observed luminosity. To check for a systematic offset between the two layers, we also estimated the extinction of each star from its own catalogued $H-K_s$ colour, with $A_H/A_{K_s} = 1.84$ as adopted by \citet{Nogueras-Lara_2021} and an intrinsic colour of 0.25~mag typical of M giants. The mean difference from the adopted values is $-0.07$~mag for the first layer and $+0.01$~mag for the second, much smaller than the scatter of 0.18 and 0.25~mag, so we find no offset between the layers. The adopted values are 1.9--3.0~mag, with uncertainties of 0.02--0.23~mag. The maps are smoothed on scales of a few arcseconds, coarser than the spectroscopic apertures, but this has little effect on the calibration, since a change in extinction moves a star along the same isochrone (an increase of 0.34~mag changes the calibrated mass by $10^{-4}\,M_\odot$ and [M/H] by 0.02~dex); it does, however, affect the observed luminosity and hence the asteroseismic predictions.
The corresponding uncertainty was propagated into
\begin{equation}
    \sigma_{M_K}^2 = \sigma_{K_s}^2 + \sigma_{A_K}^2
    + \left(\frac{5}{\ln 10}
            \frac{\sigma_d}{d}\right)^2,
    \label{eq:sigma_MK}
\end{equation}
where $\sigma_{K_s} = 0.05$~mag is the photometric uncertainty on the observed $K_s$ magnitude quoted by \citet{Rich_2017} for this sample and
$\sigma_{A_K}$ is the individual
extinction uncertainty for each star.
Sampling used the affine-invariant ensemble sampler
\textsc{emcee} \citep{Foreman-Mackey_2013} with 32
walkers, $n_{\rm burn}=500$ and $n_{\rm steps}=3000$
for all stars (starting points in
Table~\ref{tab:mcmc_inputs}), verified via trace plots
and the integrated autocorrelation time criterion
$n_{\rm steps} > 50\,\tau_{\rm int}$.
From the production chain we report the posterior
median and 16th/84th percentiles of $\log \mathrm{Age}$
and $[\mathrm{M/H}]$, together with the initial stellar
mass, $M_{\rm ini}$, recovered as a weighted mean over the
marginalisation of Eq.~(\ref{eq:lnL}), and the
derived quantities
\begin{equation}
    Z_{\rm ini} = Z_\odot \cdot 10^{[\mathrm{M/H}]},
    \qquad
    Y_{\rm ini} = 0.2485 + 1.78\,Z_{\rm ini},
    \label{eq:Zini}
\end{equation}
where $Z_\odot = 0.0152$ \citep{Bressan_2012}.
The calibrated parameters are reported in
Table~\ref{tab:mcmc_results}.
To construct the MESA models, $[\mathrm{M/H}]$ was
converted to $Z_{\rm ini}$ and $Y_{\rm ini}$ using the
equations above.
\subsection{Solar calibration of the mixing-length parameter}
\label{sec:solar_calibration}
Because the mixing-length parameter $\alpha_{\rm MLT}$
is a free calibration constant rather than a
first-principles quantity \citep{Joyce_2018}, its
appropriate value depends on the specific input physics
adopted by a given stellar evolution code and cannot be
assumed universal across codes.
The PARSEC grid used for calibration
(Sect.~\ref{sec:parsec_calibration}) uses its own
internal solar calibration of $\alpha_{\rm MLT} \approx
1.74$, obtained under different atmospheric boundary
conditions and opacity tables than the MESA setup
adopted here (Sect.~\ref{sec:mesa_models}); adopting
the MESA default of $\alpha_{\rm MLT}=2.0$ without
verification would introduce a systematic offset in the
$T_{\rm eff}$ scale of the RGB tracks relative to the
PARSEC grid from which $M_{\rm ini}$ and $Z_{\rm ini}$
were calibrated.
We therefore performed a solar calibration
using the same MESA input physics adopted for the RGB
models (Sect.~\ref{sec:mesa_models}): the
\citet{Grevesse_1998} solar composition (GS98), OPAL
high-temperature opacities \citep{Iglesias_Rogers_1996}
supplemented at $T<10^4$~K by \AE SOPUS low-temperature
opacities \citep{Marigo_2009} generated for the GS98
mixture (as no pre-computed \AE SOPUS table for this
composition is distributed with MESA), a T$-\tau$
Eddington atmospheric boundary condition with varying
opacity, and diffusive element settling.
A pre-main-sequence 1~$M_\odot$, solar-composition model
was evolved to the solar age ($4.61$~Gyr) with the
Nelder--Mead simplex algorithm implemented in MESA's
\texttt{simplex\_search} module, optimising
$\alpha_{\rm MLT}$ against the solar luminosity,
effective temperature, and surface $Z/X$ and helium
abundance.
This yields $\alpha_{\rm MLT,\odot} = 1.722$
($\chi^2 = 2.15$), which we adopt throughout the RGB
models (Sect.~\ref{sec:mesa_models}).
We additionally tested a calibration that also includes
the helioseismic constraints on the radius of the base of
the convective zone, $R_{\rm cz}$, and the RMS sound-speed
residual, $c_{\rm s,rms}$; this returns a lower value,
$\alpha_{\rm MLT,\odot} = 1.672$, but with a fit dominated almost entirely by the helioseismic terms ($\chi^2_{R_{\rm cz}} = 641$,
$\chi^2_{c_{\rm s,rms}} = 2918$), while the photospheric
observables remain well matched
($\chi^2_{\log L}=0.03$, $\chi^2_{T_{\rm eff}}=0.25$).
Since our objective is a $T_{\rm eff}$-scale correction
for RGB modelling rather than a helioseismically exact
solar interior, we adopt $\alpha_{\rm MLT,\odot} = 1.722$
as the nominal value used throughout this work.
\subsection{Stellar evolution models with MESA}
\label{sec:mesa_models}
Stellar evolution models were computed with the
Modules for Experiments in Stellar Astrophysics
\citep[\textsc{mesa};][]{Paxton2011, Paxton2013,
Paxton2015, Paxton2018, Paxton2019, Jermyn2023} (version r24.08.1, MESA SDK 24.7.1).
For each star we ran three evolutionary tracks from
the pre-main sequence through the tip of the RGB:
a nominal track using the posterior median
values $(M_{\rm ini},\,Z_{\rm ini})$, and two
boundary tracks using
$(M_{\rm ini} + \sigma_{M}^+,\,Z_{\rm ini} - \sigma_{Z}^-)$
and
$(M_{\rm ini} - \sigma_{M}^-,\,Z_{\rm ini} + \sigma_{Z}^+)$ (hereafter the high-mass and
low-mass boundary tracks, respectively),
which bracket the $1\sigma$ range of the posterior and thus
represent the range of plausible evolutionary tracks
for each star.
The initial helium abundance and metallicity for each
model follow Eq.~(\ref{eq:Zini}).
Convection was treated with the mixing-length theory
\citep[MLT;][]{Bohm-Vitense_1958}, using the
solar-calibrated value $\alpha_{\rm MLT} = 1.722$
(Sect.~\ref{sec:solar_calibration}), unless stated
otherwise (Sect.~\ref{sec:difficult_cases}).
Radiative opacities were drawn from the OPAL tables
\citep{Iglesias_Rogers_1996} with the solar abundance
mixture of \citet{Grevesse_1998} (\texttt{gs98}),
supplemented by \AE SOPUS low-temperature opacities
\citep{Marigo_2009} generated for the same GS98 mixture
(Sect.~\ref{sec:solar_calibration}); Type~2
opacities were enabled to account for enhanced C/O
abundance ratios in evolved stellar envelopes.
The default MESA nuclear reaction network
(\texttt{basic.net}), covering the pp chains and the
CNO bicycle, was used throughout; this is appropriate
for the RGB phase where only hydrogen shell burning
is active.
The atmosphere was treated with a T$-\tau$ Eddington
boundary condition with varying opacity, matching the
solar calibration setup (Sect.~\ref{sec:solar_calibration})
and chosen over a simple photosphere condition because
the emergent $T_{\rm eff}$ of cool M giants is
sensitive to the treatment of the outer boundary.
Mass loss on the RGB follows the Reimers prescription
\citep{Reimers_1975} with $\eta_{\rm Reimers} = 0.2$,
matching the value adopted in the PARSEC grid
(Sect.~\ref{sec:parsec_calibration}) to ensure
consistency between the calibration and evolution
stages.
Element diffusion was enabled only in the solar
calibration (Sect.~\ref{sec:solar_calibration}) and
disabled in the RGB tracks.
From each track we extracted the asteroseismic
observables $\nu_{\rm max}$, $\Delta\nu$, and
$\Delta\Pi$ at the evolutionary point that best
reproduces the observed position of the star in the
HR diagram.
MESA evaluates $\nu_{\rm max}$ at each timestep using the scaling relation of Eq.~(\ref{eq:scaling}), and $\Delta\nu$ from its asymptotic expression, Eq.~(\ref{eq:dnu_asymp}).
To test these asymptotic values against the separation of the radial modes, which is what an observation measures,
we computed oscillation-mode frequencies directly with
the stellar oscillation code GYRE \citep{Townsend_2013}
for all 26 catalogue entries that enter the seismic analysis, i.e. the 27 entries of Table~\ref{tab:mcmc_results} excluding NE1-1\,001, which lies outside the PARSEC grid and corresponds to the same physical star as GC25 (Sect.~\ref{sec:parsec_calibration}) (both the nominal and best-fit $\alpha_{\rm MLT}$ models for the 22 well-calibrated entries, and both the RGB and AGB
models for the four stars discussed in
Sect.~\ref{sec:bad_fits}), and compared the resulting
$\Delta\nu$, obtained from a linear fit of mode frequency
against radial order, with the asymptotic value extracted at the same evolutionary point
(Appendix~\ref{appendix:gyre_comparison}).
The results of this comparison are given in Sect.~\ref{sec:results_astero}.

The gravity-mode period spacing was computed from the
Brunt-V{\"a}is{\"a}l{\"a} frequency profile, $N(r)$,
via the asymptotic relation of Eq.~(\ref{eq:DeltaPi}).
The extraction point was identified by minimising
$\chi^2_{\rm HR}$ over all timesteps of the track,
where
\begin{equation}
    \chi^2_{\rm HR} =
    \left(\frac{\log T_{\rm eff,\,iso} - \log T_{\rm eff,\,obs}}{\sigma_{\log T_{\rm eff}}}\right)^{\!2}
    +
    \left(\frac{\log L_{\rm iso} - \log L_{\rm obs}}{\sigma_{\log L}}\right)^{\!2},
    \label{eq:chi2_HR}
\end{equation}
with $\sigma_{\log T_{\rm eff}} =
\sigma_{T_{\rm eff}}/(T_{\rm eff}\ln 10)$
propagated from the spectroscopic temperature
uncertainty.
The observed luminosity, $\log L_{\rm obs}$, was inferred
from the apparent $K_s$ magnitude, the distance
modulus $\mu = 5\log_{10}(d/10\,\mathrm{pc})$, the
individual $K$-band extinction $A_K$
(Sect.~\ref{sec:mcmc}), and a bolometric correction
$BC_K = -6.75\log_{10}(T_{\rm eff}/9500\,\mathrm{K})$
\citep{Buzzoni_2010}, adopting $d = 8275$~pc
\citep{GRAVITY_2021}.
The uncertainty, $\sigma_{\log L}$, was propagated from
the photometric, extinction, and distance uncertainties
as
\begin{equation}
    \sigma_{\log L} = \frac{1}{2.5}
    \sqrt{
        \sigma_{K_s}^2
        + \sigma_{A_K}^2
        + \left(\frac{5}{\ln 10}
                \frac{\sigma_d}{d}\right)^{\!2}
        + \left(\frac{6.75}{\ln 10}
                \frac{\sigma_{T_{\rm eff}}}{T_{\rm eff}}
          \right)^{\!2}
    },
    \label{eq:sigma_logL}
\end{equation}
adopting $\sigma_{K_s} = 0.05$~mag, the individual
$\sigma_{A_K}$ for each star, and $\sigma_d = 34$~pc.
We used the HR diagram plane rather than the
$(T_{\rm eff},\, \log g)$ plane for this minimisation
because $\log L$ is directly inferred from photometry
and is independent of the catalogue surface gravity, which carries substantial uncertainty
in the catalogues used here
(Sect.~\ref{section_2}).
For most stars, no restriction on evolutionary phase was
applied in the minimisation, since the calibrated
$(M_{\rm ini},\, Z_{\rm ini})$ values already place
the best-matching point on the upper RGB by
construction; the exceptions are discussed in
Sect.~\ref{sec:bad_fits}.
The uncertainty on each predicted asteroseismic
quantity was defined from the spread between the
boundary tracks.
Denoting the quantity extracted from the nominal,
high-mass, and low-mass tracks as $X_{\rm nom}$,
$X_{\rm hi}$, and $X_{\rm lo}$, respectively, the
asymmetric uncertainties are
\begin{equation}
    \sigma_X^+ = \left| X_{\rm nom} - X_{\rm hi}
    \right|,
    \qquad
    \sigma_X^- = \left| X_{\rm nom} - X_{\rm lo}
    \right|,
    \label{eq:astero_unc}
\end{equation}
where the absolute values account for the fact that,
depending on the quantity and evolutionary state,
either boundary track may produce the larger
deviation.
These uncertainties capture the propagation of the
posterior uncertainty in $(M_{\rm ini},\, Z_{\rm ini})$
through the stellar models into the predicted
asteroseismic observables, but do not include
systematic uncertainties arising from the adopted
MESA physics (opacity tables, mixing-length value,
mass-loss prescription).
These asteroseismic predictions are reported in
Table~\ref{tab:astero_combined}.
We adopted the GYRE-derived $\Delta\nu$ values, rather than the values computed by MESA, in Tables~\ref{tab:astero_combined}
and~\ref{tab:astero_ambiguous}, propagating the boundary-track
uncertainties by the same fractional correction
(Appendix~\ref{appendix:gyre_comparison}), except for the AGB model of GC8623, for which too few modes were available and the MESA value is retained.
We note that the $1$--$3$~percent GYRE correction is
substantially smaller than the uncertainty from the
boundary tracks themselves, so this substitution does not
materially change any of the reported values or
conclusions; we adopted it nonetheless as the more accurate
determination now available to us.
\subsection{Treatment of anomalous cases}
\label{sec:difficult_cases}
For four stars (Feld31, Feld84, GC8631, and GC8623; Sect.~\ref{sec:parsec_calibration}), the nominal MESA track does not reproduce the observed position in the HR diagram within $1\sigma$, which for the two observables fitted here ($T_{\rm eff}$ and $L$) corresponds to $\chi^2_{\rm HR} \leq 2.30$ (68\% confidence for two parameters). The reasons differ between stars and are discussed in Sect.~\ref{sec:bad_fits}.
We used the convective mixing-length parameter,
$\alpha_{\rm MLT}$, as a diagnostic tool to explore
whether the mismatch can be resolved by a
star-to-star departure from the solar-calibrated value
(Sect.~\ref{sec:solar_calibration}).
We chose $\alpha_{\rm MLT}$ because, unlike the metallicity, it cannot be measured directly, and it has the largest effect on the envelope structure on which our predictions depend. The resulting $\alpha_{\rm MLT}^*$ is an effective parameter, which may also absorb systematic errors in the spectroscopic parameters (Sect.~\ref{sec:disc_anomalous}), rather than a measurement of convective efficiency.
The solar value of $\alpha_{\rm MLT}$ is a calibration
specific to the Sun, and there is no physical reason
to expect it to be universal across the full range of
stellar masses and metallicities \citep{Joyce_2018}.
For red giants on the upper RGB, increasing
$\alpha_{\rm MLT}$ raises the efficiency of convective
energy transport, shifting the evolutionary track
towards higher effective temperatures at fixed
luminosity.
For each high-resolution star we ran a
grid of MESA models spanning
$\alpha_{\rm MLT} \in [1.4,\, 2.6]$ in steps of 0.2, extended to 2.8 for four stars (GC10195, FK5020265, Feld31, and Feld84; Appendix~\ref{appendix:alpha_chi2}),
keeping all other parameters at their calibrated
nominal values.
For each MESA track at a given $\alpha_{\rm MLT}$,
we evaluated the goodness of fit by computing
$\chi^2_{\rm HR}$ (Eq.~(\ref{eq:chi2_HR})),
minimised over all timesteps of that track.
The best-fit mixing-length parameter,
$\alpha_{\rm MLT}^*$, was then identified by fitting a
parabola through the three
grid points
closest to the $\chi^2_{\rm HR}$ minimum, $\chi^2_{\rm HR}(\alpha_{\rm MLT}) = A\,\alpha_{\rm MLT}^2 + B\,\alpha_{\rm MLT} + C$, and locating
its vertex analytically:
\begin{equation}
    \alpha_{\rm MLT}^* = -\frac{B}{2A},
    \label{eq:alpha_star}
\end{equation}
provided $A > 0$ (parabola opens upwards); if $A \leq 0$
the grid minimum was adopted instead.
Given the 0.2 grid spacing, the parabolic
interpolation, and the systematic uncertainties
discussed in Sect.~\ref{sec:solar_calibration}, we
adopted an effective uncertainty of $\pm 0.2$ on all
$\alpha_{\rm MLT}^*$ values and rounded reported values
accordingly. This value is consistent with the formal uncertainties derived from the curvature of the fitted parabola ($\Delta\chi^2_{\rm HR} = 1$), which range from 0.11 to 0.30, with a median of 0.18.
We applied this procedure to all high-resolution
stars, not only to the anomalous cases; for the
well-calibrated stars the nominal track already passes
through the observed HR-diagram position within
$1\sigma$ ($\chi^2_{\rm HR} \leq 2.11$ for all of them), so their $\alpha_{\rm MLT}^*$ values lie
close to the solar-calibrated value, even if it remains
unsurprising that adding a free parameter to a
calibration improves the fit.
From each best-fit MESA track we extracted
$\nu_{\rm max}$, $\Delta\nu$, and $\Delta\Pi$ at the
evolutionary point minimising
$\chi^2_{\rm HR}$ (Eq.~(\ref{eq:chi2_HR})),
following the same procedure described in
Sect.~\ref{sec:mesa_models} for the nominal tracks.
These values are reported for all high-resolution stars
in Table~\ref{tab:astero_combined}.
Additionally, even in the cases where this process results in a lower absolute $\chi^2$, if the nominal track already fit the observation to 1 $\sigma$, we cannot infer the new model to be strictly better, as all determinations are equally acceptable within the stated precision of the data.
\section{Results}
\label{section_4}
\subsection{The HR diagram of the NSC stellar sample}
\label{sec:hr_overview}
\begin{figure*}
    \sidecaption
    \includegraphics[width=12cm]{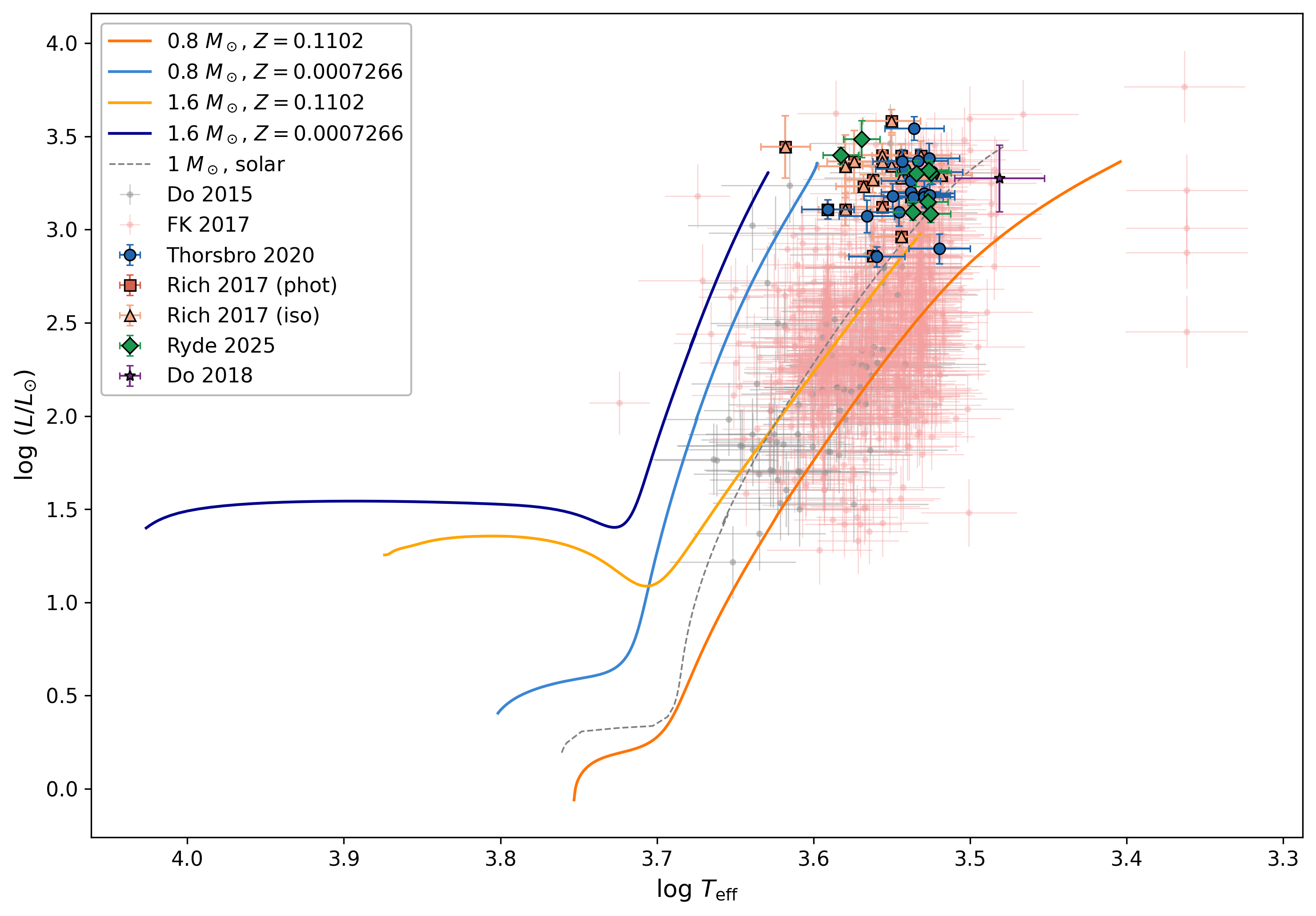}
    \caption{HR diagram of late-type giant stars in the NSC.
    Grey dots show the medium-resolution sample of \citet{Do_2015}
    and pink circles the sample of \citet{FK_2017}.
    Dark blue circles indicate the high-resolution
    stars from \citet{Rich_2017}, \citet{Do_2018},
    \citet{Thorsbro_2020}, and \citet{Ryde_2025}.
    Coloured curves show exploratory MESA tracks computed at the
    extreme metallicity values of the catalogues for two
    representative limiting masses ($M = 0.8$ and $1.6\,M_\odot$),
    labelled accordingly. The dashed grey line is a $1\,M_\odot$
    solar track for reference.}
    \label{fig:hr_overview}
\end{figure*}

As a first step, we constructed the HR
diagram of all late-type giant stars from the
catalogues described in Sect.~\ref{section_2},
shown in Fig.~\ref{fig:hr_overview}.
The observed stellar population spans effective
temperatures in the range $\sim 3000$--$4100$~K and
luminosities $\log(L/L_\odot) \sim 2.5$--$3.5$,
consistent with stars ascending the upper RGB at ages
of several Gyr.
The bulk of the sample from the medium-resolution
catalogues \citep{Do_2015, FK_2017} occupies a broad
locus in this diagram, which is
attributable to both intrinsic stellar diversity and
the larger measurement uncertainties at medium spectral
resolution.
The high-resolution sample \citep{Rich_2017, Do_2018,
Thorsbro_2020, Ryde_2025} is more compact in the HR
diagram, reflecting the smaller diversity and individual uncertainties.
Overplotted on Fig.~\ref{fig:hr_overview} are a set
of exploratory MESA evolutionary tracks computed using
the extreme metallicity values measured across the
catalogues, the highest ($Z = 0.1102$, corresponding
to the most metal-rich stars) and the lowest
($Z = 0.0007$, corresponding to the most metal-poor
stars), and two representative mass benchmarks:
$M = 0.8$ and $1.6 M_\odot$.
These tracks were not obtained from the calibration
procedure but rather computed a priori to assess
whether the observed stellar locus can be broadly
enclosed by physically plausible evolutionary models.
The comparison confirms that almost all stars in the
high-resolution sample fall within the region bounded
by these extreme tracks, demonstrating that stellar
masses in the range $\sim 0.8$--$1.6\,M_\odot$ and
metallicities spanning from sub-solar to highly
super-solar are sufficient to reproduce the observed
HR diagram distribution.
At the same time, Fig.~\ref{fig:hr_overview}
highlights a degeneracy between stellar mass and
metallicity in the HR diagram:
at a given luminosity, higher-metallicity models are cooler and higher-mass models hotter, so the same position can be matched by both lower-mass lower-metallicity or higher-mass higher-metallicity tracks.
As already noted in Sect.~\ref{sec:medium_res}, a
direct mass estimate via Eq.~(\ref{eq:mass_logg}) is
not viable for the medium-resolution sample given its
overestimated surface gravities.
Breaking the mass-metallicity degeneracy therefore
requires the Bayesian calibration against the PARSEC
isochrone grid (Sect.~\ref{sec:parsec_calibration}),
which we apply to the high-resolution sample.
\subsection{Calibrated parameters and comparison with MESA tracks}
\label{sec:good_fits}
The calibrated posterior medians and $1\sigma$
uncertainties for $\log \mathrm{Age}$,
$[\mathrm{M/H}]$, $M_{\rm ini}$, $Z_{\rm ini}$, and
$Y_{\rm ini}$ for all stars are
reported in Table~\ref{tab:mcmc_results}.
The derived initial masses span
$1.07$--$1.80\,M_\odot$ (FK16895 and GC8631, respectively, including the four anomalous stars of Sect.~\ref{sec:bad_fits}), consistent with the expected mass range
of evolved RGB stars at ages of $1.9$--$8.3$~Gyr, and
broadly enclosed by the exploratory benchmark tracks
of Fig.~\ref{fig:hr_overview}.
The calibrated metallicities span $-0.31$ to
$+0.43$~dex, broadly consistent with the input
spectroscopic $[\mathrm{Fe/H}]$ values from the
high-resolution catalogues, though in some cases
shifted by up to $\sim\!0.15$~dex relative to the
spectroscopic values, reflecting the additional
photometric constraint from $M_K$.
Of the 23 physical stars (adopting the priority Ryde
$>$ Thorsbro $>$ Do $>$ Rich for duplicate
measurements), 10 (43\%) are sub-solar,
consistent with the direct spectroscopic
findings of \citet{Thorsbro_2020} and
\citet{Ryde_2025}.
The \citet{Rich_2017} parameter variants for the same
14 physical stars, included as a secondary diagnostic,
were calibrated independently and their MESA tracks
compared directly against the primary-catalogue tracks
for the same stars.
The Rich-based tracks are systematically displaced
towards warmer effective temperatures: at matched
luminosity, the mean offset is $+67$~K (median
$+71$~K), with 23 of the 28 phot and iso combinations
(82\%) warmer than the corresponding primary-catalogue
track.
This reflects the different temperature scales of the two studies: \citet{Rich_2017} adopt temperatures based on the CO band-head strength alone, whereas \citet{Thorsbro_2020} average this estimate with an independent one based on Sc\,{\sc i} lines, which tends to yield lower values.

\begin{figure*}[h!]
    \centering
    \includegraphics[width=1.0\textwidth]
        {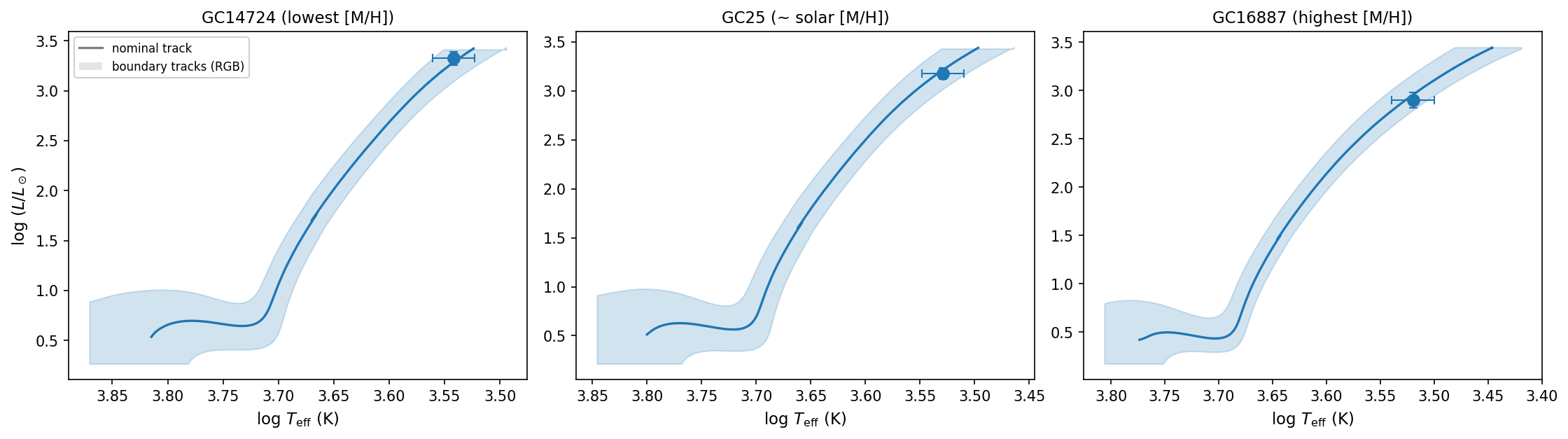}
    \caption{HR diagrams (RGB track only, solar-calibrated
             $\alpha_{\rm MLT}=1.722$) for three
             well-calibrated stars spanning the
             metallicity range of the sample: GC14724
             (lowest $[\mathrm{M/H}]$), GC25 (reference),
             and GC16887 (highest $[\mathrm{M/H}]$).
             In each panel the coloured marker shows the
             observed $(T_{\rm eff},\, \log L)$ position
             with $1\sigma$ uncertainties, and the shaded
             envelope spans the boundary tracks.}
    \label{fig:hr_wellcalibrated}
\end{figure*}

\begin{figure*}[h!]
    \centering
    \includegraphics[width=\textwidth]{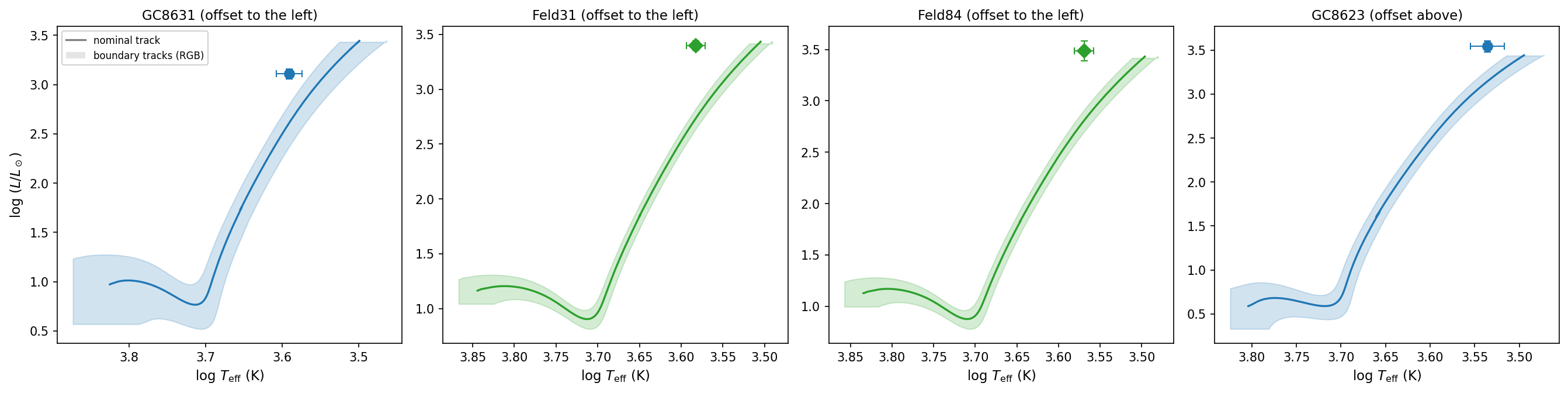}
    \caption{HR diagrams (RGB track only, solar-calibrated
             $\alpha_{\rm MLT}=1.722$, no mixing-length
             variation applied) for the four anomalous
             stars discussed in Sect.~\ref{sec:bad_fits},
             grouped by mismatch type.
             GC8631, Feld31, and Feld84 (first three panels) are
             representative of a lateral offset -- the
             observed position lies to the side of the
             nominal track at fixed luminosity.
             GC8623 (right panel) is
             representative of a luminosity excess -- the
             observed position lies above the nominal
             track at fixed effective temperature.
             In each panel the coloured marker shows the
             observed $(T_{\rm eff},\, \log L)$ position
             with $1\sigma$ uncertainties, and the shaded
             envelope spans the boundary tracks.}
    \label{fig:hr_problematic}
\end{figure*}

The corresponding calibrated parameters are listed in
Table~\ref{tab:mcmc_results_rich}
(Appendix~\ref{appendix:mcmc_rich}).
The calibrated metallicity distribution spans
$0.74$~dex in total ($[\mathrm{M/H}] = -0.31$ to
$+0.43$), with a median of $+0.01$~dex and a standard
deviation of $0.21$~dex.
Given that the systematic differences between the spectroscopic catalogues (up to 0.25~dex in the calibrated $[\mathrm{M/H}]$ of the same star; Table~\ref{tab:mcmc_results}) are comparable to the $0.21$~dex standard deviation of the distribution
itself, we do not consider this sample sufficiently
precise to test for intrinsic metallicity
dispersion beyond measurement scatter; a
maximum-likelihood estimate of the intrinsic scatter is
left for future work.
\begin{table}
\centering
\caption{Calibrated stellar parameters for the 27 primary high-resolution catalogue entries.}
\label{tab:mcmc_results}
\begin{tabular}{lcccc}
\toprule
Star & log Age & [M/H] & $M_\mathrm{ini}$ ($M_\odot$)
\\
\midrule
    GC6227$^{(1)}$ & \unc{9.688}{0.283}{0.267} & \unc{-0.261}{0.136}{0.136} & \unc{1.198}{0.273}{0.205}
    \\
    GC8623$^{(1)\dagger}$ & \unc{9.759}{0.162}{0.109} & \unc{+0.009}{0.090}{0.098} & \unc{1.205}{0.106}{0.122}
    \\
    GC8631$^{(1)\dagger}$ & \unc{9.276}{0.132}{0.054} & \unc{+0.189}{0.115}{0.135} & \unc{1.802}{0.140}{0.195}
    \\
    GC10195$^{(1)}$ & \unc{9.702}{0.238}{0.175} & \unc{-0.033}{0.134}{0.134} & \unc{1.239}{0.178}{0.181}
    \\
    GC11025$^{(1)}$ & \unc{9.644}{0.232}{0.143} & \unc{+0.203}{0.132}{0.136} & \unc{1.343}{0.162}{0.197}
    \\
    GC11532$^{(1)}$ & \unc{9.778}{0.235}{0.249} & \unc{-0.222}{0.136}{0.136} & \unc{1.137}{0.223}{0.162}
    \\
    GC13282$^{(1)}$ & \unc{9.650}{0.281}{0.227} & \unc{-0.057}{0.137}{0.141} & \unc{1.280}{0.245}{0.220}
    \\
    GC13727$^{(1)}$ & \unc{9.891}{0.165}{0.264} & \unc{-0.239}{0.137}{0.135} & \unc{1.057}{0.206}{0.112}
    \\
    GC13882$^{(1)}$ & \unc{9.828}{0.198}{0.233} & \unc{-0.149}{0.137}{0.129} & \unc{1.119}{0.197}{0.140}
    \\
    GC14024$^{(1)}$ & \unc{9.552}{0.229}{0.139} & \unc{+0.405}{0.135}{0.133} & \unc{1.469}{0.157}{0.230}
    \\
    GC14724$^{(1)}$ & \unc{9.807}{0.211}{0.239} & \unc{-0.306}{0.131}{0.128} & \unc{1.100}{0.202}{0.146}
    \\
    GC15540$^{(1)}$ & \unc{9.625}{0.252}{0.145} & \unc{-0.035}{0.132}{0.134} & \unc{1.304}{0.170}{0.205}
    \\
    GC16763$^{(1)}$ & \unc{9.805}{0.218}{0.290} & \unc{-0.133}{0.139}{0.137} & \unc{1.138}{0.259}{0.153}
    \\
    GC16867$^{(1)}$ & \unc{9.881}{0.172}{0.284} & \unc{-0.149}{0.139}{0.137} & \unc{1.082}{0.229}{0.119}
    \\
    GC16887$^{(1)}$ & \unc{9.841}{0.198}{0.257} & \unc{+0.433}{0.129}{0.139} & \unc{1.194}{0.241}{0.140}
    \\
    GC16895$^{(1)}$ & \unc{9.786}{0.233}{0.265} & \unc{-0.117}{0.140}{0.136} & \unc{1.156}{0.242}{0.163}
    \\
    FK5020265$^{(2)}$ & \unc{9.554}{0.184}{0.105} & \unc{+0.233}{0.093}{0.092} & \unc{1.444}{0.125}{0.183}
    \\
    FK15540$^{(2)}$ & \unc{9.608}{0.220}{0.131} & \unc{+0.095}{0.092}{0.092} & \unc{1.354}{0.155}{0.189}
    \\
    FK16890$^{(2)}$ & \unc{9.522}{0.182}{0.108} & \unc{+0.163}{0.093}{0.094} & \unc{1.464}{0.131}{0.186}
    \\
    FK16895$^{(2)}$ & \unc{9.919}{0.143}{0.227} & \unc{-0.092}{0.091}{0.089} & \unc{1.068}{0.172}{0.096}
    \\
    GC25$^{(1)}$ & \unc{9.825}{0.207}{0.269} & \unc{-0.054}{0.141}{0.141} & \unc{1.141}{0.237}{0.147}
    \\
    FK48$^{(2)}$ & \unc{9.740}{0.239}{0.228} & \unc{+0.121}{0.093}{0.093} & \unc{1.241}{0.232}{0.176}
    \\
    FK87$^{(2)}$ & \unc{9.882}{0.168}{0.241} & \unc{+0.114}{0.091}{0.094} & \unc{1.134}{0.199}{0.117}
    \\
    GC13727$^{(2)}$ & \unc{9.851}{0.187}{0.254} & \unc{+0.011}{0.092}{0.089} & \unc{1.137}{0.215}{0.132}
    \\
    Feld31$^{(2)\dagger}$ & \unc{9.321}{0.065}{0.035} & \unc{+0.272}{0.079}{0.074} & \unc{1.765}{0.079}{0.098}
    \\
    Feld84$^{(2)\dagger}$ & \unc{9.336}{0.079}{0.047} & \unc{+0.359}{0.090}{0.092} & \unc{1.750}{0.089}{0.112}
    \\
    NE1-1\,001$^{(3)}$ & \unc{9.839}{0.199}{0.239} & \unc{+0.219}{0.177}{0.178} & \unc{1.177}{0.214}{0.141}
    \\
\bottomrule
\end{tabular}
\tablefoot{The parameters were derived from the Bayesian MCMC
calibration against the PARSEC grid
(Sect.~\ref{sec:mcmc}), using the GRAVITY Collaboration
distance $d = 8275$~pc \citep{GRAVITY_2021} and individual
$K$-band extinctions from GALACTICNUCLEUS
\citep{Nogueras-Lara_2021}.
Uncertainties are the 16th--84th percentile half-widths
of the posterior distributions.
Columns: (1) Star name; (2) $\log_{10}(\mathrm{Age/yr})$;
(3) Metallicity [M/H]; (4) Initial mass
$M_\mathrm{ini}$ ($M_\odot$).
Source indices: $^{(1)}$~\citet{Thorsbro_2020},
$^{(2)}$~\citet{Ryde_2025}, $^{(3)}$~\citet{Do_2018}.
Stars marked $\dagger$ (Feld31, Feld84, GC8623, GC8631)
are calibrated without restricting the PARSEC grid to
evolutionary label~$=3$ (Sect.~\ref{sec:bad_fits}).
}
\end{table}
Of the 23 high-resolution physical stars in our sample,
19 yield satisfactory MESA model fits under the standard
assumptions of solar mixing length and an RGB-only calibration.
We defined a fit as satisfactory when the nominal MESA
track passes through the observed
$(T_{\rm eff},\, \log L)$ position of the star within
its $1\sigma$ uncertainties.
The remaining four stars (Feld31, Feld84,
GC8623, and GC8631) are discussed in
Sect.~\ref{sec:bad_fits}.
The remaining calibrations not shown in
Figs.~\ref{fig:hr_wellcalibrated} and~\ref{fig:hr_problematic}
are presented in
Fig.~\ref{fig:hr_all_combined} (Appendix~\ref{appendix:hr_alpha_all}),
arranged as individual panels grouped by physical star.
For the well-calibrated stars from \citet{Thorsbro_2020} and \citet{Ryde_2025} the nominal
track passes through the observed HR-diagram position
within $1\sigma$ in all cases ($\chi^2_{\rm HR} \leq 2.30$; Sect.~\ref{sec:difficult_cases}).
When a star is also present in the \citet{Rich_2017}
catalogue, the corresponding tracks are systematically
displaced towards warmer effective temperatures
relative to the primary-catalogue tracks, reflecting
the higher $T_{\rm eff}$ values reported by
\citet{Rich_2017} (see Sect.~\ref{sec:high_res}).
For this reason, $\alpha_{\rm MLT}^*$ optimisation
and asteroseismic predictions are derived exclusively
from the primary \citet{Thorsbro_2020} and
\citet{Ryde_2025} measurements.
To illustrate the range of calibration outcomes,
Fig.~\ref{fig:hr_wellcalibrated} shows three
well-calibrated stars spanning the full metallicity
range of the sample: GC14724 ($[\mathrm{M/H}]=-0.31$,
the lowest), GC25 (a reference case), and GC16887
($[\mathrm{M/H}]=+0.43$, the highest).
Figure~\ref{fig:hr_problematic} shows the four
anomalous stars discussed in Sect.~\ref{sec:bad_fits},
grouped by the type of mismatch they exhibit: GC8631,
Feld31, and Feld84 show a lateral offset, while GC8623
shows a luminosity excess.
\subsection{Asteroseismic predictions}
\label{sec:results_astero}
Table~\ref{tab:astero_combined} lists the predicted
asteroseismic observables $\nu_{\rm max}$, $\Delta\nu$,
and $\Delta\Pi$
extracted from the nominal MESA track at the
evolutionary point that best matches the observed
$(T_{\rm eff},\, \log L)$ of each star via
$\chi^2_{\rm HR}$ (Eq.~(\ref{eq:chi2_HR})).
The uncertainties on each quantity, derived from the
spread between the boundary tracks, are listed
alongside the nominal predictions.
The predicted $\nu_{\rm max}$ values span
$0.23$--$0.97\,\mu$Hz, and the large frequency
separations $\Delta\nu$ span $0.10$--$0.32\,\mu$Hz.
These ranges are consistent with the scaling
relation predictions for luminous upper-RGB stars at
the calibrated masses and luminosities.
Applying Eq.~(\ref{eq:scaling}) at the
best-matching evolutionary points, where
$\log g \approx 0.07$--$1.4$ and
$T_{\rm eff} \approx 3300$--$3900$~K, yields
$\nu_{\rm max} \approx 0.1$--$3\,\mu$Hz, in agreement with the
tabulated values.
The single most asteroseismically "accessible" star in
the sample is GC16867
($\nu_{\rm max} = 0.97^{+0.38}_{-0.20}\,\mu$Hz,
$\Delta\nu = 0.32^{+0.07}_{-0.04}\,\mu$Hz), whose
oscillation frequencies are the highest in the sample
and whose modes are therefore the easiest to resolve
photometrically. Its variability could in principle be examined with near-infrared surveys such as VVVX \citep{Saito_2024}, but our targets are close to their saturation limit in an extremely crowded field, and the \citet{Rich_2017} stars were selected to exclude known long-period variables \citep{Blum_2003, Matsunaga_2009}.
GC8631 has the highest calibrated initial mass
($1.80\,M_\odot$) among the \citet{Thorsbro_2020} and \citet{Ryde_2025} stars, a high surface gravity ($\log g = 1.41$), and also has a comparatively high
$\nu_{\rm max} = 0.73$--$0.93\,\mu$Hz.
At the other extreme, the stars with the lowest predicted values, such as
GC10195, GC11025, and GC13882, have $\nu_{\rm max} \approx 0.23$--$0.24\,\mu$Hz,
corresponding to oscillation periods of about 50~d,
which would require extended photometric baselines to
detect.
For 51 of the 52 models, the asymptotic relation overestimates $\Delta\nu$ by $1.8 \pm 0.4$~percent (median $1.7$~percent).
The remaining model, the AGB phase of GC8623, yielded too few radial modes
(5, against 10--16 for the rest of
the sample) for a robust linear fit, and shows a larger,
oppositely signed deviation; given the sparse mode
coverage, we do not consider this result reliable and
exclude it from the statistics above.
\begin{table*}
\centering
\caption{Predicted asteroseismic parameters for 22 catalogue entries corresponding to 19 distinct, well-calibrated stars.}
\label{tab:astero_combined}
\begin{tabular}{l@{\hspace{2em}}ccc@{\hspace{4em}}cccc}
\toprule
 & \multicolumn{3}{c}{Nominal ($\alpha_{\rm MLT}=1.722$)} & \multicolumn{4}{c}{Best-fit $\alpha^{*}_{\rm MLT}$} \\
Star &
$\nu_{\rm max}$ ($\mu$Hz) &
$\Delta\nu$ ($\mu$Hz) &
$\Delta\Pi_{\ell=1}$ (s) &
$\alpha^{*}_{\rm MLT}$ &
$\nu_{\rm max}$ ($\mu$Hz) &
$\Delta\nu$ ($\mu$Hz) &
$\Delta\Pi_{\ell=1}$ (s) \\
\midrule
GC25$^{(1)}$ & \unc{0.35}{0.15}{0.08} & \unc{0.15}{0.03}{0.02} & \unc{32.1}{0.7}{0.8} & 1.7 & \unc{0.35}{0.14}{0.08} & \unc{0.15}{0.03}{0.02} & \unc{32.1}{0.8}{0.7} \\
GC6227$^{(1)}$ & \unc{0.63}{0.19}{0.13} & \unc{0.23}{0.03}{0.03} & \unc{33.4}{1.0}{1.1} & 1.7 & \unc{0.63}{0.20}{0.13} & \unc{0.23}{0.04}{0.03} & \unc{33.4}{1.1}{1.0} \\
GC10195$^{(1)}$ & \unc{0.23}{0.06}{0.05} & \unc{0.11}{0.02}{0.02} & \unc{29.3}{0.9}{1.1} & 1.9 & \unc{0.25}{0.07}{0.06} & \unc{0.12}{0.02}{0.02} & \unc{28.9}{1.0}{0.8} \\
GC11025$^{(1)}$ & \unc{0.23}{0.04}{0.04} & \unc{0.10}{0.01}{0.01} & \unc{30.2}{1.1}{1.4} & 2.0 & \unc{0.25}{0.05}{0.05} & \unc{0.11}{0.01}{0.01} & \unc{29.2}{1.3}{1.0} \\
GC11532$^{(1)}$ & \unc{0.30}{0.11}{0.07} & \unc{0.14}{0.03}{0.02} & \unc{30.3}{0.8}{0.9} & 1.7 & \unc{0.30}{0.11}{0.07} & \unc{0.14}{0.03}{0.02} & \unc{30.3}{0.8}{0.8} \\
GC13282$^{(1)}$ & \unc{0.45}{0.15}{0.11} & \unc{0.17}{0.03}{0.03} & \unc{32.2}{0.9}{1.0} & 1.8 & \unc{0.46}{0.15}{0.12} & \unc{0.18}{0.03}{0.03} & \unc{32.0}{1.0}{0.8} \\
GC13727$^{(1)}$ & \unc{0.32}{0.12}{0.06} & \unc{0.13}{0.01}{0.00} & \unc{31.1}{0.7}{0.8} & 1.6 & \unc{0.28}{0.01}{0.00} & \unc{0.13}{0.01}{0.00} & \unc{30.9}{2.0}{2.3} \\
GC13882$^{(1)}$ & \unc{0.24}{0.08}{0.05} & \unc{0.12}{0.02}{0.02} & \unc{29.6}{0.8}{0.9} & 1.7 & \unc{0.24}{0.08}{0.05} & \unc{0.12}{0.02}{0.02} & \unc{29.6}{0.9}{0.8} \\
GC14024$^{(1)}$ & \unc{0.42}{0.07}{0.09} & \unc{0.15}{0.02}{0.02} & \unc{33.7}{1.0}{1.3} & 2.0 & \unc{0.46}{0.09}{0.10} & \unc{0.16}{0.02}{0.02} & \unc{33.0}{1.3}{0.9} \\
GC14724$^{(1)}$ & \unc{0.25}{0.09}{0.05} & \unc{0.12}{0.02}{0.02} & \unc{29.1}{0.7}{0.8} & 1.8 & \unc{0.26}{0.09}{0.06} & \unc{0.13}{0.02}{0.02} & \unc{29.0}{0.8}{0.7} \\
GC15540$^{(1)}$ & \unc{0.26}{0.06}{0.06} & \unc{0.12}{0.02}{0.02} & \unc{29.5}{0.9}{1.1} & 2.0 & \unc{0.29}{0.07}{0.07} & \unc{0.13}{0.02}{0.02} & \unc{28.9}{1.0}{0.7} \\
GC16763$^{(1)}$ & \unc{0.49}{0.18}{0.10} & \unc{0.19}{0.04}{0.02} & \unc{33.2}{0.8}{0.9} & 1.7 & \unc{0.48}{0.18}{0.10} & \unc{0.19}{0.04}{0.02} & \unc{33.3}{1.0}{0.9} \\
GC16867$^{(1)}$ & \unc{0.97}{0.38}{0.20} & \unc{0.32}{0.07}{0.04} & \unc{37.2}{0.5}{0.5} & 1.6 & \unc{0.91}{0.35}{0.19} & \unc{0.30}{0.06}{0.04} & \unc{37.3}{0.6}{0.6} \\
GC16887$^{(1)}$ & \unc{0.66}{0.21}{0.11} & \unc{0.22}{0.04}{0.02} & \unc{38.1}{1.2}{1.2} & 1.7 & \unc{0.65}{0.20}{0.11} & \unc{0.22}{0.04}{0.02} & \unc{38.2}{1.2}{1.2} \\
GC16895$^{(1)}$ & \unc{0.38}{0.16}{0.10} & \unc{0.16}{0.04}{0.03} & \unc{32.0}{0.6}{0.7} & 1.7 & \unc{0.38}{0.15}{0.09} & \unc{0.16}{0.04}{0.03} & \unc{32.0}{0.8}{0.7} \\
FK5020265$^{(2)}$ & \unc{0.31}{0.04}{0.05} & \unc{0.12}{0.01}{0.01} & \unc{31.2}{0.8}{1.0} & 1.9 & \unc{0.32}{0.04}{0.05} & \unc{0.13}{0.01}{0.01} & \unc{30.6}{0.9}{0.7} \\
FK15540$^{(2)}$ & \unc{0.28}{0.05}{0.05} & \unc{0.13}{0.01}{0.01} & \unc{30.4}{0.8}{1.0} & 1.8 & \unc{0.29}{0.02}{0.02} & \unc{0.12}{0.01}{0.01} & \unc{30.0}{2.3}{2.1} \\
FK16890$^{(2)}$ & \unc{0.35}{0.03}{0.04} & \unc{0.16}{0.01}{0.01} & \unc{31.3}{1.0}{1.2} & 1.9 & \unc{0.35}{0.03}{0.03} & \unc{0.14}{0.01}{0.01} & \unc{30.5}{2.2}{1.8} \\
FK16895$^{(2)}$ & \unc{0.35}{0.10}{0.06} & \unc{0.13}{0.01}{0.01} & \unc{32.3}{0.5}{0.5} & 1.7 & \unc{0.34}{0.03}{0.02} & \unc{0.14}{0.01}{0.01} & \unc{32.5}{1.8}{2.3} \\
FK48$^{(2)}$ & \unc{0.48}{0.15}{0.10} & \unc{0.18}{0.03}{0.02} & \unc{34.0}{0.7}{0.7} & 1.8 & \unc{0.49}{0.15}{0.11} & \unc{0.18}{0.03}{0.02} & \unc{33.9}{0.7}{0.6} \\
FK87$^{(2)}$ & \unc{0.42}{0.13}{0.07} & \unc{0.17}{0.03}{0.02} & \unc{34.1}{0.6}{0.6} & 1.7 & \unc{0.42}{0.13}{0.07} & \unc{0.17}{0.03}{0.02} & \unc{34.1}{0.6}{0.6} \\
GC13727$^{(2)}$ & \unc{0.33}{0.09}{0.05} & \unc{0.14}{0.02}{0.01} & \unc{32.1}{0.9}{0.8} & 1.7 & \unc{0.33}{0.09}{0.05} & \unc{0.14}{0.02}{0.01} & \unc{32.1}{0.8}{0.9} \\
\bottomrule
\end{tabular}
\tablefoot{The nominal columns ($\alpha_{\rm MLT} = 1.722$, the
solar-calibrated value; Sect.~\ref{sec:solar_calibration})
are extracted from the nominal MESA track at the
best-matching evolutionary point
(Eq.~(\ref{eq:chi2_HR})); the best-fit columns are
extracted from the best-fit $\alpha^{*}_{\rm MLT}$ tracks
(Eq.~(\ref{eq:alpha_star}), rounded to the $\pm0.2$
precision discussed in Sect.~\ref{sec:difficult_cases}).
The four stars for which the RGB fit is not satisfactory
(Feld31, Feld84, GC8623, and GC8631) are
presented separately in Table~\ref{tab:astero_ambiguous}.
Source indices: $^{(1)}$~\citet{Thorsbro_2020}, $^{(2)}$~\citet{Ryde_2025}.}
\end{table*}
For the four stars discussed in Sect.~\ref{sec:bad_fits},
for which the nominal RGB fit is not satisfactory,
Table~\ref{tab:astero_ambiguous} reports the predicted
asteroseismic parameters separately for the RGB (at the
best-fit $\alpha_{\rm MLT}^*$) and for the AGB
(at the nominal $\alpha_{\rm MLT}$).
\begin{table*}
\centering
\caption{Predicted asteroseismic parameters for the four stars discussed in Sect.~\ref{sec:bad_fits}.}
\label{tab:astero_ambiguous}
\begin{tabular}{l@{\hspace{1.5em}}cccc@{\hspace{2.5em}}ccc}
\toprule
 & & \multicolumn{3}{c}{RGB (best-fit $\alpha^{*}_{\rm MLT}$)} & \multicolumn{3}{c}{AGB (nominal)} \\
Star & $\alpha^{*}_{\rm MLT}$ &
$\nu_{\rm max}$ ($\mu$Hz) & $\Delta\nu$ ($\mu$Hz) & $\Delta\Pi_{\ell=1}$ (s) &
$\nu_{\rm max}$ ($\mu$Hz) & $\Delta\nu$ ($\mu$Hz) & $\Delta\Pi_{\ell=1}$ (s) \\
\midrule
GC8623$^{(1)}$ & 2.1 & \unc{0.21}{0.06}{0.05} & \unc{0.10}{0.02}{0.02} & \unc{27.7}{0.1}{0.1} & 0.08\tablefootmark{a} & 0.09\tablefootmark{a} & 23.7\tablefootmark{a} \\
GC8631$^{(1)}$ & 2.3 & \unc{0.93}{0.23}{0.26} & \unc{0.28}{0.04}{0.05} & \unc{34.0}{0.8}{0.6} & \unc{0.73}{0.19}{0.21} & \unc{0.23}{0.03}{0.04} & \unc{25.0}{0.4}{0.2} \\
Feld31$^{(2)}$ & 2.6 & \unc{0.50}{0.05}{0.04} & \unc{0.17}{0.01}{0.01} & \unc{29.2}{0.5}{0.4} & \unc{0.40}{0.03}{0.03} & \unc{0.14}{0.01}{0.01} & \unc{23.1}{0.2}{0.1} \\
Feld84$^{(2)}$ & 2.5 & \unc{0.42}{0.09}{0.05} & \unc{0.15}{0.02}{0.01} & \unc{29.0}{0.6}{0.0} & \unc{0.47}{0.00}{0.00} & \unc{0.16}{0.00}{0.00} & \unc{23.7}{0.4}{0.3} \\
\bottomrule
\end{tabular}
\tablefoot{For these stars, the nominal RGB track does not reproduce the observed position within $1\sigma$. For each star we report the best-fit RGB prediction (at
$\alpha^{*}_{\rm MLT}$, rounded to $\pm0.2$;
Sect.~\ref{sec:difficult_cases}) and the AGB
prediction (at the nominal $\alpha_{\rm MLT} = 1.722$).
For Feld31, Feld84, and GC8631 the RGB fit is
satisfactory ($\chi^2_{\rm HR} \leq 2.30$) and the AGB
value is shown for comparison, at the best-matching point of the AGB track, which does not reach the observed position within $1\sigma$; for GC8623
the RGB fit is not satisfactory at any $\alpha_{\rm MLT}$
($\chi^2_{\rm HR} \geq 2.6$) and the AGB value, obtained near the end of the AGB, is the
closest match.
We also examined the core-helium-burning and
post-AGB and white-dwarf phases for all four stars; the
former never improves on the RGB or AGB fit reported
here, and while the latter yields a comparably good
geometric fit for Feld84 and GC8631, it is a very
short-lived evolutionary phase and its predicted
$\Delta\Pi_{\ell=1}$ is not physically meaningful (the
gravity-mode cavity is not well defined during the rapid
post-AGB contraction), so it is not reported.
Neither the RGB nor the AGB value is presented as the
adopted parameter set for these four stars; see
Sect.~\ref{sec:bad_fits} for discussion.
Source indices: $^{(1)}$~\citet{Thorsbro_2020},
$^{(2)}$~\citet{Ryde_2025}. \tablefoottext{a}{Uncertainty not available: one of the two boundary tracks for GC8623 did not converge through the AGB phase.}}
\end{table*}
\subsection{Anomalous stars}
\label{sec:bad_fits}
The analysis in this section focuses on Feld31, Feld84,
GC8623, and GC8631 as the four cases where the
mismatch requires a modelling strategy.
As shown in Fig.~\ref{fig:hr_problematic}, the mismatch takes
one of two different forms.
Feld31, Feld84, and GC8631 have high catalogue surface gravities ($\log g \geq 1.4$) and effective
temperatures ($T_{\rm eff} \geq 3700$~K) relative to
the well-calibrated subsample, whose mean surface
gravity is $\langle \log g \rangle \approx 0.57$ and
whose effective temperatures span
$T_{\rm eff} \approx 3300$--$3700$~K.
For these three stars the observed position lies
laterally offset from the nominal RGB track: at
super-solar metallicity, the PARSEC RGB is shifted to
cooler temperatures than observed, displacing the
nominal MESA tracks away from the observed HR-diagram
position (Fig.~\ref{fig:hr_problematic}, shown for all three).
We note that a qualitatively similar lateral displacement
is seen in the \citet{Rich_2017} parameter variants for
several other stars (Sect.~\ref{sec:high_res}), where it reflects the warmer CO-based temperature scale of that study (Sect.~\ref{sec:good_fits}).
Since Feld31 and Feld84 use the parameters of \citet{Ryde_2025}, and GC8631 those of \citet{Thorsbro_2020}, this
catalogue-specific bias does not apply to them; whether their discrepancy reflects a genuine physical effect or a systematic error in the spectroscopic parameters is discussed in Sect.~\ref{sec:disc_anomalous}.
For Feld31, Feld84, and the \citet{Ryde_2025} entry of GC13727 we adopted
conservatively large $\log g$ uncertainties in
the calibration, reflecting the known difficulty of
constraining surface gravity from near-infrared K-band
spectra \citep{Do_2018, Ryde_2025}.
GC8623 presents a distinct
situation.
Its spectroscopic parameters are entirely typical of
the well-calibrated sample, and the MCMC calibration
converges normally, but the observed position in the
HR~diagram lies significantly above the
best-matching point on the nominal RGB track (Fig.~\ref{fig:hr_problematic}), at a luminosity the RGB track does not
reach at the calibrated evolutionary phase.
This mismatch cannot be attributed to the surface
gravity or metallicity being anomalous.

We used the convective mixing-length parameter,
$\alpha_{\rm MLT}$, as a diagnostic tool to test whether
either failure mode can be resolved by a star-specific
departure from the solar-calibrated value
(Sect.~\ref{sec:difficult_cases}).
Table~\ref{tab:astero_ambiguous} lists the resulting
$\alpha_{\rm MLT}^*$ values.
For Feld31, Feld84, and GC8631, whose mismatch is a
lateral offset, increasing $\alpha_{\rm MLT}$ shifts the
track to higher effective temperature at fixed
luminosity and successfully resolves the discrepancy.
For GC8623, whose mismatch is a luminosity excess, increasing $\alpha_{\rm MLT}$ improves the fit but does not fully resolve it: the best-fitting value, $\alpha_{\rm MLT}^* \approx 2.1$, reproduces the observed position only with $\chi^2_{\rm HR} = 2.7$, between the $1\sigma$ and $2\sigma$ thresholds, close to the tip of the RGB (Appendix~\ref{appendix:alpha_chi2}).

\begin{figure*}[h!]
    \centering
    \includegraphics[width=\textwidth]
        {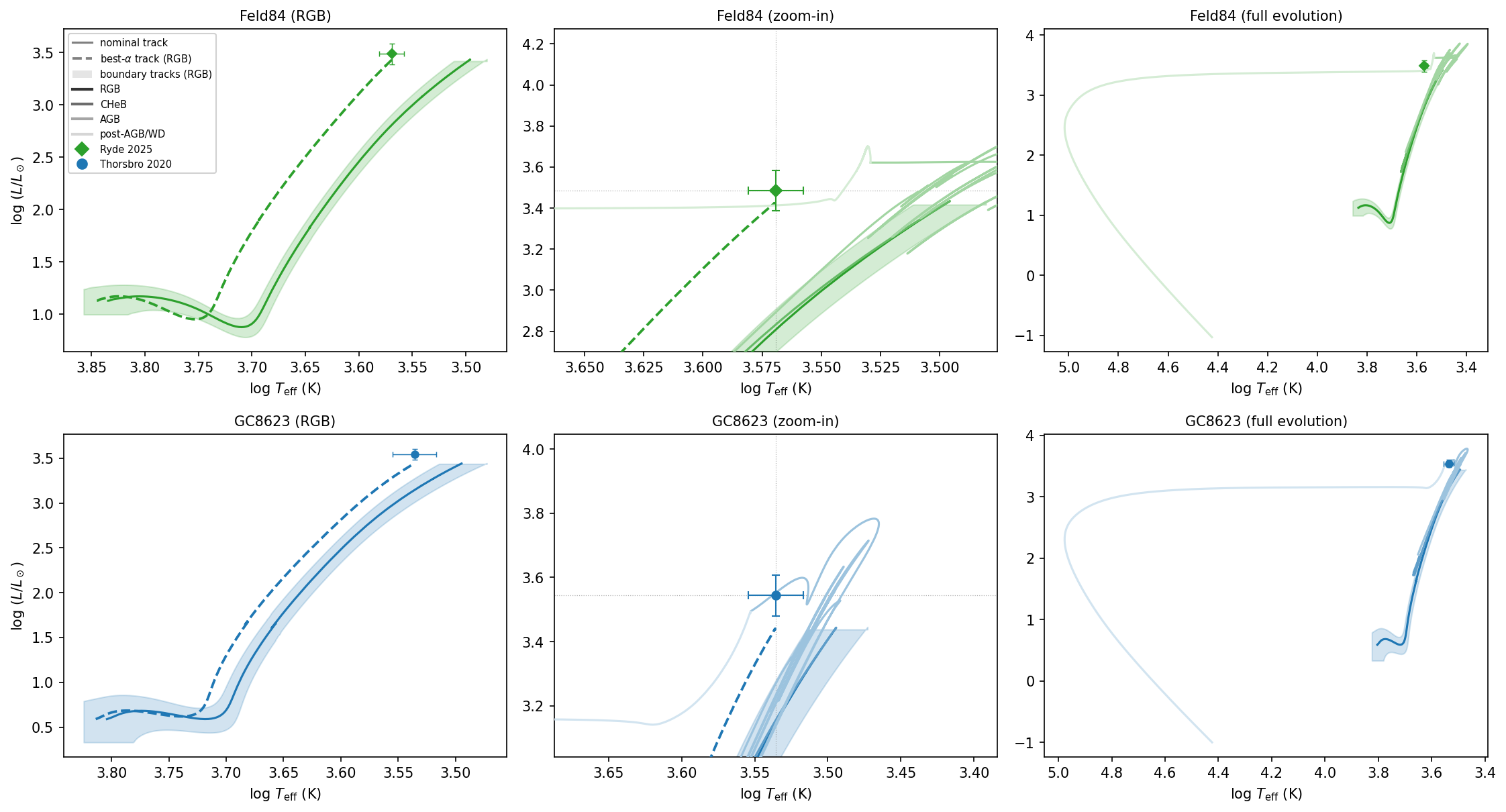}
    \caption{HR diagrams for Feld84 (top row) and
             GC8623 (bottom row).
             Left column: Nominal RGB track
             (solid) and best-fit $\alpha_{\rm MLT}^*$
             RGB track (dashed), with the observed
             position ($1\sigma$ uncertainties) and the
             boundary-track envelope.
             Centre column: Zoom on the observed
             position.
             Right column: Full evolutionary
             track (RGB through post-AGB, colour-coded by
             phase, with lighter shades for later phases) at the nominal $\alpha_{\rm MLT}$.
             For Feld84, the best-fit $\alpha_{\rm MLT}^*$
             RGB track passes through the observed
             position; for GC8623, the best-fit RGB track ends just below it,
             while the nominal track reaches it only near the end of the AGB.}
    \label{fig:alpha_grid}
\end{figure*}

This motivated us to explore an alternative explanation for these four
stars, recomputing their calibration without
restricting the PARSEC grid to evolutionary label~$=3$
(Sect.~\ref{sec:parsec_calibration}). For GC8623, the nominal track then reproduces the observed position within $1\sigma$ on the AGB ($\chi^2_{\rm HR} = 0.01$), but only at the very end of the AGB, after the thermal pulses, as the star leaves the AGB; before the first thermal pulse, the best fit is poor ($\chi^2_{\rm HR} = 6.2$).
Figure~\ref{fig:alpha_grid} illustrates the two cases
for Feld84 and GC8623.
\section{Discussion}
\label{sec:discussion}
\subsection{Interpretation of the anomalous stars}
\label{sec:disc_anomalous}
Because a change in $\alpha_{\rm MLT}$ shifts the RGB track mainly in $T_{\rm eff}$, a departure of $\alpha_{\rm MLT}^*$ from the solar value can be mimicked by a systematic error in the adopted temperature. We therefore determined for each star the offset $\Delta T_{\rm eq}$ that, applied to the observed $T_{\rm eff}$ (and propagated to $L$), brings $\alpha_{\rm MLT}^*$ back to the solar value. For the well-calibrated stars with the largest $\alpha_{\rm MLT}^*$ ($\approx$1.9--2.0), $\Delta T_{\rm eq} = -100$ to $-180$~K, that is, 0.8--1.2 times the quoted uncertainties, comparable to the differences between the two temperature indicators of \citet{Thorsbro_2020} (up to $\sim$400~K) and to the oxygen-abundance dependence found by \citet{Ryde_2025} (50--100~K); for these stars, $\alpha_{\rm MLT}^*$ cannot be distinguished from a temperature systematic. In contrast, Feld31 and Feld84 would require offsets of 455 and 475~K (4.6 and 4.8 times their uncertainties), and GC8631 of 330~K (2.2 times), although its two temperature indicators agree to within 3~K \citep{Thorsbro_2020}, so a temperature error alone is an unlikely explanation for these three stars.
These two failure modes point to a degeneracy
that cannot be broken with the data available here.
For Feld31, Feld84, and GC8631, the observed position is
consistent either with a non-solar
$\alpha_{\rm MLT}$ on the RGB or with a more
advanced evolutionary phase whose track happens to cross
the same region of the HR diagram; photometry alone
cannot distinguish between these two scenarios.
For GC8623, the two branches can be compared quantitatively. Since the probability of observing a star in a given phase is proportional to the time it spends there, we gave each branch the weight
\begin{equation}
    W = \int \exp\left(-\frac{\chi^2_{\rm HR}(t)}{2}\right) {\rm d}t ,
    \label{eq:branch_weight}
\end{equation}
which uses the same Gaussian likelihood as the calibration and therefore accounts both for the time spent near the observed position and for the quality of the fit; since the initial mass is the same on both branches, the IMF does not affect the ratio of the two weights. The AGB track reaches the observed position within $1\sigma$, but only during the last $\sim$1.5~kyr of the AGB, whereas the RGB track with $\alpha_{\rm MLT}^* = 2.1$ never comes within $1\sigma$ but stays within $2\sigma$ ($\chi^2_{\rm HR} \leq 6.18$) for the final $\sim$330~kyr before the RGB tip. With $\alpha_{\rm MLT} = 2.1$ on both branches, $W_{\rm RGB} \approx 50$~kyr and $W_{\rm AGB} \approx 14$~kyr, so the RGB solution is favoured by a factor of about 4. This preference, however, depends on accepting a super-solar mixing length: at the nominal $\alpha_{\rm MLT}$, the RGB track never comes within $2\sigma$, and the two branches have similar weights ($W_{\rm RGB}/W_{\rm AGB} \approx 0.8$). The RGB solution is thus moderately favoured, but the AGB solution cannot be excluded. The two solutions predict clearly different oscillation spectra ($\nu_{\rm max} = 0.21$ and $0.08\,\mu$Hz; Table~\ref{tab:astero_ambiguous}), so asteroseismic observations could discriminate between them.
This is precisely the kind of ambiguity that only
precision spectroscopy or resolved asteroseismology can
resolve on a star-by-star basis, and constitutes a clear
motivation for facilities such as the mid-infrared
spectrograph METIS at the Extremely Large Telescope (ELT), and for the asteroseismic
predictions presented in this work in the context of the
upcoming JASMINE and Roman missions
(Sect.~\ref{sec:disc_prospects}).
As noted in Sect.~\ref{sec:difficult_cases}, the
$\alpha_{\rm MLT}^*$ values for Feld31, Feld84, and
GC8631 remain a diagnostic rather than a physical
determination.
Their value lies in the future: once asteroseismic
detections become available from JASMINE or Roman
(Sect.~\ref{sec:disc_prospects}), the measured
$\Delta\nu$ will constrain $\alpha_{\rm MLT}$
independently and allow it to enter the MCMC
calibration directly (Sect.~\ref{sec:mcmc}), breaking
the degeneracy that currently limits these three stars.
\subsection{Prospects for asteroseismic detection}
\label{sec:disc_prospects}
The gravity-mode period spacings, $\Delta\Pi_{\ell=1}$,
span $23.1$--$38.2$~s across the sample.
These values are systematically lower than the
$\Delta\Pi \sim 50$--$80$~s range observed by
Kepler for RGB stars
\citep{Bedding_2011, Mosser_2012}, but that comparison
sample comprises predominantly less luminous red giants
ascending the lower and middle RGB, whereas the NSC stars
studied here occupy the upper RGB tip
($\log L \sim 2.8$--$3.4$).
Along the RGB, the gravity-mode period spacing
decreases monotonically with increasing luminosity as
the Brunt-V\"{a}is\"{a}l\"{a} frequency profile
evolves and the g-mode propagation cavity shrinks
\citep{Mosser_2012}.
We caution, however, that at the luminosities of our
sample ($\log L \sim 2.8$--$3.4$) the gravity-dominated
mixed modes that carry the $\Delta\Pi$ signal are
strongly damped by radiative losses in the
core-envelope coupling region, and robust $\Delta\Pi$
measurements in Kepler data are limited to
much less luminous stars \citep{Mosser_2014_delta_pi};
we therefore do not present $\Delta\Pi$ as a
near-term observational target, but retain it as a
theoretical diagnostic of internal structure.
We further note that the red-clump ambiguity that $\Delta\Pi$ is designed to resolve does not apply to this sample in any case, since our targets lie substantially above the red clump in luminosity.
These predictions provide theoretical estimates of
asteroseismic parameters for individual stellar members
of the Galactic NSC, complementing the population-level
isochrone modelling of \citet{Gallego-Cano_2026} with
star-by-star predictions, and offer concrete
observational targets for future photometric campaigns.
The upcoming near-infrared mission JASMINE
\citep{kawata2024_jasmine}, designed specifically to
observe the Galactic centre at wavelengths where the
extreme line-of-sight extinction is less severe, and
the Nancy Grace Roman Space Telescope
\citep{Huber_2023, Weiss_2025}, whose Galactic Bulge
Time-Domain Survey (GBTDS) will monitor the bulge region at high cadence and for which a field centred on the Galactic centre has also been considered \citep{Weiss_2025}, together offer a realistic
prospect of detecting solar-like oscillations in NSC
red giants.
Among our sample, GC16867 and GC8631 represent the
most promising immediate targets given their
comparatively high $\nu_{\rm max}$, compared to the
remaining stars, most of which have
$\nu_{\rm max} < 0.5\,\mu$Hz.
Confirmation of the predicted $\nu_{\rm max}$,
$\Delta\nu$, and $\Delta\Pi$ values would
simultaneously validate the calibration procedure
presented here and open a new avenue for constraining
the masses, ages, and evolutionary states of NSC stars
independently of stellar evolution models.
For most targets, the frequency resolution afforded by a
single 72-day GBTDS season,
$\Delta f \approx 1/T_{\rm obs} \approx 0.16\,\mu\mathrm{Hz}$,
is comparable to or larger than the predicted $\Delta\nu$
itself (Table~\ref{tab:astero_combined}), meaning that
only $\nu_{\rm max}$, not $\Delta\nu$, is likely to be
measurable for the majority of the sample; $\Delta\nu$
measurements will require either repeated seasons or the
most favourable targets (e.g.\ GC16867, GC8631).\footnote{\url{https://roman-docs.stsci.edu/roman-community-defined-surveys/galactic-bulge-time-domain-survey}}

\section{Conclusions}
\label{section_5}
We have presented individual MESA stellar evolution models and asteroseismic predictions for red giant stars in the NSC
of the Milky Way, combining near-infrared spectroscopic
observations from four high-resolution catalogues with
a Bayesian MCMC calibration against the PARSEC
isochrone grid and evolutionary models computed with
MESA.
Of the 23 high-resolution physical stars in the sample,
19 yield satisfactory MESA model fits under standard
assumptions, with calibrated initial masses spanning
$1.07$--$1.47\,M_\odot$, ages $3.3$--$8.3$~Gyr, and
metallicities $-0.31$ to $+0.43$~dex
(Sect.~\ref{sec:good_fits}).
Given that the systematic differences between the spectroscopic catalogues (up to 0.25~dex for the same star) are comparable to the $0.21$~dex standard deviation of the
distribution itself, we do not consider the present
sample sufficiently precise to establish intrinsic metallicity dispersion in the NSC; a statistical treatment of this question,
together with a sensitivity analysis of the adopted
age prior, is left for future work.
The remaining four stars (Feld31, Feld84,
GC8623, and GC8631) require a dedicated modelling
strategy (Sect.~\ref{sec:bad_fits}).
For Feld31, Feld84, and GC8631, the mismatch is a
lateral offset in the HR diagram that a star-specific
mixing-length parameter $\alpha_{\rm MLT}^*$ resolves.
For GC8623, the mismatch is instead a
luminosity excess, which a super-solar $\alpha_{\rm MLT}$
on the RGB reproduces only marginally and the nominal track reproduces closely only near the end of the AGB. Weighting each branch by the time spent near the observed position and by the quality of the fit, the RGB solution with a super-solar $\alpha_{\rm MLT}$ is favoured by a factor of about four; this preference disappears at the nominal $\alpha_{\rm MLT}$, so the AGB solution cannot be excluded.
Both explanations remain degenerate with the data
available here, and photometry alone cannot
distinguish a non-solar $\alpha_{\rm MLT}$ on the RGB
from a more advanced evolutionary phase.
Breaking this degeneracy, and validating the
$\alpha_{\rm MLT}^*$ values as more than a
calibration diagnostic, will require an independent
constraint on the internal structure such as a resolved
asteroseismic detection, or precision spectroscopy, motivating
facilities such as the mid-infrared spectrograph METIS
at the ELT.
From the calibrated MESA models we extract theoretical
predictions of $\nu_{\rm max}$, $\Delta\nu$, and
$\Delta\Pi$ for individual NSC red giants
(Table~\ref{tab:astero_combined}), spanning, for the nominal models,
$\nu_{\rm max} \approx 0.23$--$0.97\,\mu$Hz,
$\Delta\nu \approx 0.10$--$0.32\,\mu$Hz, and
$\Delta\Pi \approx$ $29.1$--$38.1$~s.
Given the extreme luminosities of this sample
($\log L \sim 2.8$--$3.4$), we caution that $\Delta\Pi$
is not a realistic near-term observational target: the
gravity-dominated mixed modes that carry this signal
are strongly damped at these luminosities, and robust measurements in Kepler data are limited to much less luminous giants.
We therefore retain $\Delta\Pi$ as a theoretical
diagnostic of internal structure rather than an
observational target.
The predicted $\nu_{\rm max}$ and $\Delta\nu$ values,
by contrast, provide concrete targets for the upcoming
JASMINE and Nancy Grace Roman Space Telescope missions;
among our sample, GC16867 and GC8631 are the most
promising immediate targets given their comparatively
high $\nu_{\rm max}$.
We validated the reliability of these asymptotic $\Delta\nu$ predictions by computing oscillation-mode frequencies
directly with the stellar oscillation code GYRE for the
full sample: excluding one AGB model with too few
modes for a robust fit, the asymptotic $\Delta\nu$ agrees with the GYRE value to within $1.8 \pm 0.4$~percent (Appendix~\ref{appendix:gyre_comparison}).
A limitation of the approach presented here is that
the PARSEC isochrone grid used in the calibration does
not incorporate variations of $\alpha_{\rm MLT}$: the
initial parameters are calibrated under the
solar-calibrated value, and $\alpha_{\rm MLT}^*$ is
determined only a posteriori as an independent
free parameter, which will always improve the fit to
the HR diagram.
The diagnostic value of this exercise lies in its future
application: once asteroseismic detections become
available for the stars discussed in
Sect.~\ref{sec:bad_fits}, the measured $\Delta\nu$
will constrain the internal structure and hence
$\alpha_{\rm MLT}$ independently, allowing it to be
incorporated directly into the MCMC calibration and
breaking the degeneracies identified above.
More broadly, higher signal-to-noise spectroscopy and
independent photometric age indicators for the NSC
population would significantly advance the precision of
isochrone-based modelling in this region, and progress
on both fronts can be expected from the next generation
of near-infrared space missions and extremely large
telescopes.
\paragraph{Data availability.}
All MESA and GYRE models used in this work are publicly
available on Zenodo at \url{https://doi.org/10.5281/zenodo.21989982}
(DOI: 10.5281/zenodo.21989982).

\begin{acknowledgements}

IL thanks the FCT - Fundação para a Ciência e Tecnologia, I.P. for the financial support to the Centre for Astrophysics and Gravitation (CENTRA/IST/ULisboa) through grant No.~UID/PRR/00099/2025 (DOI: 10.54499/UID/PRR/00099/2025) and grant No.~UID/00099/2025 (DOI: 10.54499/UID/00099/2025).

DC acknowledges support by FCT - Fundação para a Ciência e Tecnologia, I.P. through the grant No.~2024.01352.BD (DOI: https://doi.org/10.54499/2024.01352.BD).

This work made use of the NASA Astrophysics Data System.

\end{acknowledgements}

\bibliographystyle{aa}
\bibliography{references}

@ARTICLE{GRAVITY_2021,
  author  = {{GRAVITY Collaboration} and {Abuter}, R. and {Amorim}, A.
             and {Baub{\"o}ck}, M. and {Berger}, J.~P. and {Bonnet}, H.
             and others},
  title   = "{Improved GRAVITY astrometric accuracy from modeling of
             optical aberrations}",
  journal = {\aap},
  year    = 2021,
  volume  = {647},
  eid     = {A59},
  pages   = {A59},
  doi     = {10.1051/0004-6361/202040208},
  eprint  = {2101.12098},
  archivePrefix = {arXiv},
  primaryClass  = {astro-ph.GA},
  adsurl  = {https://ui.adsabs.harvard.edu/abs/2021A%26A...647A..59G}
}

@ARTICLE{Gallego-Cano_2026,
  author  = {{Gallego-Cano}, E. and {Sch{\"o}del}, R. and {Fritz}, T.~K.
             and {Hosek}, M.~W. and {Muzic}, K. and {Feldmeier-Krause}, A.
             and {Mart{\'i}nez-Arranz}, {\'A}. and {Nogueras-Lara}, F.},
  title   = "{Age and metallicity of the Milky Way's nuclear star
             cluster studied at 3 pc from Sagittarius A*}",
  journal = {\aap},
  year    = 2026,
  volume  = 708,
  pages   = {A77},
  doi     = {10.1051/0004-6361/202558555}
}

@ARTICLE{Schodel_2014,
  author  = {{Sch{\"o}del}, R. and {Feldmeier}, A. and {Kunneriath}, D. and {Stolovy}, S. and {Neumayer}, N. and {Amaro-Seoane}, P. and {Nishiyama}, S.},
  title   = "{Surface brightness profile of the Milky Way's nuclear star cluster}",
  journal = {\aap}, year = 2014, volume = {566}, pages = {A47},
  doi     = {10.1051/0004-6361/201423481}
}

@ARTICLE{Fritz_2016,
  author  = {{Fritz}, T.~K. and {Chatzopoulos}, S. and {Gerhard}, O. and others},
  title   = "{The Nuclear Cluster of the Milky Way: Total Mass and Luminosity}",
  journal = {\apj}, year = 2016, volume = {821}, number = {1}, pages = {44},
  doi     = {10.3847/0004-637X/821/1/44}
}

@ARTICLE{Nogueras-Lara_2018,
  author = {{Nogueras-Lara}, F. and {Gallego-Calvente}, A.~T. and {Dong}, H. and {Gallego-Cano}, E. and {Girard}, J.~H.~V. and {Hilker}, M. and {de Zeeuw}, P.~T. and {Feldmeier-Krause}, A. and {Nishiyama}, S. and {Najarro}, F. and {Neumayer}, N. and {Sch{\"o}del}, R.},
  title = "{GALACTICNUCLEUS: A high angular resolution JHKs imaging survey of the Galactic centre - I. Methodology, performance, and near-infrared extinction towards the Galactic centre}",
  journal = {\aap},
  year = 2018,
  volume = 610,
  pages = "A83",
  doi = "10.1051/0004-6361/201732002"
}

@ARTICLE{Thorsbro_2020,
  author = {{Thorsbro}, B. and {Ryde}, N. and {Rich}, R.~M. and {Schultheis}, M. and {Renaud}, F. and {Spitoni}, E. and {Fritz}, T.~K. and {Mastrobuono-Battisti}, A. and {Origlia}, L. and {Matteucci}, F. and {Sch{\"o}del}, R.},
  title = "{Detailed Abundances in the Galactic Center: Evidence of a Metal-rich Alpha-enhanced Stellar Population}",
  journal = {\apj},
  year = 2020,
  month = may,
  volume = {894},
  number = {1},
  eid = {26},
  pages = {26},
  doi = {10.3847/1538-4357/ab8226},
  archivePrefix = {arXiv},
  eprint = {2003.11085},
  primaryClass = {astro-ph.GA},
  adsurl = {https://ui.adsabs.harvard.edu/abs/2020ApJ...894...26T}
}

@ARTICLE{FK_2017,
  author = {{Feldmeier-Krause}, A. and {Kerzendorf}, W. and {Neumayer}, N. and {Sch{\"o}del}, R. and {Nogueras-Lara}, F. and {Do}, T. and {de Zeeuw}, P.~T. and {Kuntschner}, H.},
  title = "{KMOS view of the Galactic Centre - II. Metallicity distribution of late-type stars}",
  journal = {\mnras},
  year = 2017,
  month = jan,
  volume = {464},
  number = {1},
  pages = {194-209},
  doi = {10.1093/mnras/stw2339},
  archivePrefix = {arXiv},
  eprint = {1610.01623},
  primaryClass = {astro-ph.GA},
  adsurl = {https://ui.adsabs.harvard.edu/abs/2017MNRAS.464..194F}
}

@ARTICLE{Chen_2023,
  author = {{Chen}, Zhuo and {Do}, Tuan and {Ghez}, Andrea M. and {Hosek}, Matthew W. and {Feldmeier-Krause}, Anja and {Chu}, Devin S. and {Bentley}, Rory O. and {Lu}, Jessica R. and {Morris}, Mark R.},
  title = "{The Star Formation History of the Milky Way's Nuclear Star Cluster}",
  journal = {\apj},
  year = 2023,
  month = feb,
  volume = {944},
  number = {1},
  eid = {79},
  pages = {79},
  doi = {10.3847/1538-4357/aca8ad},
  archivePrefix = {arXiv},
  eprint = {2212.01397},
  primaryClass = {astro-ph.GA},
  adsurl = {https://ui.adsabs.harvard.edu/abs/2023ApJ...944...79C}
}

@ARTICLE{Schodel_2020,
  author  = {{Sch{\"o}del}, R. and {Nogueras-Lara}, F. and {Gallego-Cano}, E. and {Shahzamanian}, B. and {Gallego-Calvente}, A.~T. and {Gardini}, A.},
  title   = "{The Milky Way's nuclear star cluster: Old, metal-rich, and cuspy. Structure and star formation history from deep imaging}",
  journal = {\aap}, year = 2020, volume = {641}, pages = {A102},
  doi     = {10.1051/0004-6361/201936688}
}

@ARTICLE{Nogueras-Lara_2023,
  author  = {{Nogueras-Lara}, F. and {Feldmeier-Krause}, A. and {Sch{\"o}del}, R. and {Sormani}, M.~C. and others},
  title   = "{Smooth kinematic and metallicity gradients reveal that the Milky Way's nuclear star cluster and disc might be part of the same structure}",
  journal = {\aap}, year = 2023, volume = {680}, pages = {A75},
  doi     = {10.1051/0004-6361/202347421}
}

@ARTICLE{Schodel_2010,
  author    = {{Sch{\"o}del}, R. and {Najarro}, F. and {Muzic}, K.
               and {Eckart}, A.},
  title     = "{Peering through the veil: near-infrared photometry and
               extinction for the Galactic nuclear star cluster. Accurate
               near infrared H, K$_s$, and L' photometry and the
               near-infrared extinction-law toward the central parsec of
               the Galaxy}",
  journal   = {\aap},
  year      = 2010,
  volume    = {511},
  eid       = {A18},
  pages     = {A18},
  doi       = {10.1051/0004-6361/200913183},
  eprint    = {0912.1273},
  archivePrefix = {arXiv},
  primaryClass  = {astro-ph.GA},
  adsurl    = {https://ui.adsabs.harvard.edu/abs/2010A%26A...511A..18S}
}

@ARTICLE{Do_2015,
  author  = {{Do}, T. and {Kerzendorf}, W. and {Winsor}, N. and others},
  title   = "{Discovery of Low-metallicity Stars in the Central Parsec of the Milky Way}",
  journal = {\apj},
  year    = 2015,
  volume  = 809,
  pages   = {143},
  doi     = {10.1088/0004-637X/809/2/143},
  eprint  = {1506.07891},
  archivePrefix = {arXiv},
  primaryClass  = {astro-ph.GA}
}

@ARTICLE{Rich_2017,
  author = {{Rich}, R.~M. and {Ryde}, N. and {Thorsbro}, B. and {Fritz}, T.~K. and {Schultheis}, M. and {Origlia}, L. and {J{\"o}nsson}, H.},
  title = "{Detailed Abundances for the Old Population near the Galactic Center. I. Metallicity Distribution of the Nuclear Star Cluster}",
  journal = {\aj},
  year = 2017,
  month = dec,
  volume = {154},
  number = {6},
  eid = {239},
  pages = {239},
  doi = {10.3847/1538-3881/aa970a},
  archivePrefix = {arXiv},
  eprint = {1710.08477},
  primaryClass = {astro-ph.GA},
  adsurl = {https://ui.adsabs.harvard.edu/abs/2017AJ....154..239R}
}

@ARTICLE{Do_2018,
  author = {{Do}, Tuan and {Kerzendorf}, Wolfgang and {Konopacky}, Quinn and {Marcinik}, Joseph M. and {Ghez}, Andrea and {Lu}, Jessica R. and {Morris}, Mark R.},
  title = "{Super-solar Metallicity Stars in the Galactic Center Nuclear Star Cluster: Unusual Sc, V, and Y Abundances}",
  journal = {\apjl},
  year = 2018,
  month = mar,
  volume = {855},
  number = {1},
  eid = {L5},
  pages = {L5},
  doi = {10.3847/2041-8213/aaaec3},
  archivePrefix = {arXiv},
  eprint = {1802.08270},
  primaryClass = {astro-ph.GA},
  adsurl = {https://ui.adsabs.harvard.edu/abs/2018ApJ...855L...5D}
}

@ARTICLE{Ryde_2025,
  author = {{Ryde}, Nils and {Nandakumar}, Govind and {Schultheis}, Mathias and {Kordopatis}, Georges and {di Matteo}, Paola and {Haywood}, Misha and {Sch{\"o}del}, Rainer and {Nogueras-Lara}, Francisco and {Rich}, R. Michael and {Thorsbro}, Brian and {Mace}, Gregory N. and {Agertz}, Oscar and {Amarsi}, Anish M. and {Kocher}, Jessica and {Molero}, Marta and {Origlia}, Livia and {Pagnini}, Giulia and {Spitoni}, Emanuele},
  title = "{Chemical Abundances in the Nuclear Star Cluster of the Milky Way: Alpha-element Trends and Their Similarities with the Inner Bulge}",
  journal = {\apj},
  year = 2025,
  month = feb,
  volume = {979},
  number = {2},
  eid = {174},
  pages = {174},
  doi = {10.3847/1538-4357/ad9b2b},
  archivePrefix = {arXiv},
  eprint = {2412.04528},
  primaryClass = {astro-ph.GA},
  adsurl = {https://ui.adsabs.harvard.edu/abs/2025ApJ...979..174R}
}

@ARTICLE{Brown_1991,
  author    = {{Brown}, T.~M. and {Gilliland}, R.~L.
               and {Noyes}, R.~W. and {Ramsey}, L.~W.},
  title     = "{Detection of possible p-mode oscillations on Procyon}",
  journal   = {\apj},
  year      = 1991,
  volume    = {368},
  pages     = {599--609},
  doi       = {10.1086/169725},
  adsurl    = {https://ui.adsabs.harvard.edu/abs/1991ApJ...368..599B}
}

@ARTICLE{Kjeldsen_1995,
  author    = {{Kjeldsen}, H. and {Bedding}, T.~R.},
  title     = "{Amplitudes of stellar oscillations: the implications
               for asteroseismology}",
  journal   = {\aap},
  year      = 1995,
  volume    = {293},
  pages     = {87--106},
  eprint    = {astro-ph/9403015},
  archivePrefix = {arXiv},
  adsurl    = {https://ui.adsabs.harvard.edu/abs/1995A%26A...293...87K}
}

@ARTICLE{Huber_2011,
  author  = {{Huber}, D. and {Bedding}, T.~R. and {Stello}, D. and
             {Hekker}, S. and {Mathur}, S. and {Mosser}, B. and others},
  title   = "{Testing Scaling Relations for Solar-like Oscillations
             from the Main Sequence to Red Giants Using Kepler Data}",
  journal = {\apj},
  year    = 2011,
  volume  = {743},
  eid     = {143},
  pages   = {143},
  doi     = {10.1088/0004-637X/743/2/143},
  eprint  = {1109.3460},
  archivePrefix = {arXiv},
  primaryClass  = {astro-ph.SR},
  adsurl  = {https://ui.adsabs.harvard.edu/abs/2011ApJ...743..143H}
}

@BOOK{Asteroseismology_book,
  author    = {{Aerts}, C. and {Christensen-Dalsgaard}, J{\o}rgen and {Kurtz}, D.~W.},
  title     = "{Asteroseismology}",
  year      = 2010,
  publisher = {Springer},
  address   = {Dordrecht},
  series    = {Astronomy and Astrophysics Library},
  doi       = {10.1007/978-1-4020-5803-5}
}

@ARTICLE{Bedding_2011,
  author    = {{Bedding}, T.~R. and {Mosser}, B. and {Huber}, D.
               and {Montalb{\'a}n}, J. and {Beck}, P. and {Christensen-Dalsgaard}, J.
               and {Elsworth}, Y.~P. and {Garc{\'i}a}, R.~A.
               and {Miglio}, A. and {Stello}, D. and others},
  title     = "{Gravity modes as a way to distinguish between hydrogen-
               and helium-burning red giant stars}",
  journal   = {\nat},
  year      = 2011,
  volume    = {471},
  pages     = {608--611},
  doi       = {10.1038/nature09935},
  eprint    = {1103.5805},
  archivePrefix = {arXiv},
  primaryClass  = {astro-ph.SR},
  adsurl    = {https://ui.adsabs.harvard.edu/abs/2011Natur.471..608B}
}

@ARTICLE{Mosser_2012,
  author  = {{Mosser}, B. and {Goupil}, M.~J. and {Belkacem}, K. and
             {Michel}, E. and {Stello}, D. and {Marques}, J.~P. and
             others},
  title   = "{Probing the core structure and evolution of red giants using
             gravity-dominated mixed modes observed with Kepler}",
  journal = {\aap},
  year    = 2012,
  volume  = {540},
  eid     = {A143},
  pages   = {A143},
  doi     = {10.1051/0004-6361/201118519}
}

@ARTICLE{Tassoul_1980,
  author  = {{Tassoul}, M.},
  title   = "{Asymptotic approximations for stellar nonradial pulsations}",
  journal = {\apjs},
  year    = 1980,
  volume  = {43},
  pages   = {469--490},
  doi     = {10.1086/190678},
  adsurl  = {https://ui.adsabs.harvard.edu/abs/1980ApJS...43..469T}
}

@ARTICLE{Borucki_2010,
  author    = {{Borucki}, William J. and {Koch}, David and {Basri}, Gibor
               and {Batalha}, Natalie and {Brown}, Timothy and others},
  title     = "{Kepler Planet-Detection Mission: Introduction and
               First Results}",
  journal   = {Science},
  year      = 2010,
  volume    = {327},
  number    = {5968},
  pages     = {977},
  doi       = {10.1126/science.1185402},
  adsurl    = {https://ui.adsabs.harvard.edu/abs/2010Sci...327..977B},
  bibcode   = {2010Sci...327..977B}
}

@ARTICLE{Howell_2014,
  author    = {{Howell}, Steve B. and {Sobeck}, Charlie and {Haas}, Michael
               and {Still}, Martin and {Barclay}, Thomas and others},
  title     = "{The K2 Mission: Characterization and Early Results}",
  journal   = {PASP},
  year      = 2014,
  volume    = {126},
  number    = {938},
  pages     = {398--408},
  doi       = {10.1086/676406},
  adsurl    = {https://ui.adsabs.harvard.edu/abs/2014PASP..126..398H},
  bibcode   = {2014PASP..126..398H}
}

@ARTICLE{Ricker_2015,
  author    = {{Ricker}, George R. and {Winn}, Joshua N. and
               {Vanderspek}, Roland and {Latham}, David W. and others},
  title     = "{Transiting Exoplanet Survey Satellite}",
  journal   = {JATIS},
  year      = 2015,
  volume    = {1},
  number    = {1},
  pages     = {014003},
  doi       = {10.1117/1.JATIS.1.1.014003},
  adsurl    = {https://ui.adsabs.harvard.edu/abs/2015JATIS...1a4003R},
  bibcode   = {2015JATIS...1a4003R}
}

@ARTICLE{Chaplin_2013,
  author    = {{Chaplin}, William J. and {Miglio}, Andrea},
  title     = "{Asteroseismology of Solar-Type and Red-Giant Stars}",
  journal   = {ARA\&A},
  year      = 2013,
  volume    = {51},
  pages     = {353--392},
  doi       = {10.1146/annurev-astro-082812-140938},
  adsurl    = {https://ui.adsabs.harvard.edu/abs/2013ARA&A..51..353C},
  bibcode   = {2013ARA&A..51..353C}
}

@ARTICLE{Miglio_2013,
  author    = {{Miglio}, A. and {Chiappini}, C. and {Morel}, T. and
               {Barbieri}, M. and {Chaplin}, W.~J. and {Girardi}, L. and
               {Montalb{\'a}n}, J. and {Valentini}, M. and
               {Mosser}, B. and {Baudin}, F. and {Casagrande}, L. and
               {Fossati}, L. and {Silva Aguirre}, V. and {Baglin}, A.},
  title     = "{Galactic archaeology: mapping and dating stellar
               populations with asteroseismology of red-giant stars}",
  journal   = {MNRAS},
  year      = 2013,
  volume    = {429},
  number    = {1},
  pages     = {423--428},
  doi       = {10.1093/mnras/sts345},
  adsurl    = {https://ui.adsabs.harvard.edu/abs/2013MNRAS.429..423M},
  bibcode   = {2013MNRAS.429..423M}
}

@ARTICLE{kawata2024_jasmine,
  author  = {{Kawata}, D. and {Kawahara}, H. and {Gouda}, N. and others},
  title   = "{JASMINE: Near-infrared astrometry and time-series photometry science}",
  journal = {\pasj},
  year    = 2024,
  volume  = {76},
  number  = {3},
  pages   = {386--425},
  doi     = {10.1093/pasj/psae020},
  eprint  = {2307.05666},
  archivePrefix = {arXiv},
  primaryClass  = {astro-ph.IM}
}

@ARTICLE{Huber_2023,
  author    = {{Huber}, Daniel and {Pinsonneault}, Marc and
               {Beck}, Paul and {Bedding}, Timothy R. and
               {Bland-Hawthorn}, Joss and others},
  title     = "{Asteroseismology with the Roman Galactic Bulge
               Time-Domain Survey}",
  journal   = {arXiv e-prints},
  year      = 2023,
  eid       = {arXiv:2307.03237},
  pages     = {arXiv:2307.03237},
  doi       = {10.48550/arXiv.2307.03237},
  adsurl    = {https://ui.adsabs.harvard.edu/abs/2023arXiv230703237H}
}

@ARTICLE{Weiss_2025,
  author  = {{Weiss}, T.~J. and {Downing}, N.~J. and {Pinsonneault}, M.~H.
             and {Zinn}, J.~C. and {Stello}, D. and {Bedding}, T.~R.
             and {Cao}, K. and {Hon}, M. and {Reyes}, C. and {Gaudi}, B.~S.
             and {Wilson}, R.~F. and {Huber}, D. and {Sharma}, S.},
  title   = "{Modeling Asteroseismic Yields for the Roman Galactic Bulge Time-Domain Survey}",
  journal = {\apj},
  year    = 2025,
  volume  = {987},
  number  = {2},
  eid     = {181},
  pages   = {181},
  doi     = {10.3847/1538-4357/adde5b},
  eprint  = {2503.04999},
  archivePrefix = {arXiv},
  primaryClass  = {astro-ph.SR}
}

@ARTICLE{Paxton2011,
  author  = {{Paxton}, B. and {Bildsten}, L. and {Dotter}, A. and others},
  title   = "{Modules for Experiments in Stellar Astrophysics (MESA)}",
  journal = {\apjs},
  year    = 2011,
  volume  = 192,
  pages   = {3},
  doi     = {10.1088/0067-0049/192/1/3}
}

@ARTICLE{Paxton2013,
  author  = {{Paxton}, B. and {Cantiello}, M. and {Arras}, P. and others},
  title   = "{Modules for Experiments in Stellar Astrophysics (MESA): Planets, Oscillations, Rotation, and Massive Stars}",
  journal = {\apjs},
  year    = 2013,
  volume  = 208,
  pages   = {4},
  doi     = {10.1088/0067-0049/208/1/4}
}

@ARTICLE{Paxton2015,
  author  = {{Paxton}, B. and {Marchant}, P. and {Schwab}, J. and others},
  title   = "{Modules for Experiments in Stellar Astrophysics (MESA): Binaries, Pulsations, and Explosions}",
  journal = {\apjs},
  year    = 2015,
  volume  = 220,
  pages   = {15},
  doi     = {10.1088/0067-0049/220/1/15}
}

@ARTICLE{Paxton2018,
  author  = {{Paxton}, B. and {Schwab}, J. and {Bauer}, E.~B. and others},
  title   = "{Modules for Experiments in Stellar Astrophysics (MESA): Convective Boundaries, Element Diffusion, and Massive Star Explosions}",
  journal = {\apjs},
  year    = 2018,
  volume  = 234,
  pages   = {34},
  doi     = {10.3847/1538-4365/aaa5a8}
}

@ARTICLE{Paxton2019,
  author  = {{Paxton}, B. and {Smolec}, R. and {Schwab}, J. and others},
  title   = "{Modules for Experiments in Stellar Astrophysics (MESA): Pulsating Variable Stars, Rotation, Convective Boundaries, and Energy Conservation}",
  journal = {\apjs},
  year    = 2019,
  volume  = 243,
  pages   = {10},
  doi     = {10.3847/1538-4365/ab2241}
}

@ARTICLE{Jermyn2023,
  author  = {{Jermyn}, Adam S. and {Bauer}, Evan B. and {Schwab}, Josiah and {Farmer}, R. and {Ball}, Warrick H. and {Bellinger}, Earl P. and {Dotter}, Aaron and {Joyce}, Meridith and {Marchant}, Pablo and {Mombarg}, Joey S.~G. and {Wolf}, William M. and {Sunny Wong}, Tin Long and {Cinquegrana}, Giulia C. and {Farrell}, Eoin and {Smolec}, R. and {Thoul}, Anne and {Cantiello}, Matteo and {Herwig}, Falk and {Toloza}, Odette and {Bildsten}, Lars and {Townsend}, Richard H.~D. and {Timmes}, F.~X.},
  title   = "{Modules for Experiments in Stellar Astrophysics (MESA): Time-dependent Convection, Energy Conservation, Automatic Differentiation, and Infrastructure}",
  journal = {\apjs},
  year    = 2023,
  volume  = {265},
  number  = {1},
  pages   = {15},
  doi     = {10.3847/1538-4365/acae8d}
}

@ARTICLE{Bressan_2012,
  author    = {{Bressan}, A. and {Marigo}, P. and {Girardi}, L.
               and {Salasnich}, B. and {Dal Cero}, C.
               and {Rubele}, S. and {Nanni}, A.},
  title     = "{PARSEC: stellar tracks and isochrones with the PAdova
               and TRieste Stellar Evolution code}",
  journal   = {\mnras},
  year      = 2012,
  volume    = {427},
  pages     = {127--145},
  doi       = {10.1111/j.1365-2966.2012.21948.x},
  eprint    = {1208.4498},
  archivePrefix = {arXiv},
  primaryClass  = {astro-ph.SR},
  adsurl    = {https://ui.adsabs.harvard.edu/abs/2012MNRAS.427..127B}
}

@ARTICLE{Townsend_2013,
  author  = {{Townsend}, R.~H.~D. and {Teitler}, S.~A.},
  title   = "{GYRE: an open-source stellar oscillation code based on
             a new Magnus Multiple Shooting scheme}",
  journal = {\mnras},
  year    = 2013,
  volume  = {435},
  pages   = {3406--3418},
  doi     = {10.1093/mnras/stt1533},
  eprint  = {1308.2965},
  archivePrefix = {arXiv},
  primaryClass  = {astro-ph.SR},
  adsurl  = {https://ui.adsabs.harvard.edu/abs/2013MNRAS.435.3406T}
}

@ARTICLE{Joyce_2018,
  author    = {{Joyce}, M. and {Chaboyer}, B.},
  title     = "{Not All Stars Are the Sun: Empirical Calibration of the
               Mixing Length for Metal-poor Stars Using
               One-dimensional Stellar Evolution Models}",
  journal   = {\apj},
  year      = 2018,
  volume    = {856},
  eid       = {10},
  pages     = {10},
  doi       = {10.3847/1538-4357/aab200},
  eprint    = {1712.05082},
  archivePrefix = {arXiv},
  primaryClass  = {astro-ph.SR},
  adsurl    = {https://ui.adsabs.harvard.edu/abs/2018ApJ...856...10J}
}

@INPROCEEDINGS{McGregor_2003,
  author    = {{McGregor}, P.~J. and {Hart}, J. and {Conroy}, P.~G.
               and {Pfitzner}, M.~L. and {Bloxham}, G.~J.
               and {Jones}, D.~J. and {Downing}, M.~D.
               and {Dawson}, M. and {Young}, P. and {Jarnyk}, M.
               and {Van Harmelen}, J.},
  title     = "{Gemini near-infrared integral field spectrograph (NIFS)}",
  booktitle = {Instrument Design and Performance for Optical/Infrared
               Ground-based Telescopes},
  series    = {Proc.\ SPIE},
  volume    = {4841},
  pages     = {1581--1591},
  year      = 2003,
  doi       = {10.1117/12.459448},
  adsurl    = {https://ui.adsabs.harvard.edu/abs/2003SPIE.4841.1581M}
}

@ARTICLE{Sharples_2013,
  author    = {{Sharples}, R. and {Bender}, R. and {Agudo Berbel}, A.
               and {Bezawada}, N. and {Castillo}, R. and {Cirasuolo}, M.
               and {Davidson}, G. and {Davies}, R. and {Dubbeldam}, M.
               and {Fairley}, A. and {Finger}, G.
               and {F{\"o}rster Schreiber}, N. and {Gonte}, F.
               and {Hess}, A. and {Jung}, I. and {Lewis}, I.
               and {Lizon}, J.-L. and {Muschielok}, B. and {Pasquini}, L.
               and {Pirard}, J. and {Popovic}, D. and {Ramsay}, S.
               and {Rees}, P. and {Richter}, J. and {Riquelme}, M.
               and {Rodrigues}, M. and {Saviane}, I. and {Schlichter}, J.
               and {Schmidtobreick}, L. and {Segovia}, A. and {Smette}, A.
               and {Szeifert}, T. and {van Kesteren}, A. and {Wegner}, M.
               and {Wiezorrek}, E.},
  title     = "{First Light for the KMOS Multi-Object Integral-Field Spectrometer}",
  journal   = {The Messenger},
  year      = 2013,
  volume    = {151},
  pages     = {21--23},
  adsurl    = {https://ui.adsabs.harvard.edu/abs/2013Msngr.151...21S}
}

@INPROCEEDINGS{McLean_1998,
  author    = {{McLean}, I.~S. and {Becklin}, E.~E. and {Bendiksen}, O.
               and {Brims}, G. and {Canfield}, J. and {Figer}, D.~F.
               and {Graham}, J.~R. and {Hare}, J. and {Lacayanga}, F.
               and {Larkin}, J.~E. and {Larson}, S.~B.
               and {Levenson}, N. and {Magnone}, N. and {Teplitz}, H.
               and {Wong}, W.},
  title     = "{Design and development of NIRSPEC: a near-infrared echelle
               spectrograph for the Keck II telescope}",
  booktitle = {Infrared Astronomical Instrumentation},
  series    = {Proc.\ SPIE},
  volume    = {3354},
  pages     = {566--578},
  year      = 1998,
  doi       = {10.1117/12.317283},
  adsurl    = {https://ui.adsabs.harvard.edu/abs/1998SPIE.3354..566M}
}

@ARTICLE{Fritz_2021,
  author  = {{Fritz}, T.~K. and {Patrick}, L.~R. and {Feldmeier-Krause}, A. and {Sch{\"o}del}, R. and {Schultheis}, M. and others},
  title   = "{A KMOS survey of the nuclear disk of the Milky Way. I. Survey design and metallicities}",
  journal = {\aap}, year = 2021, volume = {649}, pages = {A83},
  doi     = {10.1051/0004-6361/202040026}
}

@ARTICLE{Schultheis_2026,
  author  = {{Schultheis}, M. and {Serrano}, L. and {Thorsbro}, B. and {Nogueras-Lara}, F. and others},
  title   = "{Chemical analysis of the Milky Way's nuclear star cluster. Evidence for a metallicity gradient}",
  journal = {\aap}, year = 2026, volume = {705}, pages = {A235},
  doi     = {10.1051/0004-6361/202557723}
}

@INPROCEEDINGS{Park_2014,
  author    = {{Park}, C. and {Jaffe}, D.~T. and {Yuk}, I.-S.
               and {Chun}, M.-Y. and {Pak}, S. and {Kim}, K.-M.
               and {Pavel}, M. and {Lee}, H. and {Oh}, H.
               and {Jeong}, U. and {Sim}, C.~K. and {Lee}, H.-I.
               and {Nguyen Le}, H.~A. and {Strubhar}, J.
               and {Gully-Santiago}, M. and {Oh}, J.~S.
               and {Cha}, S.-M. and {Moon}, B. and {Park}, K.
               and {Brooks}, C. and {Ko}, K. and {Han}, J.-Y.
               and {Nah}, J. and {Hill}, P.~C. and {Lee}, S.
               and {Barnes}, S. and {Yu}, Y.~S. and {Kaplan}, K.
               and {Mace}, G. and {Kim}, H. and {Lee}, J.-J.
               and {Hwang}, N. and {Park}, B.-G.},
  title     = "{Design and early performance of IGRINS
               (Immersion Grating Infrared Spectrometer)}",
  booktitle = {Ground-based and Airborne Instrumentation for Astronomy V},
  series    = {Proc.\ SPIE},
  volume    = {9147},
  eid       = {91471D},
  pages     = {91471D},
  year      = 2014,
  doi       = {10.1117/12.2056431},
  adsurl    = {https://ui.adsabs.harvard.edu/abs/2014SPIE.9147E..1DP}
}

@ARTICLE{Schultheis_2021,
  author  = {{Schultheis}, M. and {Fritz}, T.~K. and {Nandakumar}, G. and others},
  title   = "{The nuclear stellar disc of the Milky Way: A dynamically cool and metal-rich component possibly formed from the central molecular zone}",
  journal = {\aap}, year = 2021, volume = {650}, pages = {A191},
  doi     = {10.1051/0004-6361/202140499}
}

@ARTICLE{Nogueras-Lara_2022,
  author  = {{Nogueras-Lara}, F.},
  title   = "{Kinematic data rebuild the nuclear star cluster as the most metal-rich region of the Galaxy}",
  journal = {\aap}, year = 2022, volume = {666}, pages = {A72},
  doi     = {10.1051/0004-6361/202244411}
}

@ARTICLE{St_stad_2015,
  author  = {{St{\o}stad}, M. and {Do}, T. and {Murray}, N. and others},
  title   = "{Mapping the Outer Edge of the Young Stellar Cluster in the Galactic Center}",
  journal = {\apj},
  year    = 2015,
  volume  = 808,
  pages   = {106},
  doi     = {10.1088/0004-637X/808/2/106}
}

@ARTICLE{Nogueras-Lara_2021c,
  author  = {{Nogueras-Lara}, F. and {Sch{\"o}del}, R. and {Neumayer}, N.},
  title   = "{The nuclear star cluster and nuclear stellar disc of the Milky Way: Different Stellar Populations and Star Formation Histories}",
  journal = {\apj}, year = 2021, volume = {920}, number = {2}, pages = {97},
  doi     = {10.3847/1538-4357/ac185e}
}

@ARTICLE{Tang_2014,
  author  = {{Tang}, J. and {Bressan}, A. and {Rosenfield}, P.
             and {Slemer}, A. and {Marigo}, P. and {Girardi}, L.
             and {Bianchi}, L.},
  title   = "{New PARSEC evolutionary tracks of massive stars at low
             metallicity: testing canonical stellar evolution in
             nearby star-forming dwarf galaxies}",
  journal = {\mnras},
  year    = 2014,
  volume  = {445},
  pages   = {4287--4305},
  doi     = {10.1093/mnras/stu2029},
  eprint  = {1410.1745},
  archivePrefix = {arXiv},
  adsurl  = {https://ui.adsabs.harvard.edu/abs/2014MNRAS.445.4287T}
}

@ARTICLE{Chen_2014,
  author  = {{Chen}, Y. and {Girardi}, L. and {Bressan}, A.
             and {Marigo}, P. and {Barbieri}, M. and {Kong}, X.},
  title   = "{Improving PARSEC models for very low mass stars}",
  journal = {\mnras},
  year    = 2014,
  volume  = {444},
  pages   = {2525--2543},
  doi     = {10.1093/mnras/stu1605},
  eprint  = {1409.0322},
  archivePrefix = {arXiv},
  adsurl  = {https://ui.adsabs.harvard.edu/abs/2014MNRAS.444.2525C}
}

@ARTICLE{Chen_2015,
  author  = {{Chen}, Y. and {Bressan}, A. and {Girardi}, L.
             and {Marigo}, P. and {Kong}, X. and {Lanza}, A.},
  title   = "{PARSEC evolutionary tracks of massive stars up to 350 M$_\odot$ at metallicities 0.0001 $\leq$ Z $\leq$ 0.04}",
  journal = {\mnras},
  year    = 2015,
  volume  = {452},
  pages   = {1068--1080},
  doi     = {10.1093/mnras/stv1281},
  eprint  = {1506.01681},
  archivePrefix = {arXiv},
  adsurl  = {https://ui.adsabs.harvard.edu/abs/2015MNRAS.452.1068C}
}

@ARTICLE{Reimers_1975,
  author  = {{Reimers}, D.},
  title   = "{Circumstellar absorption lines and mass loss
             from red giants}",
  journal = {M\'emoires de la Soci\'et\'e Royale des Sciences
             de Li\`ege},
  year    = 1975,
  volume  = {8},
  pages   = {369--382},
  adsurl  = {https://ui.adsabs.harvard.edu/abs/1975MSRSL...8..369R}
}

@ARTICLE{Cohen_2003,
  author  = {{Cohen}, M. and {Wheaton}, W.~A. and {Megeath}, S.~T.},
  title   = "{Spectral Irradiance Calibration in the Infrared.
             XIV. The Absolute Calibration of 2MASS}",
  journal = {\aj},
  year    = 2003,
  volume  = {126},
  pages   = {1090--1096},
  doi     = {10.1086/376474},
  adsurl  = {https://ui.adsabs.harvard.edu/abs/2003AJ....126.1090C}
}

@ARTICLE{Marigo_2008,
  author  = {{Marigo}, P. and {Girardi}, L. and {Bressan}, A.
             and {Groenewegen}, M.~A.~T. and {Silva}, L.
             and {Granato}, G.~L.},
  title   = "{Evolution of asymptotic giant branch stars.
             II. Optical to far-infrared isochrones with
             improved TP-AGB models}",
  journal = {\aap},
  year    = 2008,
  volume  = {482},
  pages   = {883--905},
  doi     = {10.1051/0004-6361:20078467},
  adsurl  = {https://ui.adsabs.harvard.edu/abs/2008A%26A...482..883M}
}

@ARTICLE{Trabucchi_2021,
  author  = {{Trabucchi}, M. and {Wood}, P.~R. and {Mowlavi}, N.
             and {Pastorelli}, G. and {Marigo}, P. and {Girardi}, L.
             and {Lebzelter}, T.},
  title   = "{Modelling long-period variables -- II. Fundamental mode
             pulsation in the non-linear regime}",
  journal = {\mnras},
  year    = 2021,
  volume  = {500},
  pages   = {1575--1591},
  doi     = {10.1093/mnras/staa3356},
  eprint  = {2010.13654},
  archivePrefix = {arXiv},
  adsurl  = {https://ui.adsabs.harvard.edu/abs/2021MNRAS.500.1575T}
}

@ARTICLE{Kroupa_2001,
  author  = {{Kroupa}, P.},
  title   = "{On the variation of the initial mass function}",
  journal = {\mnras},
  year    = 2001,
  volume  = {322},
  pages   = {231--246},
  doi     = {10.1046/j.1365-8711.2001.04022.x},
  eprint  = {astro-ph/0009005},
  archivePrefix = {arXiv},
  adsurl  = {https://ui.adsabs.harvard.edu/abs/2001MNRAS.322..231K}
}

@ARTICLE{Salaris_1993,
  author  = {{Salaris}, M. and {Chieffi}, A. and {Straniero}, O.},
  title   = "{The alpha-enhanced isochrones and their impact on the fits to the Galactic globular cluster system}",
  journal = {\apj}, year = 1993, volume = {414}, pages = {580},
  doi     = {10.1086/173105}
}

@ARTICLE{Hosek_2019,
  author  = {{Hosek}, Jr., Matthew W. and {Lu}, Jessica R. and {Anderson}, Jay and {Najarro}, Francisco and {Ghez}, Andrea M. and others},
  title   = "{The Unusual Initial Mass Function of the Arches Cluster}",
  journal = {\apj}, year = 2019, volume = {870}, number = {1}, pages = {44},
  doi     = {10.3847/1538-4357/aaef90}
}

@ARTICLE{Nogueras-Lara_2021,
  author  = {{Nogueras-Lara}, F. and {Sch{\"o}del}, R. and {Neumayer}, N.},
  title   = "{GALACTICNUCLEUS: A high-angular-resolution $JHK_s$ imaging
             survey of the Galactic centre. IV. Extinction maps and
             de-reddened photometry}",
  journal = {\aap},
  year    = 2021,
  volume  = {653},
  eid     = {A133},
  pages   = {A133},
  doi     = {10.1051/0004-6361/202140996},
  eprint  = {2107.00021},
  archivePrefix = {arXiv},
  primaryClass  = {astro-ph.GA},
  adsurl  = {https://ui.adsabs.harvard.edu/abs/2021A%26A...653A.133N}
}

@ARTICLE{Foreman-Mackey_2013,
  author  = {{Foreman-Mackey}, D. and {Hogg}, D.~W.
             and {Lang}, D. and {Goodman}, J.},
  title   = "{emcee: The MCMC Hammer}",
  journal = {\pasp},
  year    = 2013,
  volume  = {125},
  pages   = {306--312},
  doi     = {10.1086/670067},
  eprint  = {1202.3665},
  archivePrefix = {arXiv},
  primaryClass  = {astro-ph.IM},
  adsurl  = {https://ui.adsabs.harvard.edu/abs/2013PASP..125..306F}
}

@ARTICLE{Grevesse_1998,
  author  = {{Grevesse}, N. and {Sauval}, A.~J.},
  title   = "{Standard Solar Composition}",
  journal = {Space Science Reviews},
  year    = 1998,
  volume  = {85},
  pages   = {161--174},
  doi     = {10.1023/A:1005161325181},
  adsurl  = {https://ui.adsabs.harvard.edu/abs/1998SSRv...85..161G}
}

@ARTICLE{Iglesias_Rogers_1996,
  author  = {{Iglesias}, C.~A. and {Rogers}, F.~J.},
  title   = "{Updated Opal Opacities}",
  journal = {\apj},
  year    = 1996,
  volume  = {464},
  pages   = {943--953},
  doi     = {10.1086/177381},
  adsurl  = {https://ui.adsabs.harvard.edu/abs/1996ApJ...464..943I}
}

@ARTICLE{Marigo_2009,
  author  = {{Marigo}, P. and {Aringer}, B.},
  title   = "{Low-temperature gas opacity. \AE SOPUS: a versatile and
             quick computational tool}",
  journal = {\aap},
  year    = 2009,
  volume  = {508},
  pages   = {1539--1569},
  doi     = {10.1051/0004-6361/200912598},
  eprint  = {0907.3248},
  archivePrefix = {arXiv},
  primaryClass  = {astro-ph.SR},
  adsurl  = {https://ui.adsabs.harvard.edu/abs/2009A%26A...508.1539M}
}

@ARTICLE{Bohm-Vitense_1958,
  author  = {{B{\"o}hm-Vitense}, E.},
  title   = "{{\"U}ber die Wasserstoffkonvektionszone in Sternen
             verschiedener Effektivtemperaturen und Leuchtkr{\"a}fte}",
  journal = {Zeitschrift f{\"u}r Astrophysik},
  year    = 1958,
  volume  = {46},
  pages   = {108},
  adsurl  = {https://ui.adsabs.harvard.edu/abs/1958ZA.....46..108B}
}

@ARTICLE{Buzzoni_2010,
  author  = {{Buzzoni}, A. and {Patelli}, L. and {Bellazzini}, M.
             and {Fusi Pecci}, F. and {Oliva}, E.},
  title   = "{Bolometric correction and spectral energy distribution of
             cool stars in Galactic clusters}",
  journal = {\mnras},
  year    = 2010,
  volume  = 403,
  pages   = {1592--1610},
  doi     = {10.1111/j.1365-2966.2009.16223.x}
}

@ARTICLE{Saito_2024,
  author  = {{Saito}, R.~K. and {Hempel}, M. and {Alonso-Garc{\'\i}a}, J. and {Lucas}, P.~W. and {Minniti}, D. and others},
  title   = "{The VISTA Variables in the V{\'\i}a L{\'a}ctea extended (VVVX) ESO public survey: Completion of the observations and legacy}",
  journal = {\aap}, year = 2024, volume = {689}, pages = {A148},
  doi     = {10.1051/0004-6361/202450584}
}

@ARTICLE{Blum_2003,
  author    = {{Blum}, R.~D. and {Ram{\'i}rez}, S.~V. and {Sellgren}, K.
               and {Olsen}, K.},
  title     = "{Really Cool Stars and the Star Formation History at the
               Galactic Center}",
  journal   = {\apj},
  year      = 2003,
  volume    = {597},
  pages     = {323--346},
  doi       = {10.1086/378380},
  eprint    = {astro-ph/0307291},
  archivePrefix = {arXiv},
  primaryClass  = {astro-ph},
  adsurl    = {https://ui.adsabs.harvard.edu/abs/2003ApJ...597..323B}
}

@ARTICLE{Matsunaga_2009,
  author    = {{Matsunaga}, N. and {Kawadu}, T. and {Nishiyama}, S.
               and {Nagayama}, T. and {Hatano}, H. and {Tamura}, M.
               and {Glass}, I.~S. and {Nagata}, T.},
  title     = "{A near-infrared survey of Miras and the distance to the
               Galactic Centre}",
  journal   = {\mnras},
  year      = 2009,
  volume    = {399},
  pages     = {1709--1729},
  doi       = {10.1111/j.1365-2966.2009.15393.x},
  eprint    = {0907.2761},
  archivePrefix = {arXiv},
  primaryClass  = {astro-ph.GA},
  adsurl    = {https://ui.adsabs.harvard.edu/abs/2009MNRAS.399.1709M}
}

@ARTICLE{Mosser_2014_delta_pi,
  author  = {{Mosser}, B. and {Benomar}, O. and {Belkacem}, K.
             and {Goupil}, M.~J. and {Lagarde}, N. and {Michel}, E.
             and {Lebreton}, Y. and {Stello}, D. and {Vrard}, M.
             and {Barban}, C. and {Bedding}, T.~R. and others},
  title   = "{Mixed modes in red giants: A window on stellar evolution}",
  journal = {\aap},
  year    = 2014,
  volume  = {572},
  eid     = {L5},
  pages   = {L5},
  doi     = {10.1051/0004-6361/201425039},
  eprint  = {1411.1082},
  archivePrefix = {arXiv},
  primaryClass  = {astro-ph.SR},
  adsurl  = {https://ui.adsabs.harvard.edu/abs/2014A%26A...572L...5M}
}

\newpage
\begin{appendix}
\nolinenumbers
\section{Calibrated parameters for the \citet{Rich_2017} variants}
\label{appendix:mcmc_rich}
As described in Sect.~\ref{sec:high_res}, each star with a
counterpart in the \citet{Rich_2017} catalogue was also
calibrated independently using that survey's photometric
($_{\rm phot}$) and isochrone-based ($_{\rm iso}$) parameter
sets. Table~\ref{tab:mcmc_results_rich} lists the resulting
posterior medians and $1\sigma$ uncertainties for
$\log\mathrm{Age}$, $[\mathrm{M/H}]$, and
$M_{\rm ini}$. As discussed in Sect.~\ref{sec:good_fits},
the resulting tracks are systematically displaced towards
warmer effective temperatures relative to the primary-catalogue
calibrations, a displacement in the same direction as that seen for
Feld31 and Feld84 (Sect.~\ref{sec:bad_fits}), though,
as discussed there, this catalogue-specific bias does not
apply to those two stars, which are measured on the
\citet{Ryde_2025} spectroscopic scale.
\begin{table}
\small
\centering
\caption{Calibrated stellar parameters for the \citet{Rich_2017} parameter variants.}
\label{tab:mcmc_results_rich}
\begin{tabular}{lccc}
\toprule
Star & log Age & [M/H] & $M_\mathrm{ini}$ ($M_\odot$)
\\
\midrule
GC6227$_{\text{phot}}^{(4)}$ & \unc{9.780}{0.237}{0.299} & \unc{-0.620}{0.136}{0.140} & \unc{1.064}{0.253}{0.158}
\\
GC6227$_{\text{iso}}^{(4)}$ & \unc{9.637}{0.307}{0.254} & \unc{-0.392}{0.137}{0.136} & \unc{1.212}{0.262}{0.226}
\\
GC8631$_{\text{phot}}^{(4)}$ & \unc{9.473}{0.268}{0.165} & \unc{+0.050}{0.136}{0.137} & \unc{1.493}{0.201}{0.276}
\\
GC8631$_{\text{iso}}^{(4)}$ & \unc{9.349}{0.140}{0.071} & \unc{+0.200}{0.126}{0.126} & \unc{1.676}{0.094}{0.168}
\\
GC10195$_{\text{phot}}^{(4)}$ & \unc{9.733}{0.236}{0.185} & \unc{-0.091}{0.136}{0.128} & \unc{1.201}{0.177}{0.176}
\\
GC10195$_{\text{iso}}^{(4)}$ & \unc{9.621}{0.231}{0.129} & \unc{-0.016}{0.134}{0.133} & \unc{1.314}{0.152}{0.193}
\\
GC11025$_{\text{phot}}^{(4)}$ & \unc{9.671}{0.235}{0.150} & \unc{+0.253}{0.134}{0.134} & \unc{1.325}{0.166}{0.192}
\\
GC11025$_{\text{iso}}^{(4)}$ & \unc{9.612}{0.217}{0.125} & \unc{+0.203}{0.133}{0.133} & \unc{1.374}{0.147}{0.194}
\\
GC11532$_{\text{phot}}^{(4)}$ & \unc{9.884}{0.168}{0.250} & \unc{-0.436}{0.135}{0.131} & \unc{1.022}{0.190}{0.109}
\\
GC11532$_{\text{iso}}^{(4)}$ & \unc{9.895}{0.160}{0.250} & \unc{-0.501}{0.130}{0.130} & \unc{1.004}{0.187}{0.104}
\\
GC13282$_{\text{phot}}^{(4)}$ & \unc{9.632}{0.295}{0.223} & \unc{-0.178}{0.137}{0.136} & \unc{1.265}{0.243}{0.227}
\\
GC13282$_{\text{iso}}^{(4)}$ & \unc{9.503}{0.255}{0.155} & \unc{-0.111}{0.137}{0.137} & \unc{1.411}{0.185}{0.241}
\\
GC13882$_{\text{phot}}^{(4)}$ & \unc{9.947}{0.123}{0.211} & \unc{-0.209}{0.117}{0.114} & \unc{1.024}{0.160}{0.085}
\\
GC13882$_{\text{iso}}^{(4)}$ & \unc{9.947}{0.125}{0.213} & \unc{-0.214}{0.118}{0.113} & \unc{1.023}{0.161}{0.087}
\\
GC14024$_{\text{phot}}^{(4)}$ & \unc{9.594}{0.288}{0.187} & \unc{-0.059}{0.136}{0.138} & \unc{1.329}{0.215}{0.236}
\\
GC14024$_{\text{iso}}^{(4)}$ & \unc{9.446}{0.159}{0.088} & \unc{+0.280}{0.131}{0.129} & \unc{1.575}{0.106}{0.182}
\\
GC14724$_{\text{phot}}^{(4)}$ & \unc{9.863}{0.183}{0.247} & \unc{-0.432}{0.125}{0.119} & \unc{1.036}{0.194}{0.121}
\\
GC14724$_{\text{iso}}^{(4)}$ & \unc{9.796}{0.215}{0.225} & \unc{-0.455}{0.126}{0.120} & \unc{1.080}{0.187}{0.149}
\\
GC15540$_{\text{phot}}^{(4)}$ & \unc{9.732}{0.236}{0.197} & \unc{-0.236}{0.132}{0.126} & \unc{1.170}{0.184}{0.176}
\\
GC15540$_{\text{iso}}^{(4)}$ & \unc{9.627}{0.236}{0.141} & \unc{-0.203}{0.130}{0.128} & \unc{1.263}{0.157}{0.191}
\\
GC16763$_{\text{phot}}^{(4)}$ & \unc{9.862}{0.185}{0.292} & \unc{-0.238}{0.138}{0.139} & \unc{1.076}{0.239}{0.125}
\\
GC16763$_{\text{iso}}^{(4)}$ & \unc{9.761}{0.246}{0.278} & \unc{-0.281}{0.140}{0.139} & \unc{1.138}{0.257}{0.171}
\\
GC16867$_{\text{phot}}^{(4)}$ & \unc{9.912}{0.152}{0.269} & \unc{-0.282}{0.135}{0.136} & \unc{1.034}{0.202}{0.101}
\\
GC16867$_{\text{iso}}^{(4)}$ & \unc{9.892}{0.165}{0.281} & \unc{-0.246}{0.138}{0.139} & \unc{1.053}{0.222}{0.110}
\\
GC16887$_{\text{phot}}^{(4)}$ & \unc{9.772}{0.243}{0.264} & \unc{+0.227}{0.142}{0.139} & \unc{1.230}{0.268}{0.175}
\\
GC16887$_{\text{iso}}^{(4)}$ & \unc{9.658}{0.278}{0.229} & \unc{+0.379}{0.137}{0.139} & \unc{1.351}{0.255}{0.228}
\\
GC16895$_{\text{phot}}^{(4)}$ & \unc{9.871}{0.180}{0.282} & \unc{-0.153}{0.136}{0.134} & \unc{1.088}{0.230}{0.123}
\\
GC16895$_{\text{iso}}^{(4)}$ & \unc{9.780}{0.233}{0.266} & \unc{-0.095}{0.139}{0.135} & \unc{1.163}{0.246}{0.165}
\\
\bottomrule
\end{tabular}
\tablefoot{The calibration was applied to the \citet{Rich_2017} photometric ($_{\rm phot}$) and isochrone-based ($_{\rm iso}$) parameter sets (source index $^{(4)}$), for the stars also present in the primary \citet{Thorsbro_2020} and \citet{Ryde_2025} samples (Table~\ref{tab:mcmc_results}). As discussed in Sect.~\ref{sec:high_res}, these
calibrations are not used for the
$\alpha_{\rm MLT}^*$ optimisation or the asteroseismic
predictions of Table~\ref{tab:astero_combined}; they
are included here as a secondary diagnostic
(Sect.~\ref{sec:good_fits}).
Columns and uncertainties follow
Table~\ref{tab:mcmc_results}.}
\end{table}
\section{Calibrated and best-fit $\alpha_{\rm MLT}^*$ tracks for all stars}
\label{appendix:hr_alpha_all}
Figure~\ref{fig:hr_all_combined} presents the calibrated
MESA tracks for the 20 remaining physical stars not shown in Fig.~\ref{fig:hr_wellcalibrated}, in the order in which they first
appear in Table~\ref{tab:mcmc_results}.
For the three physical stars observed in both
\citet{Thorsbro_2020} and \citet{Ryde_2025} (GC13727, GC15540/FK15540, and GC16895/FK16895 in Table~\ref{tab:mcmc_results}), only
the higher-priority \citet{Ryde_2025} measurement is
shown, following the priority order Ryde $>$ Thorsbro
$>$ Do $>$ Rich adopted throughout this work.
For Feld31 and GC8631, the full evolutionary
track is shown together with an inset zooming on the
observed position.
\begin{figure*}[h!]
    \centering
    \includegraphics[width=\textwidth]
        {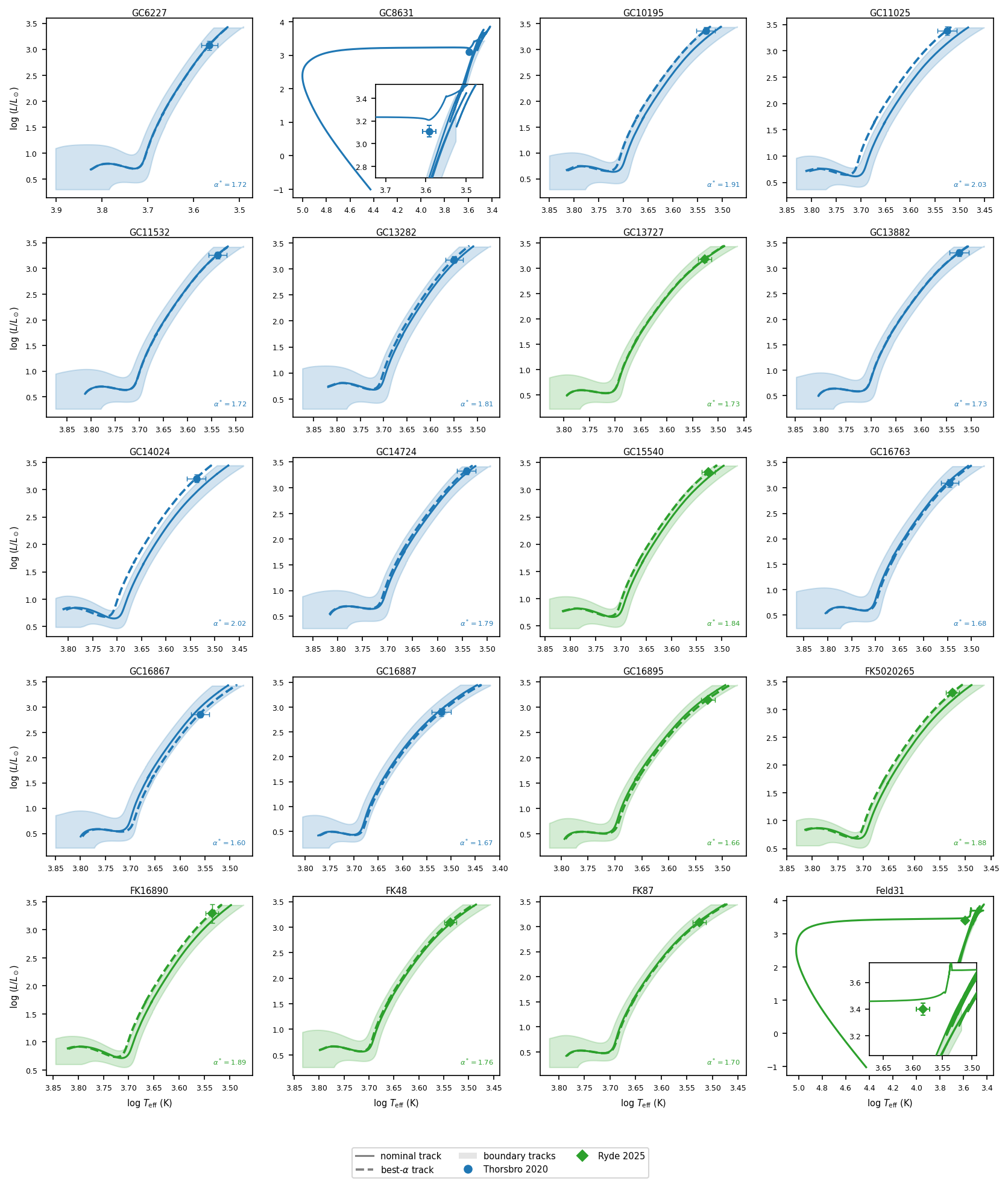}
    \caption{HR diagrams for the remaining 20
             high-resolution physical stars, in the order
             they appear in Table~\ref{tab:mcmc_results}.
             Solid curves show the nominal MESA track;
             dashed curves show the best-fit
             $\alpha_{\rm MLT}^*$ track (not shown for
             Feld31 and GC8631, whose full
             evolutionary track is shown instead, with an
             inset zooming on the observed position).
             Shaded envelopes span the boundary tracks.
             Marker colour and shape indicate the source
             catalogue.}
    \label{fig:hr_all_combined}
\end{figure*}
\section{MCMC posterior distributions}
\label{appendix:corners}
Figure~\ref{fig:corners} shows corner plots of the MCMC
posterior distributions for the four stars
discussed in Sect.~\ref{sec:bad_fits} (GC8631, Feld31, Feld84, and GC8623, in the same arrangement as in Figs.~\ref{fig:hr_problematic} and~\ref{fig:summary_6stars_3x2}).
For these four stars, the posteriors remain
well-converged and unimodal despite the mismatch with
the nominal RGB track discussed in
Sect.~\ref{sec:bad_fits}, supporting the robustness of
the calibrated $(M_{\rm ini},\, Z_{\rm ini})$ values
independently of the mixing-length or evolutionary-phase
diagnostic applied afterwards.
Corner plots for all 27 catalogue entries are available upon
request.

\begin{figure*}[h!]
    \centering
    \includegraphics[width=0.93\textwidth]
        {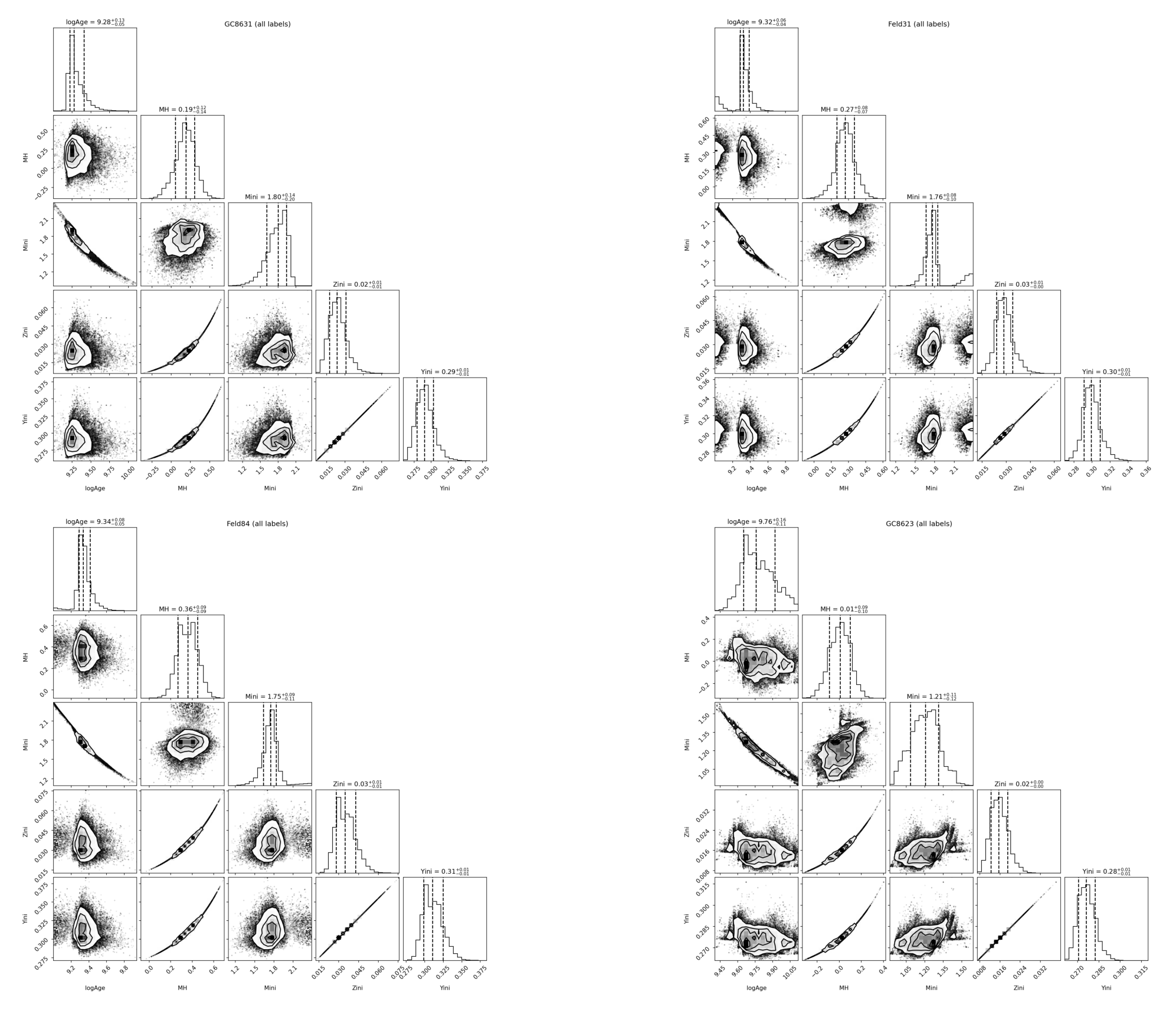}
    \caption{MCMC posterior distributions
             ($M_{\rm ini}$, $[\mathrm{M/H}]$, and
             derived quantities) for the four
             stars discussed in Sect.~\ref{sec:bad_fits}:
             GC8631, Feld31, Feld84, and GC8623.}
    \label{fig:corners}
\end{figure*}
\section{Best-fit mixing-length determination}
\label{appendix:alpha_chi2}
Figure~\ref{fig:summary_6stars_3x2} shows the minimum
$\chi^2_{\rm HR}$ as a function of $\alpha_{\rm MLT}$ on
the RGB for the four anomalous stars, together with the parabolic
fits used to determine $\alpha_{\rm MLT}^*$
(Eq.~(\ref{eq:alpha_star})).
For the well-calibrated stars (not shown), such as GC25 ($\alpha_{\rm MLT}^* = 1.69$), the curve has a well-defined minimum close to
the solar-calibrated value, confirming that the nominal
track already provides a good fit.
For Feld31, Feld84, and GC8631, the minimum is
displaced to a super-solar value
($\alpha_{\rm MLT}^* = 2.61$, $2.54$, and $2.33$,
respectively) with $\chi^2_{\rm HR} \lesssim 0.4$ at the
minimum, showing that a mixing-length adjustment on the
RGB fully resolves the mismatch for these three stars.
For GC8623, by contrast, the minimum of the fitted parabola is $\chi^2_{\rm HR} = 2.6$, just above the $1\sigma$ threshold, so a mixing-length adjustment on the RGB improves the fit without fully resolving the mismatch (Sect.~\ref{sec:bad_fits}).
\begin{figure*}[h!]
    \centering
    \includegraphics[width=\textwidth]{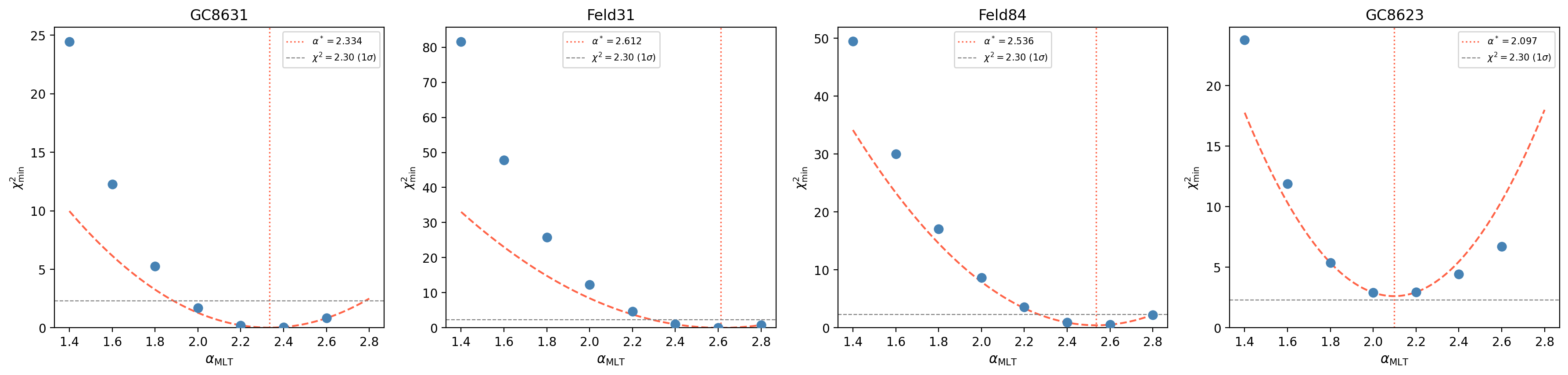}
    \caption{Minimum $\chi^2_{\rm HR}$ in
             $(\log T_{\rm eff},\, \log L)$ space as a
             function of $\alpha_{\rm MLT}$, evaluated on
             the RGB, for the four anomalous
             stars discussed in Sect.~\ref{sec:bad_fits}, in the same arrangement as in Fig.~\ref{fig:hr_problematic}.
             Filled circles show the values from the MESA
             grid runs; the dashed red curve is the
             parabolic fit used to determine
             $\alpha_{\rm MLT}^*$ (vertical dotted line).
             The horizontal dashed grey line marks
             $\chi^2_{\rm HR} = 2.30$, corresponding to a $1\sigma$ (68\%) match for two fitted quantities.}
    \label{fig:summary_6stars_3x2}
\end{figure*}

\section{MCMC calibration input parameters}
\label{appendix:mcmc_inputs}

Table~\ref{tab:mcmc_inputs} lists the full set of input
parameters used in the MCMC calibration for each star:
the spectroscopic observables adopted from the source
catalogues, the photometric $K_s$ magnitudes, the
$\log g$ uncertainties adopted in the calibration
(Sect.~\ref{sec:mcmc}), and the chain lengths and
initial walker positions used in \textsc{emcee}.
Stars are grouped by source catalogue.
Feld31, Feld84, GC8623, and GC8631 (marked $\dagger$ in
Table~\ref{tab:mcmc_results}) are
calibrated without restricting the PARSEC grid to
evolutionary label~$=3$, as described in
Sect.~\ref{sec:bad_fits}.

\begin{table*}
\centering
\caption{Input parameters of the MCMC calibration for the 27 primary high-resolution catalogue entries.}
\label{tab:mcmc_inputs}
\small
\begin{tabular}{lcccccccccrr}
\toprule
Star & $T_{\rm eff}$ & $\sigma_{T}$ & $[\rm Fe/H]$ &
$\sigma_{[\rm Fe/H]}$ & $\log g$ & $\sigma_{\log g}$ &
$K_s$ & $n_{\rm burn}$ & $n_{\rm steps}$ &
$\log\mathrm{Age}_0$ & $[\mathrm{M/H}]_0$ \\
\midrule
GC6227$^{(1)}$ & 3681 & 150 & -0.27 & 0.15 & 0.92 & 0.3 & 11.82 & 500 & 3000 & 9.8 & -0.27 \\
GC8623$^{(1)}$ & 3433 & 150 & -0.12 & 0.15 & 0.38 & 0.3 & 10.53 & 500 & 3000 & 9.8 & -0.12 \\
GC8631$^{(1)}$ & 3899 & 150 & +0.27 & 0.15 & 1.41 & 0.3 & 11.56 & 500 & 3000 & 9.8 & +0.27 \\
GC10195$^{(1)}$ & 3414 & 150 & -0.01 & 0.15 & 0.55 & 0.3 & 10.67 & 500 & 3000 & 9.8 & -0.01 \\
GC11025$^{(1)}$ & 3359 & 150 & +0.25 & 0.15 & 0.64 & 0.3 & 10.41 & 500 & 3000 & 9.8 & +0.25 \\
GC11532$^{(1)}$ & 3450 & 150 & -0.23 & 0.15 & 0.48 & 0.3 & 10.90 & 500 & 3000 & 9.8 & -0.23 \\
GC13282$^{(1)}$ & 3543 & 150 & -0.04 & 0.15 & 0.77 & 0.3 & 10.99 & 500 & 3000 & 9.8 & -0.04 \\
GC13727$^{(1)}$ & 3382 & 150 & -0.28 & 0.15 & 0.31 & 0.3 & 10.82 & 500 & 3000 & 9.8 & -0.28 \\
GC13882$^{(1)}$ & 3344 & 150 & -0.17 & 0.15 & 0.35 & 0.3 & 10.56 & 500 & 3000 & 9.8 & -0.17 \\
GC14024$^{(1)}$ & 3450 & 150 & +0.49 & 0.15 & 0.96 & 0.3 & 10.99 & 500 & 3000 & 9.8 & +0.49 \\
GC14724$^{(1)}$ & 3482 & 150 & -0.32 & 0.15 & 0.45 & 0.3 & 10.56 & 500 & 3000 & 9.8 & -0.32 \\
GC15540$^{(1)}$ & 3496 & 150 & +0.01 & 0.15 & 0.61 & 0.3 & 10.49 & 500 & 3000 & 9.8 & +0.01 \\
GC16763$^{(1)}$ & 3510 & 150 & -0.15 & 0.15 & 0.58 & 0.3 & 10.66 & 500 & 3000 & 9.8 & -0.15 \\
GC16867$^{(1)}$ & 3628 & 150 & -0.18 & 0.15 & 0.76 & 0.3 & 11.77 & 500 & 3000 & 9.8 & -0.18 \\
GC16887$^{(1)}$ & 3308 & 150 & +0.41 & 0.15 & 0.77 & 0.3 & 11.63 & 500 & 3000 & 9.8 & +0.41 \\
GC16895$^{(1)}$ & 3438 & 150 & -0.13 & 0.15 & 0.50 & 0.3 & 10.75 & 500 & 3000 & 9.8 & -0.13 \\
GC25$^{(1)}$ & 3383 & 150 & -0.08 & 0.15 & 0.38 & 0.3 & 10.44 & 500 & 3000 & 9.6 & -0.08 \\
FK5020265$^{(2)}$ & 3350 & 100 & +0.26 & 0.1 & 0.60 & 0.2 & 9.91 & 500 & 3000 & 9.8 & +0.26 \\
FK15540$^{(2)}$ & 3361 & 100 & +0.11 & 0.1 & 0.50 & 0.2 & 10.49 & 500 & 3000 & 9.8 & +0.11 \\
FK16890$^{(2)}$ & 3423 & 100 & +0.19 & 0.1 & 0.70 & 0.2 & 10.72 & 500 & 3000 & 9.8 & +0.19 \\
FK16895$^{(2)}$ & 3366 & 100 & -0.12 & 0.1 & 0.40 & 0.2 & 10.75 & 500 & 3000 & 9.8 & -0.12 \\
FK48$^{(2)}$ & 3440 & 100 & +0.12 & 0.1 & 0.60 & 0.2 & 10.76 & 500 & 3000 & 9.5 & +0.12 \\
FK87$^{(2)}$ & 3352 & 100 & +0.09 & 0.1 & 0.50 & 0.2 & 10.75 & 500 & 3000 & 9.6 & +0.09 \\
Feld31$^{(2)}$ & 3825 & 100 & +0.42 & 0.1 & 1.50 & 0.8 & 10.61 & 500 & 3000 & 9.2 & +0.42 \\
Feld84$^{(2)}$ & 3709 & 100 & +0.48 & 0.1 & 1.40 & 0.8 & 10.66 & 500 & 3000 & 9.2 & +0.48 \\
GC13727$^{(2)}$ & 3356 & 100 & +0.00 & 0.1 & 0.40 & 0.5 & 10.82 & 500 & 3000 & 9.8 & +0.00 \\
NE1-1\,001$^{(3)}$ & 3029 & 200 & +0.14 & 0.2 & 1.00 & 0.5 & 10.40 & 500 & 3000 & 9.9 & +0.14 \\
\bottomrule
\end{tabular}
\tablefoot{Spectroscopic and photometric inputs, together with the MCMC chain lengths and initial walker positions. Input parameters for the \citet{Rich_2017} photometric
and isochrone-based variants (secondary diagnostic,
Sect.~\ref{sec:high_res}) are not repeated here.
The surface gravity uncertainty $\sigma_{\log g}$ was set
individually per star: $\sigma_{\log g} = 0.8$ for
Feld31 and Feld84 (highly uncertain K-band $\log g$ at
high metallicity), $\sigma_{\log g} = 0.5$ for the \citet{Ryde_2025} entry of GC13727 and for NE1-1\,001, $\sigma_{\log g} = 0.3$ for the
\citet{Thorsbro_2020} sample, and $\sigma_{\log g} = 0.2$
for the \citet{Ryde_2025} sample
\citep{FK_2017, Do_2018}.
Source indices: $^{(1)}$~\citet{Thorsbro_2020},
$^{(2)}$~\citet{Ryde_2025}, $^{(3)}$~\citet{Do_2018}.}
\end{table*}

\section{Isochrone interpolation scheme}
\label{appendix:interpolation}
Since the PARSEC grid only exists at discrete
$(\log \mathrm{Age},\, [\mathrm{M/H}])$ values, but the
MCMC sampler proposes continuous values, we pre-built a
dictionary of RGB tracks indexed by
$(\log \mathrm{Age},\, [\mathrm{M/H}])$ to enable
efficient interpolation during sampling.
Each track was stored as a set of 1D linear interpolating
functions in $\log g$, evaluated over a uniform grid of
200 points spanning the common $\log g$ range of the
four nearest grid corners.
At arbitrary $(\log \mathrm{Age},\, [\mathrm{M/H}])$
values, any track quantity was then recovered by bilinear
interpolation between those four corners.
This scheme avoids repeated disc access to the PARSEC
grid files during sampling while preserving the full
resolution of the downloaded isochrones.

\section{Comparison between asymptotic and GYRE $\Delta\nu$}
\label{appendix:gyre_comparison}

Table~\ref{tab:gyre_comparison} lists, for all 52 models
attempted, the asymptotic $\Delta\nu$ computed by MESA alongside the value
computed directly from the GYRE mode frequencies, as
described in Sect.~\ref{sec:mesa_models}.
One model (the AGB phase of GC8623) returned too few modes for a robust linear fit
and is excluded from the summary statistics quoted in
the caption and in Sect.~\ref{sec:results_astero}, though
its individual value is still listed for
completeness.

\begin{table*}
\centering
\caption{Comparison between the asymptotic $\Delta\nu$ computed by MESA and the $\Delta\nu$ obtained from GYRE.}
\label{tab:gyre_comparison}
\begin{tabular}{llccccc}
\toprule
Star & Model & $N_{\rm modes}$ & $\Delta\nu_{\rm asymp}$ ($\mu$Hz) & $\Delta\nu_{\rm GYRE}$ ($\mu$Hz) & Difference (\%) \\
\midrule
    GC25$^{(1)}$ & nominal & 11 & 0.1536 & 0.1506 & +2.0 \\
    GC25$^{(1)}$ & best-fit $\alpha_{\rm MLT}$ & 11 & 0.1512 & 0.1484 & +1.9 \\
    GC6227$^{(1)}$ & nominal & 13 & 0.2329 & 0.2270 & +2.6 \\
    GC6227$^{(1)}$ & best-fit $\alpha_{\rm MLT}$ & 13 & 0.2329 & 0.2270 & +2.6 \\
    GC10195$^{(1)}$ & nominal & 10 & 0.1099 & 0.1085 & +1.2 \\
    GC10195$^{(1)}$ & best-fit $\alpha_{\rm MLT}$ & 10 & 0.1178 & 0.1156 & +1.9 \\
    GC11025$^{(1)}$ & nominal & 11 & 0.1047 & 0.1037 & +0.9 \\
    GC11025$^{(1)}$ & best-fit $\alpha_{\rm MLT}$ & 11 & 0.1124 & 0.1108 & +1.5 \\
    GC11532$^{(1)}$ & nominal & 11 & 0.1405 & 0.1385 & +1.5 \\
    GC11532$^{(1)}$ & best-fit $\alpha_{\rm MLT}$ & 11 & 0.1403 & 0.1382 & +1.5 \\
    GC13282$^{(1)}$ & nominal & 13 & 0.1748 & 0.1718 & +1.7 \\
    GC13282$^{(1)}$ & best-fit $\alpha_{\rm MLT}$ & 13 & 0.1800 & 0.1766 & +1.9 \\
    GC13727$^{(1)}$ & nominal & 10 & 0.1286 & 0.1268 & +1.4 \\
    GC13727$^{(1)}$ & best-fit $\alpha_{\rm MLT}$ & 10 & 0.1351 & 0.1330 & +1.5 \\
    GC13882$^{(1)}$ & nominal & 10 & 0.1176 & 0.1162 & +1.2 \\
    GC13882$^{(1)}$ & best-fit $\alpha_{\rm MLT}$ & 10 & 0.1179 & 0.1165 & +1.2 \\
    GC14024$^{(1)}$ & nominal & 13 & 0.1527 & 0.1502 & +1.7 \\
    GC14024$^{(1)}$ & best-fit $\alpha_{\rm MLT}$ & 14 & 0.1675 & 0.1644 & +1.9 \\
    GC14724$^{(1)}$ & nominal & 10 & 0.1261 & 0.1243 & +1.4 \\
    GC14724$^{(1)}$ & best-fit $\alpha_{\rm MLT}$ & 10 & 0.1293 & 0.1272 & +1.7 \\
    GC15540$^{(1)}$ & nominal & 11 & 0.1187 & 0.1173 & +1.2 \\
    GC15540$^{(1)}$ & best-fit $\alpha_{\rm MLT}$ & 11 & 0.1295 & 0.1271 & +1.9 \\
    GC16763$^{(1)}$ & nominal & 12 & 0.1946 & 0.1901 & +2.4 \\
    GC16763$^{(1)}$ & best-fit $\alpha_{\rm MLT}$ & 12 & 0.1920 & 0.1878 & +2.2 \\
    GC16867$^{(1)}$ & nominal & 15 & 0.3230 & 0.3152 & +2.5 \\
    GC16867$^{(1)}$ & best-fit $\alpha_{\rm MLT}$ & 15 & 0.3064 & 0.2997 & +2.2 \\
    GC16887$^{(1)}$ & nominal & 14 & 0.2258 & 0.2209 & +2.2 \\
    GC16887$^{(1)}$ & best-fit $\alpha_{\rm MLT}$ & 14 & 0.2220 & 0.2174 & +2.1 \\
    GC16895$^{(1)}$ & nominal & 12 & 0.1630 & 0.1603 & +1.7 \\
    GC16895$^{(1)}$ & best-fit $\alpha_{\rm MLT}$ & 12 & 0.1603 & 0.1579 & +1.5 \\
    FK5020265$^{(2)}$ & nominal & 12 & 0.1254 & 0.1238 & +1.3 \\
    FK5020265$^{(2)}$ & best-fit $\alpha_{\rm MLT}$ & 12 & 0.1301 & 0.1281 & +1.6 \\
    FK15540$^{(2)}$ & nominal & 12 & 0.1333 & 0.1316 & +1.3 \\
    FK15540$^{(2)}$ & best-fit $\alpha_{\rm MLT}$ & 11 & 0.1207 & 0.1189 & +1.5 \\
    FK16890$^{(2)}$ & nominal & 13 & 0.1609 & 0.1580 & +1.8 \\
    FK16890$^{(2)}$ & best-fit $\alpha_{\rm MLT}$ & 12 & 0.1407 & 0.1380 & +1.9 \\
    FK16895$^{(2)}$ & nominal & 10 & 0.1320 & 0.1298 & +1.7 \\
    FK16895$^{(2)}$ & best-fit $\alpha_{\rm MLT}$ & 11 & 0.1392 & 0.1375 & +1.2 \\
    FK48$^{(2)}$ & nominal & 13 & 0.1834 & 0.1800 & +1.9 \\
    FK48$^{(2)}$ & best-fit $\alpha_{\rm MLT}$ & 13 & 0.1865 & 0.1828 & +2.0 \\
    FK87$^{(2)}$ & nominal & 12 & 0.1708 & 0.1677 & +1.9 \\
    FK87$^{(2)}$ & best-fit $\alpha_{\rm MLT}$ & 12 & 0.1690 & 0.1660 & +1.8 \\
    GC13727$^{(2)}$ & nominal & 11 & 0.1444 & 0.1420 & +1.7 \\
    GC13727$^{(2)}$ & best-fit $\alpha_{\rm MLT}$ & 11 & 0.1447 & 0.1422 & +1.7 \\
    GC8623$^{(1)}$ & RGB, best-fit $\alpha_{\rm MLT}$ & 10 & 0.1051 & 0.1037 & +1.4 \\
    GC8623$^{(1)}$ & AGB, nominal & 5 & 0.0868 & 0.0931 & -6.8 \\
    GC8631$^{(1)}$ & RGB, best-fit $\alpha_{\rm MLT}$ & 16 & 0.2829 & 0.2755 & +2.7 \\
    GC8631$^{(1)}$ & AGB, nominal & 15 & 0.2337 & 0.2284 & +2.3 \\
    Feld31$^{(2)\dagger}$ & RGB, best-fit $\alpha_{\rm MLT}$ & 14 & 0.1759 & 0.1716 & +2.5 \\
    Feld31$^{(2)\dagger}$ & AGB, nominal & 14 & 0.1439 & 0.1416 & +1.6 \\
    Feld84$^{(2)\dagger}$ & RGB, best-fit $\alpha_{\rm MLT}$ & 14 & 0.1546 & 0.1514 & +2.1 \\
    Feld84$^{(2)\dagger}$ & AGB, nominal & 14 & 0.1634 & 0.1602 & +2.0 \\
\bottomrule
\end{tabular}
\tablefoot{The GYRE values are obtained directly from oscillation-mode calculations \citep{Townsend_2013}, for all 52 models
attempted (22 well-calibrated entries $\times$ nominal and
best-fit $\alpha_{\rm MLT}$, plus the 4 stars discussed
in Sect.~\ref{sec:bad_fits} $\times$ RGB and AGB).
The GYRE value is the slope of a linear fit of mode
frequency against radial order $n_{\rm pg}$.
For the AGB model of GC8623
($N_{\rm modes} = 5$), too few
modes were available for a reliable fit; this
entry is retained for completeness but excluded from
the statistics below.
Excluding it, the asymptotic relation overestimates $\Delta\nu$ by +1.8\%\ on average (standard deviation 0.4\%, median +1.7\%).
Source indices: $^{(1)}$~\citet{Thorsbro_2020},
$^{(2)}$~\citet{Ryde_2025}.}
\end{table*}

\end{appendix}

\end{document}